\documentclass[11pt]{article}
\pdfoutput=1
\usepackage{jheppub}
\usepackage{bm,bbm}
\usepackage{amsmath}
\edef\restoreparindent{\parindent=\the\parindent\relax}
\usepackage{parskip}
\restoreparindent
\usepackage{ascmac}
\usepackage{tikz}
\usetikzlibrary{intersections, calc,positioning,decorations.pathreplacing,decorations.pathmorphing}
\tikzset{>=latex}

\def\d{{\rm d}}

\def\i{{\rm i}}

\def\rD{{\rm{D}}}

\def\rL{{\rm{L}}}

\def\CB{{\cal B}}

\def\CD{{\cal D}}

\def\CM{{\cal M}}

\def\CO{{\cal O}}

\def\CT{{\cal T}}

\def\BB{\mathbbm{B}}
\def\Bb{\mathbbm{b}}

\def\bL{\mathbb{L}}

\def\BH{\mathbb{H}}

\def\BR{\mathbb{R}}

\def\BZ{\mathbb{Z}}

\def\b0{\bm{0}_\perp}
\def\bx{\bm{x}}
\def\bL{\bold{L}}

\def\SO{\mathrm{SO}}

\DeclareFontFamily{U}{mathx}{\hyphenchar\font45}
\DeclareFontShape{U}{mathx}{m}{n}{
      <5> <6> <7> <8> <9> <10>
      <10.95> <12> <14.4> <17.28> <20.74> <24.88>
      mathx10
      }{}
\DeclareSymbolFont{mathx}{U}{mathx}{m}{n}
\DeclareFontSubstitution{U}{mathx}{m}{n}
\DeclareMathAccent{\widecheck}{0}{mathx}{"71}
\title{Regge OPE blocks and light-ray operators}

\author[a,b]{Nozomu Kobayashi,}
\author[a]{Tatsuma Nishioka}
\author[a]{and Yoshitaka Okuyama}

\affiliation[a]{Department of Physics, Faculty of Science,
The University of Tokyo,\\
Bunkyo-Ku, Tokyo 113-0033, Japan}
\affiliation[b]{Kavli Institute for the Physics and Mathematics of the Universe (WPI), \\
The University of Tokyo Institutes for Advanced Study, The University of Tokyo, \\
Kashiwa, Chiba 277-8583, Japan}

\abstract{
We consider the structure of the operator product expansion (OPE) in conformal field theory by employing the OPE block formalism.
The OPE block acted on the vacuum is promoted to an operator and its implications are examined on a non-vacuum state.
We demonstrate that the OPE block is dominated by a light-ray operator in the Regge limit, which reproduces precisely the Regge behavior of conformal blocks when used inside scalar four-point functions.
Motivated by this observation, we propose a new form of the OPE block, called the light-ray channel OPE block that has a well-behaved expansion dominated by a light-ray operator in the Regge limit.
We also show that the two OPE blocks have the same asymptotic form in the Regge limit and confirm the assertion that the Regge limit of a pair of spacelike-separated operators in a Minkowski patch is equivalent to the OPE limit of a pair of timelike-separated operators associated with the original pair in a different Minkowski patch.
}

\preprint{IPMU20-0055}

\begin{document}
\maketitle

\section{Introduction}

An operator product expansion (OPE) is one of the most fundamental postulates in local quantum field theories \cite{Wilson:1969zs,Wilson:1972ee}.
For a given pair of operators $\CO_i(x_1)$ and $\CO_j(x_2)$ their operator product may be expanded into the form:
\begin{align}\label{OPE_intro}
    \CO_i(x_1)\, \CO_j(x_2) 
        = \sum_k\,
            \CB_k^{ij} (x_1, x_2) \ ,
\end{align}
where $k$ labels a complete set of operators.
Let $\Delta_i$ be the dimension of an operators $\CO_i$, then the OPE asserts that if theories have well-behaved ultraviolet structures the bi-local function $\CB_k^{ij} (x_1, x_2)$ has an asymptotic expansion in the short distant limit:
\begin{align}
    \CB_k^{ij} (x_1, x_2) \underset{x_1 \to x_2}{\sim} \frac{1}{|x_{12}|^{\Delta_i + \Delta_j - \Delta_k}}\, \CO_k(x_2) + \cdots \ .
\end{align}
The expansion may involve an infinite number of composite operators associated with the operator product \eqref{OPE_intro}, preventing us from determining less divergent terms in practice.

The structure of the bi-local function $\CB_k^{ij}(x_1, x_2)$ may be constrained by the symmetries of any kind in the theory.
In conformal field theory (CFT), the complete set of operators is given by conformal primary fields $\CO_{\Delta, J}$ with conformal dimension $\Delta$ and spin $J$, and it is sufficient to consider the OPE \eqref{OPE_intro} for $i,j,k$ primary operators (see \cite{Poland:2018epd} for a review).
The bi-local function $\CB_k^{ij}(x_1, x_2)$ is referred to as \emph{the OPE block} in CFT.
The behavior of the OPE block is determined considerably by conformal symmetry to all orders, and the OPE can be shown to converge on the vacuum $|\Omega\rangle$ as an asymptotic expansion \cite{Mack:1976pa}.
The vacuum OPE block $\CB_k^{ij}(x_1, x_2)\,|\Omega\rangle$ with $i,j$ scalar primary used to constitute an integral part of the studies in CFT a long time ago \cite{Ferrara:1971vh,Ferrara:1972ay,Ferrara:1972uq,Dobrev:1975ru,Dobrev:1977qv} with a view to bootstrapping higher-point correlation functions from three-point functions.
The non-perturbative form organizing all contributions from the conformal multiplet was given in an integral representation by means of the shadow formalism, which is instrumental in constructing conformal blocks while keeping the conformal invariance and the analytic structure manifest \cite{Dolan:2011dv,SimmonsDuffin:2012uy}.

The objective of this paper is to explore the structure of the OPE block that holds on general states away from the vacuum.
Compared to the vacuum case, the determination of the non-vacuum OPE block is more complicated, and has attracted less attention until now.
In local quantum field theories, any non-vacuum state can be well-approximated by a state created by acting a local operator on the vacuum due to the Reeh-Schlieder property \cite{Reeh:1961ujh}.
Hence in CFT it is enough to determine the OPE block acted on a primary state of the form, $\CB_{k}^{ij}(x_1, x_2)\,|\CO_l\rangle$.
One can proceed in the same way as the vacuum OPE block with using the shadow formalism to reduce the problem of fixing $\CB_{k}^{ij}(x_1, x_2)\,|\CO_l\rangle$ to a calculation involving four-point functions, which however requires more efforts than fixing the vacuum OPE block that can be fixed solely by three-point functions \cite{Schroer:1974de}.
Thus we do not follow this straightforward-looking but cumbersome strategy.

In this paper we attempt to promote the vacuum OPE block in CFT to an operator identity by invoking the operator-state correspondence and examine the structure and the validity on a non-vacuum state.
We will deal for simplicity with the OPE of two scalar primary operators $\CO_1(x_1)$ and $\CO_2(x_2)$ and assume the operator identity takes the form:
\begin{align}\label{OPE}
    \CO_1(x_1)\, \CO_2(x_2) 
        = \sum_{[\Delta, J]}\,
            \CB_{\Delta, J} (x_1, x_2) + \text{(non-vacuum part)}\ ,
\end{align}
where $\CB_{\Delta, J} (x_1, x_2)$ is the vacuum OPE block exchanging the operator $\CO_{\Delta, J}$ (we omit the superscript indicating the dependence on the two operators $\CO_1, \CO_2$).
The non-vacuum part is a possible contribution from operators that annihilate the vacuum but become non-vanishing on a non-vacuum state.
The existence of such an additional contribution is suggested by \cite{Kravchuk:2018htv}.
Meanwhile, we will be focused on the vacuum OPE blocks and their implications on a non-vacuum state, leaving the determination of the non-vacuum part for future studies.

We employ a new representation of the vacuum OPE block which has a geometric interpretation as an AdS propagating field smeared over the geodesic between the boundary points $x_1$ and $x_2$ in an AdS spacetime, initially obtained for a scalar channel in \cite{deBoer:2016pqk,Czech:2016xec,daCunha:2016crm} and generalized to any channel recently in \cite{Chen:2019fvi}.
We use the new representation inside four-point functions to see if it leads to the known behaviors of conformal blocks.

\begin{figure}[htbp]
    \centering
    \includegraphics[width=7cm]{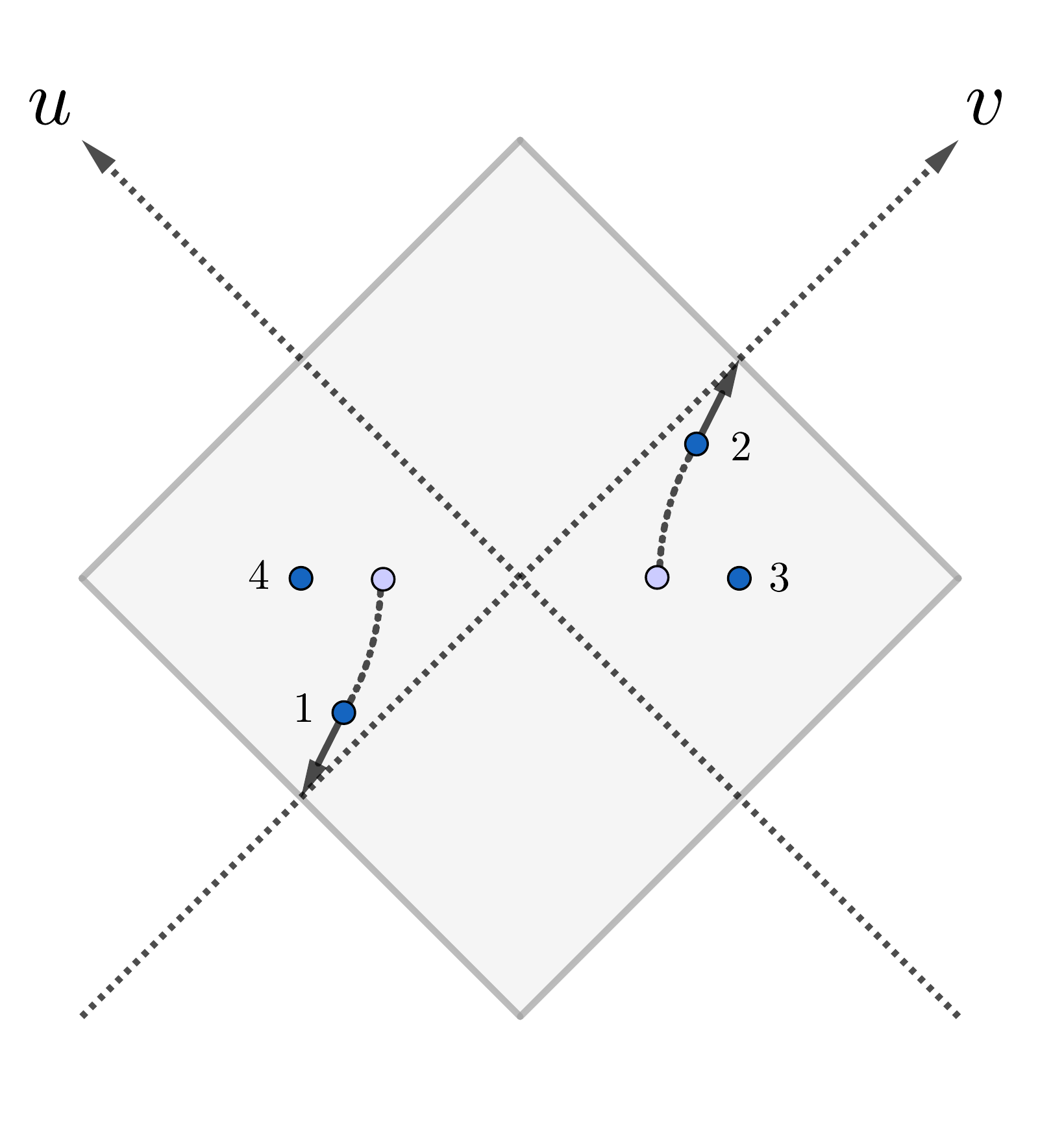}
    \caption{The Regge configuration of four points.}
    \label{fig:Regge1}
\end{figure}

Among four-point functions of various operator orderings, a particularly interesting one is the correlator
\begin{align}
    \langle\,\CO_4 (x_4)\, \CO_1(x_1)\, \CO_2(x_2)\,\CO_3(x_3)\,\rangle \ ,
\end{align}
which has a well-behaved OPE in the channel $\CO_1\times \CO_2$ when the four operators are all spacelike-separate, but exhibits a peculiar behavior in the so-called Regge limit where the pair of points $1$ and $4$ and the pair of points $2$ and $3$ become timelike-separated as in figure \ref{fig:Regge1}.
The latter behavior is dominated by an operator of unusual conformal dimension $1-J$ and spin $1-\Delta$ when the former is governed by an operator $\CO_{\Delta, J}$ as we will review in section \ref{sec:Lorentzian conformal block in Regge regime}.

One may wonder if a non-vacuum contribution in the OPE \eqref{OPE_intro} is responsible for the Regge behavior of the conformal block.
Indeed the Regge behavior is seen to be dominated by a non-local operator $\bL[\CO_{\Delta, J}]$ generated by acting on $\CO_{\Delta, J}$ with the light transform $\bL$ changing the quantum dimensions \cite{Kravchuk:2018htv}:
\begin{align}
    \bL: (\Delta, J) \to (1-J, 1-\Delta) \ .
\end{align}
Nevertheless we demonstrate in section \ref{sec:Regge conformal block via Lorentzian OPE} that the Regge behavior can be recovered precisely from the holographic representation of the vacuum OPE block.
Along the way we show the vacuum OPE block $\CB_{\Delta, J}$ approaches a light-ray operator $\bL[\CO_{\Delta, J}]$ in the Regge limit as is consistent with the observation above.

Our derivation closely follows the relevant works \cite{Afkhami-Jeddi:2017rmx,Hartman:2016lgu} where similar results were obtained in a slightly different way.
These works started with a pair of timelike-separated operators, took the Regge-like limit of the timelike OPE block $\CB_{\Delta,J}^\diamondsuit$ proposed by \cite{Czech:2016xec,deBoer:2016pqk}, which differs from the spacelike OPE block $\CB_{\Delta, J}$ of \cite{Chen:2019fvi} we use in this paper, and then analytically continued the result to the spacelike configuration.
To fill the gap between the two approaches, in section \ref{sec:timelike-block}, we compare $\CB_{\Delta,J}^\diamondsuit$ with another form of the timelike OPE block $\CB_{\Delta,J}^\text{T}$ obtained by analytically continuing the spacelike one $\CB_{\Delta, J}$.
We derive the precise relation between the two blocks $\CB_{\Delta,J}^\diamondsuit$ and $\CB_{\Delta,J}^\text{T}$ by matching their OPE limits and show that $\CB_{\Delta,J}^\text{T}$ also derives the Regge behavior of the conformal block, confirming the validity of the results in \cite{Afkhami-Jeddi:2017rmx,Hartman:2016lgu}.

The emergence of light-ray operators in the Regge limit was envisaged already in \cite{Caron-Huot:2013fea,Kravchuk:2018htv}, where the Regge limit of a pair of spacelike-separated operators in a Minkowski patch is interpreted as the ordinary OPE limit of a pair of timelike-separated operators associated to the original pair in a different Minkowski patch with a light-ray operator exchanged in the timelike OPE channel.
To make this statement in a concrete form, in section \ref{sec:Light-ray channel OPE block}, we propose a new OPE block $\BB_{\bL[\Delta, J]}$ named the light-ray channel OPE block that exchanges a light-ray operator in the OPE.
We prove the weak version of the statement that the spacelike OPE block $\CB_{\Delta, J}$ approaches $\BB_{\bL[\Delta, J]}$ in the Regge limit.

In section \ref{sec:Discussions} we discuss whether the statement can be formulated as an operator identity relating the spacelike OPE block to the light-ray channel OPE block.
We speculate a possible form of such an identity based on the consistency with the behavior of the conformal blocks of different operator orderings, leaving further analysis for future works.
Appendix \ref{sec:Notations and Normalization} summarizes our notations and the normalization of correlation functions in this paper.
The rest of the appendices contain some technical details skipped in the main text.

\section{Regge limit in CFT}
This section reviews background material on the Regge limit and conformal blocks in CFT.
After defining the Regge configuration for four-point functions in section \ref{ss:Regge_config} we introduce the adapted coordinates that are suitable for describing the Regge limit in section \ref{ss:Adapted_coord}.
The adapted coordinates are uplifted to the embedding space in section \ref{ss:adapted_embed}.
Four-point functions of various operator orderings and their Regge behaviors are explained in section \ref{sec:Lorentzian conformal block in Regge regime}.
Finally the Regge behavior is associated with light-ray operators in section \ref{ss:Regge_light-ray}.

\subsection{Regge configuration}\label{ss:Regge_config}
Consider the conformal block decomposition of a scalar four-point function with all operators spacelike-separated:
\begin{align}\label{4pt_CPW}
    \langle\,\CO_{1}(x_1)\,\CO_{2}(x_2)\,\CO_{3}(x_3)\,\CO_{4}(x_4)\,\rangle= T_{\{\Delta_i\}}(x_i)\, g(z,\bar{z}) \ .
\end{align}
Here $T_{\{\Delta_i\}}(x_i)$ transforms in the same way as the four-point function defined by
\begin{align}
   T_{\{\Delta_i\}}(x_i)=\frac{1}{(x_{12}^2)^\frac{\Delta_{12}^+}{2}\,(x_{34}^2)^\frac{\Delta_{34}^+}{2}}\,\left( \frac{x_{24}^2}{x_{14}^2}\right)^\frac{\Delta_{12}^-}{2}\,\left( \frac{x_{14}^2}{x_{13}^2}\right)^\frac{\Delta_{34}^-}{2}\ , \qquad \Delta^\pm_{i j}=\Delta_i\pm \Delta_j\ ,
\end{align}
while $g(z,\bar{z})$, which is conformally invariant, can be decomposed into the conformal blocks using the OPE of the products $\CO_{1}(x_1)\,\CO_{2}(x_2)$ and $\CO_{4}(x_4)\,\CO_{3}(x_3)$ (see figure \ref{fig:CPWE-normal}):
\begin{align}
    g(z,\bar{z}) &=\sum_{[\Delta, J]}\,c_{12,[\Delta,J]}\, c_{43,[\Delta,J]} \,G_{\Delta, J}(z, \bar z) \ .
\end{align}
Here $c_{12, [\Delta, J]}$ is the OPE coefficient associated with the three-point function $\langle\, \CO_1\,\CO_2\, \CO_{\Delta, J}\,\rangle$ as in \eqref{3-pt-phys-ast}, and the cross ratios are defined by
\begin{align}
        \mathfrak{u} = z\, \bar z = \frac{x_{12}^2\,x_{34}^2}{x_{13}^2\, x_{24}^2} \ , \qquad
        \mathfrak{v} = (1-z) (1- \bar z) = \frac{x_{14}^2\,x_{23}^2}{x_{13}^2\, x_{24}^2} \ .
\end{align}
When all operators are spacelike-separated, the conformal block $G_{\Delta,J}(z,\bar{z})$ is normalized to have the following asymptotic form:\footnote{This asymptotic form is determined by taking the OPE explicitly.
The overall factor is sensitive to the normalization of two-/three-point functions and the definition of the conformal blocks.
A nice summary of various normalizations used in literature is given in TABLE I of \cite{Poland:2018epd}.
Our convention and notations are relegated to appendix \ref{sec:Notations and Normalization}.}
\begin{align}\label{normalization-CPW}
   G_{\Delta,J}(z,\bar{z})~\to ~ 2^{-J}z^{\frac{\Delta-J}{2}}\bar{z}^{\frac{\Delta+J}{2}}\ , \qquad 0\ll z\ll\bar{z}\ll 1 \ .
\end{align}

\begin{figure}[htbp]
    \centering
    \tikzset{every picture/.style={line width=0.75pt}}
    \begin{tikzpicture}
        \draw (-2,-0.3) node [font=\LARGE]  {$\displaystyle g ( z,\bar{z}) = \sum_{[\Delta, J]}$};
    
        \draw (0.5, 1) node [above, font=\large] {$\CO_{1}$} -- (1.5, 0) -- (0.5, -1) node [below, font=\large] {$\CO_{2}$};
        \draw (1.5, 0) -- node [below, midway, font=\large] {$\CO_{\Delta, J}$} (3.5, 0);
        \draw (4.5, 1) node [above, font=\large] {$\CO_{4}$} -- (3.5, 0) -- (4.5, -1) node [below, font=\large] {$\CO_{3}$};
    \end{tikzpicture}
    \caption{A diagrammatic description of the conformal block decomposition. A local primary operators $\CO_{\Delta,J}$ appears in the intermediate state.}
    \label{fig:CPWE-normal}
\end{figure}

It will be convenient to introduce the lightcone coordinates by
\begin{align}\label{LC_coord}
   x^\mu=(u,v,\bm{x}_{\perp}) \ , \qquad u=t-x^1 \ , \qquad v = t + x^1  \ ,\qquad \bm{x}_{\perp} \in \BR^{d-2} \ , 
\end{align}
where $t$ is the Lorentzian time and $x^1$ is a spacial coordinate.
Using conformal symmetry we can locate the four points on the same two-dimensional subspace $\BR^{1,1} \in \BR^{1,d-1}$.
Without loss of generality the four points can be arranged as follows \cite{Cornalba:2007fs,Kravchuk:2018htv}:
\begin{align}\label{Regge_parametrization}
    \begin{aligned}
        x_1 &= (\rho, -\bar \rho, \b0) \ , \\
        x_2 &= (-\rho, \bar \rho, \b0) \ , \\
        x_3 &= (-1, 1, \b0) \ , \\
        x_4 &= (1, -1, \b0) \ ,
    \end{aligned}
\end{align}
where $\b0 \in \BR^{d-2}$ is the zero vector in the transverse space to the two-dimensional plane $\BR^{1,1}$.
With this parametrization the cross ratios become
\begin{align}\label{cross_ratio}
    z = \frac{4\rho}{(\rho + 1)^2} \ , \qquad \bar z = \frac{4\bar \rho}{(\bar \rho + 1)^2} \ .
\end{align}

We suppose the initial configuration is given by \eqref{Regge_parametrization} with $0\le \rho,\bar{\rho} < 1$ so that all the operators are spacelike-separated.
Hence the four-point function \eqref{4pt_CPW} takes the same form as the Euclidean correlator.
The Regge limit is achieved by taking $\rho \to 0$ then letting $\bar \rho \to \infty$ as in figure \ref{fig:Regge1}:
\begin{align}
    \rho \to 0 \ , \qquad \bar \rho \to \infty\ ,\qquad  \rho\, \bar\rho: \text{fixed} \ .
\end{align}
We will also use another parametrization, 
\begin{align}\label{Regge_polar}
    \rho = r\, e^{-t_{\mathrm{R}}} \ , \qquad \bar \rho = r\, e^{t_{\mathrm{R}}} \ ,
\end{align}
where the Regge limit is given by
\begin{align}\label{Regge_limit}
    \qquad t_{\mathrm{R}} \to \infty \ , \qquad  r: \text{fixed}\qquad (0<r<1) \ .
\end{align}
In the Regge limit the operator $1$ ($2$) moves into the past (future) lightcone of the operator $4$ ($3$) while they are spacelike-separated otherwise.
Using the notation \eqref{causal-rel} used in \cite{Kravchuk:2018htv}, it means
\begin{align}
    1 < 4 \ , \qquad 3 < 2 \ .
\end{align}
In particular the operators $1$ and $2$ are always spacelike-separated:
\begin{align}
    1 \approx 2 \ ,
\end{align}
and far apart along the lightlike $\bar \rho$ coordinate.

\subsection{Adapted coordinates and light-ray translation}\label{ss:Adapted_coord}

\begin{figure}[htbp]
    \centering
    \begin{minipage}[c]{0.45\hsize}
    \centering
    \includegraphics[width=7cm]{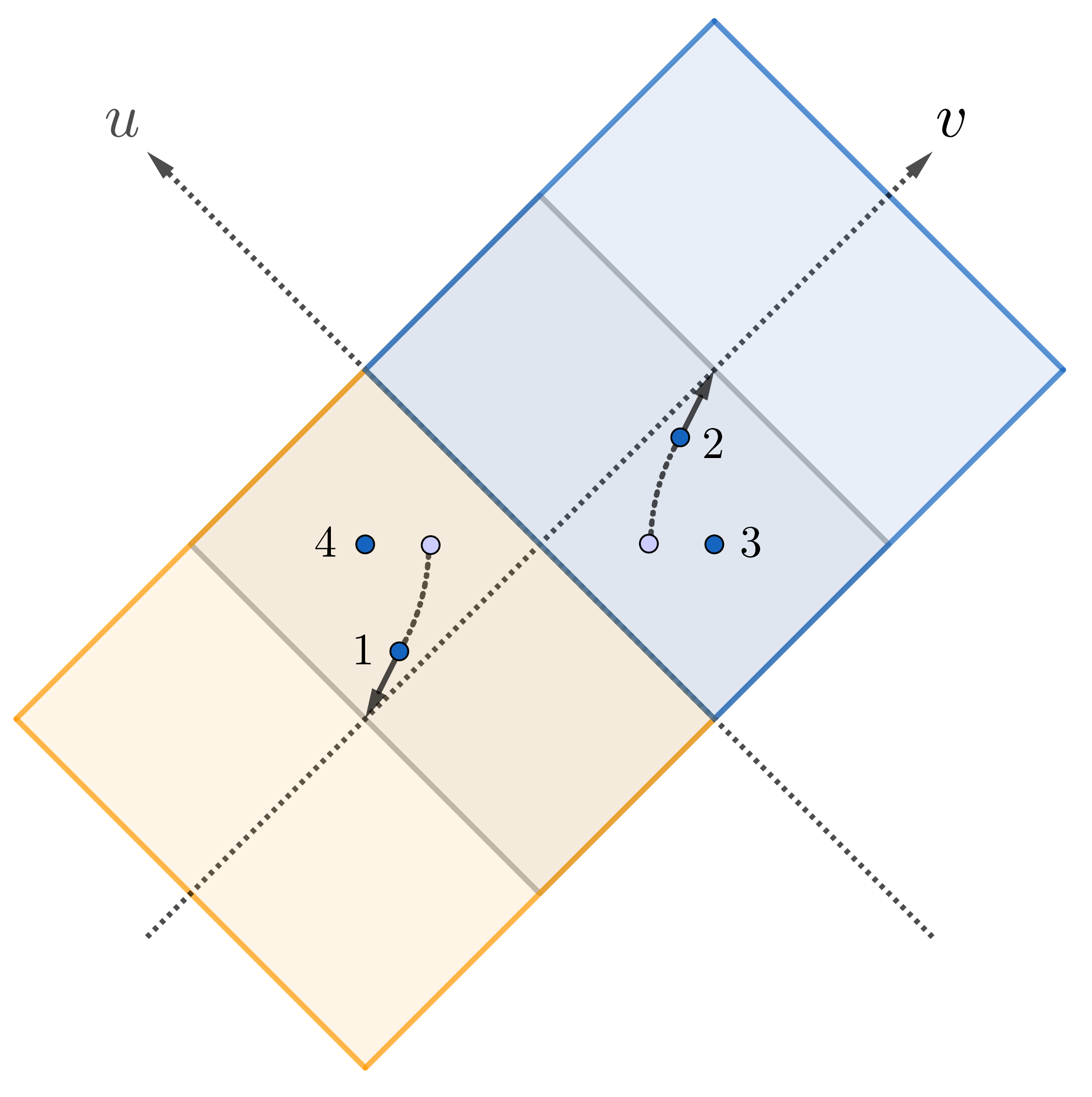}
    \end{minipage}
    \qquad
    \begin{minipage}[c]{0.45\hsize}
    \centering
    \includegraphics[width=6cm]{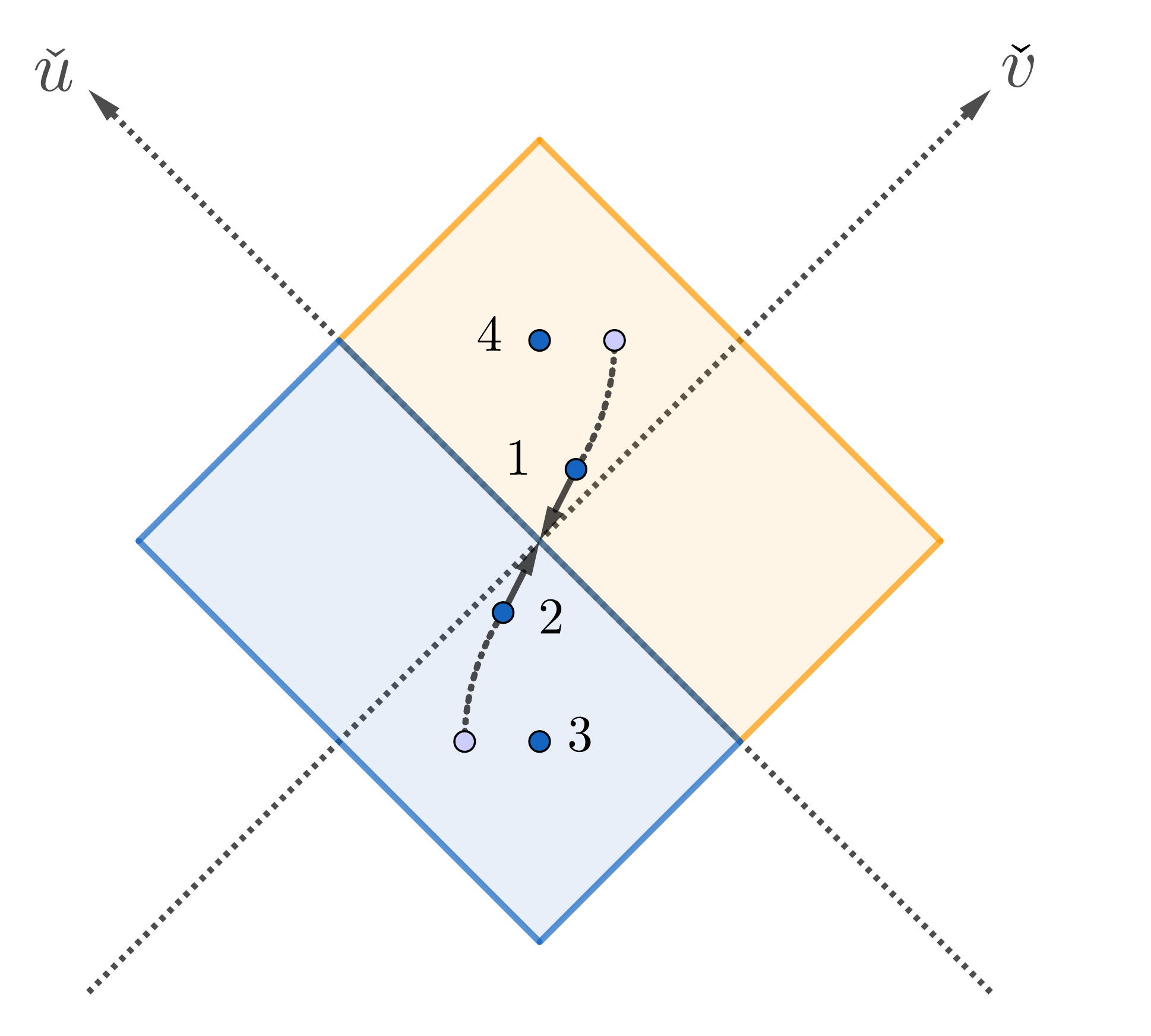}
    \end{minipage}
    \caption{
    [Left] The two different Minkowski patches covering the $v>0$ (blue) and $v<0$ (orange) regions, which are adapted to the Regge limit where the points $1$ and $2$ approach the origins in the each patch respectively.
    [Right]
    The Minkowski patch in the new coordinates glues upper-right and lower-left halves of the two Minkowski patches in the original coordinates.
    The points $1$ and $2$ are timelike-separated and approach the origin.
    }
    \label{fig:Regge_adapted1}
\end{figure}

In the Regge limit described above it is not obvious whether the OPE between $\CO_{1}$ and $\CO_{2}$ converges and the validity of the s-channel decomposition holds.
The notion of causality, however, is subtle in CFT as any pair of spacelike-separated points can be mapped to a pair of timelike-separated points by a conformal transformation if we describe the points in a particular Minkowski patch $\CM_d$ while the causal ordering is still well-defined in the universal cover $\widetilde \CM_d$ \cite{Luscher:1974ez}.
To illustrate this point more concretely we switch from the original coordinates $x^\mu$ to new coordinates $\check{x}^\mu$ as follows \cite{Cornalba:2007fs,Cornalba:2009ax}:
\begin{align}\label{Adapted_coord}
    \check{x} = 
    (\check u, \check{v}, \check\bx_\perp) = -\,\frac{1}{v}\,( x^2, 1, \bx_\perp) \ .
\end{align}
This is a conformal transformation as seen from the transformation law of the line element:
\begin{align}
    -\d \check u\, \d \check{v} + \d \check \bx_\perp^2 = \frac{1}{v^2}\,[ -\d u\, \d v + \d \bx_\perp^2] \ .
\end{align}
It is discontinuous at $v=0$ and maps the two different Minkowski patches covering the $v < 0$ and $v>0$ regions to one Minkowski patch in the new coordinates (see figure \ref{fig:Regge_adapted1}). 

Choosing a pair of points $1$ and $2$ as
\begin{align}\label{original_pt}
    x_1 = - x_2 = (u, v, \b0) \ ,
\end{align}
then they are mapped by the transformation \eqref{Adapted_coord} to the points
\begin{align}\label{mapped_pt}
    \check{x}_1 = - \check{x}_2 = \left( u, -\frac{1}{v}, \b0\right) \ .
\end{align}
The distance between the two points in the new coordinates becomes
\begin{align}
    \check{x}_{12}^2 =  4\,\frac{u}{v}= - \frac{x_{12}^2}{v^2} \ ,
\end{align}
so they are timelike/spacelike-separated in the new coordinates if they are spacelike/timelike to each other in the original coordinates.
For the Regge configuration \eqref{Regge_parametrization} in the parametrization \eqref{Regge_polar}, we choose $u = \rho = r\,e^{-t_{\mathrm{R}}}$ and $v = -\bar\rho = -r\,e^{t_{\mathrm{R}}}$.
The points $1$ and $2$, which are spacelike-separated in the original coordinates, become timelike in the new coordinates, and the distance between the two goes to zero in the Regge limit:
\begin{align}
    \check{x}_{12}^2 = - 4\,e^{-2t_{\mathrm{R}}} \underset{t_{\mathrm{R}}\to \infty}{\longrightarrow} 0 \ .
\end{align}
Thus in the new coordinates the Regge limit looks like the ordinary OPE limit between a pair of points, but in different Minkowski patches \cite{Caron-Huot:2013fea}.

\begin{figure}[htbp]
    \centering
    \includegraphics[width=10cm]{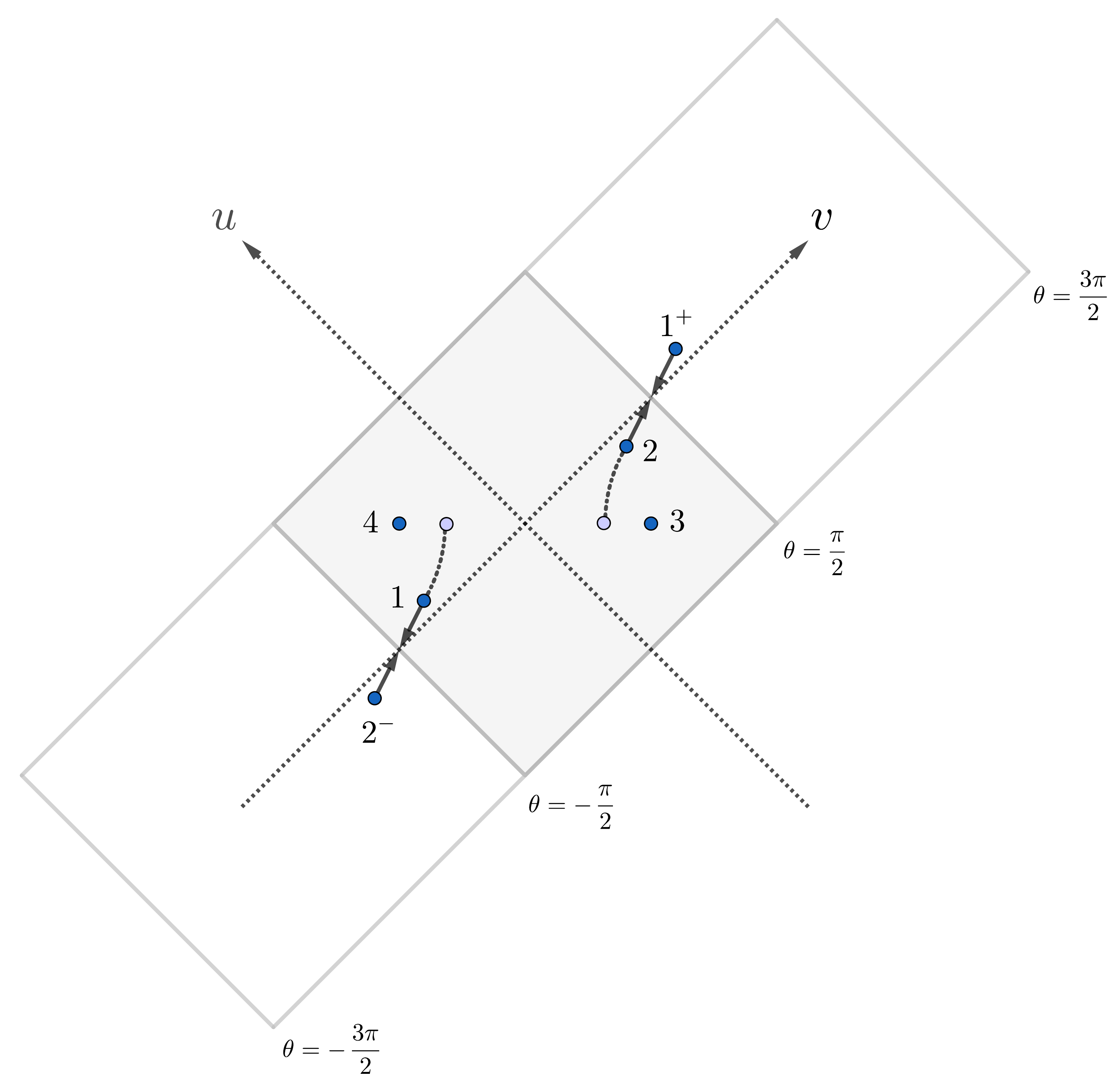}
    \caption{The Regge parametrization and its null-translated configurations by the $\CT$ symmetry.
    The points $1^+$ ($2^-$) is the null-translated point of $1$ ($2$) in the future (past) direction.
    They describe the same point in $\CM_d$ but in a different Minkowski patch on $\widetilde \CM_d$.
    }
    \label{fig:Regge_LT}
\end{figure}

To get more insight into this transformation, we parametrize the $v$ coordinate by
\begin{align}
    v = \tan \theta \ ,
\end{align}
then the Minkowski patch $\CM_d$ is covered by the range $-\pi /2 \le \theta \le \pi/2$.
It is seen from \eqref{original_pt} and \eqref{mapped_pt} that the coordinate transformation \eqref{Adapted_coord} on the $\bx_\perp = \b0$ plane induces the shift
\begin{align}
    \theta \to \theta + \frac{\pi}{2} \ ,
\end{align}
for the $v<0$ region ($-\pi/2 \le \theta < 0$) and
\begin{align}
    \theta \to \theta - \frac{\pi}{2} \ ,
\end{align}
for the $v>0$ region ($0 < \theta\le \pi/2$), while leaving $u$ fixed.
We can continue the Minkowski patch to different patches in the universal cover $\widetilde \CM_d$ by extending $\theta$ to an arbitrary value.
From the viewpoint of $\widetilde \CM_d$ the above transformation is equivalent to mapping one patch to the next patch by shifting $\theta \to \theta + \pi$.
The shift in $\theta$ is a symmetry of $\CM_d$ as points with $\theta$ differing by $n\,\pi$ ($n\in \BZ$) in the universal cover $\widetilde \CM_d$ represent the same point in $\CM_d$.
More generally there exists the $\CT$ symmetry that sends a point $p\in \widetilde \CM_d$ to the associated point $\CT p$ in a different patch by translating along a light-ray in the future direction \cite{Kravchuk:2018htv}.
We will denote the future and past null-translated points by $p^+ \equiv \CT p$ and $p^- \equiv \CT^{-1} p$ (see figure \ref{fig:Regge_LT}).
In this description it is clear from the figure that if a pair of points are spacelike-separated and one of the points is acted by $\CT$, then the resulting pair becomes timelike-separated.
The points $p$ and $\CT p$ are the same point in $\CM_d$, so $\CT$ commutes with an infinitesimal conformal transformation.
The $\CT$ symmetry is associated with the light transform $\bL$, an $\BZ_2$ element of the restricted Weyl group of the Lorentzian conformal group $\SO(2,d)$, which will play a crucial role in the following discussion.

\subsection{Adapted coordinates in embedding space}\label{ss:adapted_embed}
While the coordinate transformation from the original to the adapted coordinates given by \eqref{Adapted_coord} is discontinuous and is hard to find at first sight, it has a simple description as a rotation if the physical spacetime is uplifted into a pseudo-Riemannian manifold $\BR^{d,2}$ called the embedding space.

To set the stage, we first review the embedding space formalism \cite{Costa:2011mg}, where a primary operator $\CO^{\mu_1\cdots \mu_J}(x)$ with conformal dimension $\Delta$ and spin $J$ is encoded in a homogeneous function $\CO_{\Delta,J}(P,Z)$:
\begin{align}
    &\CO_{\Delta,J}(\lambda P,\alpha Z+\beta P)=\lambda^{-\Delta}\,\alpha^J\,\CO_{\Delta,J}(P,Z)\ .
\end{align}
$P^A~(A = 0, 1, \cdots, d+1)$ is a vector on the projective null cone and $Z^A$ is called a polarization vector in the embedding space $\BR^{d,2}$, subject to the conditions:
\begin{align}\label{PP=ZZ=PZ=0}
    P\cdot P=P\cdot Z=Z\cdot Z=0 \ , \qquad P,\,Z\in\BR^{d,2} \ ,
\end{align}
which enjoys the ``gauge" symmetry, $P^A\sim \lambda\,P^A$ for $\lambda \in \BR$.
We regain the encoding polynomial $\CO_{\Delta,J}(x,z)$ in physical space\footnote{In Euclidean case, $z$'s and $Z$'s must be complex in order to keep the condition $Z\cdot Z=z\cdot z=0$ non-trivial.}
\begin{align}
    &\CO_{\Delta,J}(x,z)=\CO_{\Delta}^{\mu_1\cdots\mu_J}(x)\,z_{\mu_1}\cdots z_{\mu_J}\ ,\qquad z^\mu z_\mu=0\ ,
\end{align}
by taking the Poincar\'e section with the gauge condition $P^+=1$ in the embedding space,\footnote{We use the same symbol $z$ both for the polarization vector and the cross ratios \eqref{cross_ratio}, but their distinction should be clear from context.
}
\begin{align}\label{Poincare-section}
    \begin{alignedat}{2}
        P^A & = (P^+,P^-,P^\mu)&& = (1,x^2,x^\mu)\ , \\
        Z^A & =(Z^+,Z^-,Z^\mu)&& = (0,2z\cdot x,z^\mu)\ ,
    \end{alignedat}
\end{align}
written in the lightcone coordinates:
\begin{align}
    \d P^2=-\d P^+\d P^- +\d P^\mu \d P_\mu \ , \qquad \d Z^2=-\d Z^+\d Z^- +\d Z^\mu \d Z_\mu\ .
\end{align}
Given the encoding polynomial $\CO_{\Delta,J}(x,z)$ one can recover the tensor structure $\CO_{\Delta,\mu_1\cdots\mu_J}(x)$ by pulling off $z$'s and subtracting traces.

For later purpose, we record the relation between an embedding space polynomial and the physical space counterpart for a more general gauge choice of $P^+$:
\begin{align}\label{general-gaugeP^+}
\begin{aligned}
    &\CO_{\Delta,J}(P,Z)=|P^+|^{-\Delta}\,\CO_{\Delta,J}\left(x^\mu=\frac{P^\mu}{P^+},\, z^\mu=Z^\mu-\frac{Z^+}{P^+}\,P^\mu\right) \ .
\end{aligned}
\end{align}
Notably, the generator of the conformal group $g\in\mathrm{SO}(d,2)$ acts linearly on embedding space vectors:
\begin{align}\label{linear-SO(d,2)}
    &\CO(P,Z)~ \xrightarrow[g\in\mathrm{SO}(d,2)]{}~\CO(g P,g Z) \ ,
\end{align}
which is one of the advantages of lifting operators to the projective null cone in $\BR^{d,2}$.

The conformal transformation \eqref{Adapted_coord} is also linearly realized in the embedding space just as $\pi/2$ rotations on the $(P^-,P^u)$ and $(P^+,P^v)$ planes \cite{Hofman:2008ar}:
\begin{align}\label{Adapted-embedding-transf}
    \begin{alignedat}{2}
        \check{P}^A         &=(\check{P}^+,\check{P}^-,\check{P}^u,\check{P}^v,\check{\bm{P}}_\perp)
        && =(-P^v,-P^u,P^-,P^+,\bm{P}_\perp)\ ,\\
        \check{Z}^A
        &=(\check{Z}^+,\check{Z}^-,\check{Z}^u,\check{Z}^v,\check{\bm{Z}}_\perp)
        && =(-Z^v,-Z^u,Z^-,Z^+,\bm{Z}_\perp)\ .
    \end{alignedat}
\end{align}
Indeed combining \eqref{Poincare-section}, \eqref{general-gaugeP^+} and \eqref{Adapted-embedding-transf} one reproduces the transformation law \eqref{Adapted_coord}:
\begin{align}
    \check{x}^\mu=(\check{u},\check{v},\tilde{\bm{x}}_\perp)=\frac{\check{P}^\mu}{\check{P}^+}=-\frac{1}{v}\,(x^2,1,\bm{x}_\perp)\ ,
\end{align}
where $\check{P}^A$ is the embedding vector for $\check{x}^\mu$ gauge equivalent to the Poincar\'e section:
\begin{align}\label{Adapted-section}
    &\check{P}^A=(\check{P}^+,\check{P}^-,\check{P}^u,\check{P}^v,\check{\bm{P}}_\perp)=-v\,(1,\check{x}^2,\check{u},\check{v},\check{\bm{x}}_\perp)\ .
\end{align}
The embedding space encoding polynomial in the adapted coordinates is defined through the relation:
\begin{align}\label{def-of-tildeO}
    &\widecheck{\CO}_{\Delta,J}(\check{P},\check{Z})=\CO_{\Delta,J}(P,Z)\ ,
\end{align}
which is related to its physical space counterpart $\widecheck{\CO}_{\Delta,J}(\check{x},\check{z})$ by \eqref{general-gaugeP^+}:
\begin{align}\label{adapted-section-E-P}
    \widecheck{\CO}_{\Delta,J}(\check{P},\check{Z})
        =|v|^{-\Delta}\,\widecheck{\CO}_{\Delta,J}\left(\check{x},\check{z}\right)
        =|\check{v}|^{\Delta}\,\widecheck{\CO}_{\Delta,J}\left(\check{x},\check{z}\right)\ .
\end{align}
Projecting to the physical space, we obtain the following Weyl transformation law of a primary operator:
\begin{align}\label{Weyl-adapted}
    \CO_{\Delta,J}\left(x,z\right)
        =|v|^{-\Delta}\,\widecheck{\CO}_{\Delta,J}\left(\check{x},\check{z}\right)
        =|\check{v}|^{\Delta}\,\widecheck{\CO}_{\Delta,J}\left(\check{x},\check{z}\right)\ .
\end{align}
Furthermore it is clear from \eqref{Poincare-section} and \eqref{Adapted-section} that the distances between two points in the $x^\mu$-coordinate and the $\check{x}^\mu$-coordinate are related by
\begin{align}\label{distance-rel}
    &x^2_{12}=-2P_1\cdot P_2= -2\check{P}_{1}\cdot \check{P}_2=v_1\,v_2\,\check{x}_{12}^2\ .
\end{align}

\subsection{Lorentzian conformal block in Regge regime}\label{sec:Lorentzian conformal block in Regge regime}
According to the Osterwalder-Schrader reconstruction theorem \cite{Osterwalder:1973dx,Osterwalder:1974tc}, Lorentzian correlation functions are derivable from the Euclidean counterpart by an analytic continuation with an appropriate $\i\,\epsilon$-prescription.\footnote{More precisely, the Osterwalder-Schrader reconstruction theorem states that the Wightman functions satisfying the Wightman axioms \cite{Streater:1989vi} can be reconstructed from reflection positive Schwinger functions (Euclidean correlators) obeying a growth condition. This theorem has many subtleties, but we do not get into details here.}
An $n$-point Lorentzian correlator $\langle\,\CO_1 (x_1)\cdots \CO_n (x_n)\,\rangle$ can be calculated in the following steps:
\begin{itemize}
    \item Start with the configuration where all operators are mutually spacelike (including Euclidean configuration with $t_i=0$):
    \begin{align}
    x_{j}=x_{\mathrm{in},j}=(t_{\mathrm{in},j},\vec{x}_{\mathrm{in},j})\ , \qquad \vec{x}_j\in\BR^{d-1}\ , \qquad x_{i j}^2>0 \mathrm{\ for \  any \ }i\neq j\ ,
    \end{align}
    where we use the Cartesian coordinate.
    \item Add the time components of $x_j$s infinitesimal negative imaginary parts:\footnote{More generally, we can give $x_j$s infinitesimal past directed imaginary coordinate $\zeta_j=\mathrm{Im}\, x_j<0$ \cite{Haag:1992hx}. In that case the operator ordering of the Wightman function agrees with the order of $-\mathrm{Im}\,\zeta_j$s.}
    \begin{align}
         x_{j}=(t_{\mathrm{in},j}-\i\,\epsilon_j,\vec{x}_{\mathrm{in},j})\ , \qquad \epsilon_j>0\ .
    \end{align}
    \item Continue the real parts of $x_j$s to the desired values:
    \begin{align}
        x_j=(t_j-\i\,\epsilon_j,\vec{x}_j )\ .
    \end{align}
    \item Take all the $\epsilon_j$s to zero while keeping the ordering of $\epsilon_j$s.
\end{itemize}
The operators in the resulting Wightman function is aligned in the descending order of $\epsilon_j$s. For example, when we choose the ordering of $\epsilon_j$s as
$\epsilon_1>\epsilon_2>\cdots> \epsilon_n>0$, the Wightman function becomes
\begin{align}
\langle\Omega |\,\CO_1 (x_1)\,\CO_2 (x_2)\,\cdots\, \CO_n (x_n) \, | \Omega \rangle
\end{align}
Conversely, when $\epsilon_n>\epsilon_{n-1}>\cdots> \epsilon_1>0$ the resulting Wightman function is
\begin{align}
\langle\Omega |\,\CO_n (x_n)\,\CO_{n-1} (x_{n-1})\,\cdots\, \CO_1 (x_1)\,  | \Omega \rangle \ .
\end{align}
The operators that are spacelike-separated commute to each other and the order of the $\epsilon_j$s does not matter.\footnote{See e.g. section 3 of \cite{Hartman:2015lfa} for more detailed arguments and examples}

In the configuration \eqref{Regge_parametrization} we have in mind, the four operators are no longer mutually spacelike-separated when $\bar\rho >1$ and there are four types of Wightman correlation functions depending on the operator ordering:
\begin{align}\label{Wightman_orderings}
    \begin{aligned}
        &\bullet\, \langle\Omega |\,\CO_{4}(x_4)\,\CO_{1}(x_1)\,\CO_{2}(x_2)\,\CO_{3}(x_3)\,| \Omega \rangle \\
        &\bullet\, \langle\Omega |\,\CO_{4}(x_4)\,\CO_{1}(x_1)\,\CO_{3}(x_3)\,\CO_{2}(x_2)\,| \Omega \rangle \\
        &\bullet\, \langle\Omega |\,\CO_{1}(x_1)\,\CO_{4}(x_4)\,\CO_{2}(x_2)\,\CO_{3}(x_3)\,| \Omega \rangle\\
        &\bullet\, \langle\Omega |\,\CO_{1}(x_1)\,\CO_{4}(x_4)\,\CO_{3}(x_3)\,\CO_{2}(x_2)\,| \Omega \rangle 
    \end{aligned}
\end{align}
Note that there are more correlators with different operator orderings, but they fall into one of the above orderings up to the exchange of spacelike-separated operators.
The first ordering is the Lorentzian time-ordered correlator and more natural than the others as it follows from the path integral formalism.
In the Regge limit, the time-ordered correlator exhibits a characteristic behavior, which will be of our particular interest in this paper.
The fourth ordering is the anti-time-ordered correlator and shows a similar behavior to the time-ordered one.

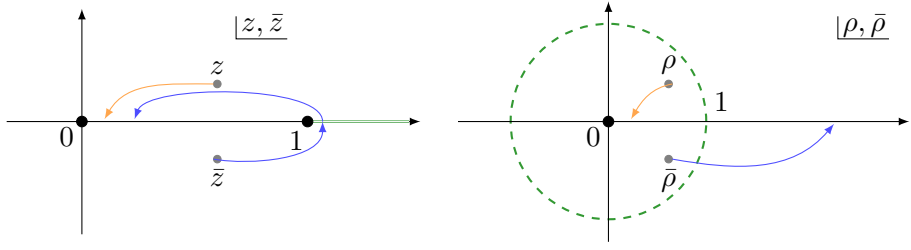
\begin{figure}[ht]
	\centering
	\begin{tikzpicture}
        \draw[] (2.05,1.4) -- (2.05,1.1) -- (2.73,1.1);
        \filldraw[gray] (1.8,-0.5) circle (0.05);
        \filldraw[gray] (1.8,0.5) circle (0.05);
        \draw[->] (-1,0) to (4.5,0);
        \draw[->] (0,-1.5) to (0,1.5);
        \draw[double distance=0.05pt,green!50!black!80] (3.05,0) to (4.35,0);
        \draw[->][orange!70] (1.75,0.5) to[out=180,in=60] (0.3,0.03);
        \draw[->][blue!70] (1.75,-0.5) to[out=-10,in=-90,distance=0.5cm] (3.2,0);
        \draw[->][blue!70] (3.2,0) to[out=90,in=70,distance=0.5cm] (0.7,0.03);
        \draw[fill=black] (0,0) circle (0.07);
        \draw[fill=black] (3,0) circle (0.07);
	    \node[above] at (2.4,1) {$z,\bar{z}$}; 
	    \node at (-0.2,-0.2) {$0$};
	    \node[below] at (2.85,0) {$1$};
	    \node[below] at (1.8,-0.5) {$\bar{z}$};
	    \node[above] at (1.8,0.5) {$z$};

        \draw[fill=black] (7,0) circle (0.07);
	    \node at (6.8,-0.2) {$0$};
	    \draw[dashed, thick, green!50!black!80] (7,0) circle (1.3);
	    \draw[->] (5,0) to (11,0);
	    \draw[->] (7,-1.6) to (7,1.6);
	    \node[above] at (8.5,0) {$1$};
	    \draw[] (10.05,1.4) -- (10.05,1.1) -- (10.73,1.1);
	    \node[above] at (10.4,1) {$\rho,\bar{\rho}$};
	    \filldraw[gray] (7.8,-0.5) circle (0.05);
        \filldraw[gray] (7.8,0.5) circle (0.05);
        \node[below] at (7.8,-0.5) {$\bar{\rho}$};
	    \node[above] at (7.8,0.5) {$\rho$};
	    \draw[->][orange!70] (7.85,0.5) to[out=190,in=60] (7.3,0.03);
		\draw[->][blue!70] (7.85,-0.5) to[out=-10,in=-130] (10,-0.03);
	\end{tikzpicture}
	\caption{The paths of $(z,\bar{z})$ and $(\rho,\bar{\rho})$ under the analytic continuation to $\bar{\rho}>1$ regime for the time-ordered correlator. When $\bar{\rho}$ goes out of the dashed unit sphere in the complex $\rho$-plane, $\bar{z}$ crosses the branch cut on the positive real half line $[1,\infty)$ in the complex $z$-plane.}
	\label{fig:z-1-crossratios}
\end{figure}

We will be interested in the time-ordered correlator (the first ordering in \eqref{Wightman_orderings}) as a most non-trivial case and relegate the argument of the second and third ordering to appendix \ref{sec:Analytic continuation of the second and third ordering}.
In this case the operators $4$ and $2$ are on the left of the operators $1$ and $3$ respectively and the operators $1$ and $2$ are always spacelike-separated before and after taking the Regge limit.
The $\i\,\epsilon$-prescription for the time-ordered correlator amounts to
\begin{align}
 t_{41}\to t_{41}-\i\,\epsilon\ , \qquad  t_{23}\to t_{23}-\i\,\epsilon   \qquad\quad (\epsilon >0) \ .
\end{align}
In the lightcone coordinates, this is equivalent to
\begin{align}
  \rho \to \rho + \i\,\epsilon  \ ,\qquad  \bar\rho \to \bar\rho - \i\,\epsilon  \ . 
\end{align}
We choose the initial configuration in the lightcone coordinate \eqref{LC_coord} as
\begin{align}
    x_2=-x_1=(-r,r,\bm{0}_\perp) \ , \qquad  x_3=-x_4=(-1,1,\bm{0}_\perp) \qquad (0< r < 1) \ ,
\end{align}
and take the Regge limit \eqref{Regge_limit}.
In view of the relation \eqref{cross_ratio} the cross ratios $z$ and $\bar z$ are initially located at slightly above and below the interval $[0,1]$ respectively in the complex plane (see figure \ref{fig:z-1-crossratios}).
In taking the Regge limit, $z$ approaches zero while $\bar z$ starts from a point slightly below the real axis, goes around $\bar z = 1$ in counter-clockwise and approaches zero:
\begin{align}\label{Regge_cross_ratio}
    z\underset{\rho \to 0}{\longrightarrow} 0 \ , \qquad \bar z = \frac{4(\bar \rho - \i\,\epsilon)}{(\bar \rho + 1 - \i\,\epsilon )^2}\quad \underset{\bar\rho \to 1}{\longrightarrow}\quad 1 + \frac{\epsilon^2}{4} + O(\epsilon^3)\quad \underset{\bar\rho \to \infty}{\longrightarrow}\quad 0 \ .
\end{align}

For the anti-time-ordered correlator (the fourth ordering in \eqref{Wightman_orderings}), the $\i \, \epsilon$-prescription is performed in the opposite way to the time-ordered one, resulting in $\bar z$ encircling around $\bar z =1$ clockwise.
The correlators in the second and third orderings do not cross the branch cut around $\bar z =1$ and remain the same as the Euclidean correlator (see appendix \ref{sec:Analytic continuation of the second and third ordering}).

To sum up the results the Wightman functions are given by 
\begin{itemize}
    \item $\langle\Omega |\,\CO_{4}(x_4)\,\CO_{1}(x_1)\,\CO_{2}(x_2)\,\CO_{3}(x_3)\,| \Omega \rangle=e^{- \frac{\i\pi}{2}(\Delta^-_{12}+\Delta^-_{43})}|T_{\{\Delta_i\}}(x_i)|\,g^{\circlearrowleft}(z,\bar{z})$,
    \item $\langle\Omega |\,\CO_{4}(x_4)\,\CO_{1}(x_1)\,\CO_{3}(x_3)\,\CO_{2}(x_2)\,| \Omega \rangle=e^{-\frac{\i\pi}{2}(\Delta^-_{12}+\Delta^-_{43})}\,|T_{\{\Delta_i\}}(x_i)|\,g(z,\bar{z})$,
    \item $\langle\Omega |\,\CO_{1}(x_1)\,\CO_{4}(x_4)\,\CO_{2}(x_2)\,\CO_{3}(x_3)\,| \Omega \rangle=e^{\frac{\i\pi}{2}(\Delta^-_{12}+\Delta^-_{43})}\,|T_{\{\Delta_i\}}(x_i)|\,g(z,\bar{z})$,
    \item $\langle\Omega |\,\CO_{1}(x_1)\,\CO_{4}(x_4)\,\CO_{3}(x_3)\,\CO_{2}(x_2)\,| \Omega \rangle=e^{\frac{\i\pi}{2}(\Delta^-_{12}+\Delta^-_{43})}\,|T_{\{\Delta_i\}}(x_i)|\,g^{\circlearrowright}(z,\bar{z})$,
\end{itemize}
where $\circlearrowright$ ($\circlearrowleft$) stands for the analytic continuation of $\bar z$ around $\bar z = 1$ in (counter)clockwise.
The phase factors $e^{\pm\frac{\i\pi}{2}(\Delta^-_{12}+\Delta^-_{43})}$ come from the analytic continuation of $T_{\{\Delta_i\}}(x_i)$ under $t_{41}\to  t_{41}-\i\, \epsilon$.
The absolute value $|T_{\{\Delta_i\}}(x_i)|$ follows from the scale dependent factor defined in \eqref{4pt_CPW}, whose explicit form in the polar parametrization \eqref{Regge_parametrization} and \eqref{Regge_polar} is given by
\begin{align}
    &|T_{\{\Delta_i\}}(x_i)|
        =\frac{1}{2^{\Delta^+_{12}+\Delta^+_{43}}\, r^{\Delta^+_{12}}}\left[\frac{(1+r\, e^{-t_{\mathrm{R}}})(1+ e^{-t_{\mathrm{R}}}/r)}{(1-r\, e^{-t_{\mathrm{R}}})(1- e^{-t_{\mathrm{R}}}/r)}\right]^{\frac{1}{2}(\Delta^-_{12}+\Delta^-_{43})}  \ .
\end{align}

\subsection{Regge limit and light transform}\label{ss:Regge_light-ray}
The details of the analytic continuation of the conformal block for the (anti-)time-ordered correlator are relegated to appendix \ref{sec:Regge-deriv}.
The leading contribution of the conformal block in the Regge limit \eqref{Regge_limit} is given by \eqref{Regge-confpw-ast}:
\begin{align}\label{ReggeCPW-LT}
    &G^{\circlearrowleft,\circlearrowright}_{\Delta,J} (z, \bar z)\xrightarrow[t_{\mathrm{R}}\to\infty,\, r\to0]{}\mp\frac{\i}{\pi}\frac{e^{\pm \frac{\i\pi}{2}(\Delta^-_{12}+\Delta^-_{43})}}{\kappa_{\Delta + J}}\, 2^{-J}z^{\frac{(1-J)-(1-\Delta)}{2}}\bar{z}^{\frac{(1-J)+(1-\Delta)}{2}}\,{}_2 F_1\left[\frac{d}{2}-1,\Delta-1,\Delta+1-\frac{d}{2},\frac{z}{\bar{z}}\right]\ .
\end{align}
Correspondingly, the time-ordered correlator becomes
\begin{align}\label{Regge-dom-time-order}
    \begin{aligned}
    \langle\Omega |\, &\CO_{4}(x_4)\,\CO_{1}(x_1)\, \CO_{2}(x_2)\,\CO_{3}(x_3)\, | \Omega \rangle \\
    &=e^{ -\frac{\i\pi}{2}(\Delta^-_{12}+\Delta^-_{43})}\,|T_{\{\Delta_i\}}(x_i)|\,g^{\circlearrowleft}(z,\bar{z})\\
    &=-\frac{\i}{\pi}\sum_{[\Delta,J]}   \frac{ c_{12,[\Delta,J]}c_{43,[\Delta,J]}}{2^{\Delta^+_{12}+\Delta^+_{43} + 3J - 2}\,\kappa_{\Delta + J}} \\
    &\qquad \quad \times e^{(J-1)t_{\mathrm{R}}}\,r^{\Delta-\Delta^+_{12}-1}  \left[ {}_2 F_1\left[\frac{d}{2}-1,\Delta-1,\Delta+1-\frac{d}{2},r^2\right]+ O(e^{-t_{\mathrm{R}}})\right]  \ ,
    \end{aligned}
\end{align}
where we used the polar variables.\footnote{The relation between the cross ratios $(z,\bar{z})$ and the polar variables $(r,t_{\mathrm{R}})$ in the Regge limit is \eqref{Regge-crossR-2}.}
The anti-time-ordered correlator has a similar form to \eqref{Regge-dom-time-order}, except for the minus sign in the third line.

Using the cross ratios, the conformal block \eqref{ReggeCPW-LT} is seen to have the asymptotic behavior:
\begin{align}\label{ReggeCPW_asymptotic}
    &G^{\circlearrowleft,\circlearrowright}_{\Delta,J} (z, \bar z)\sim   z^{\frac{(1-J)-(1-\Delta)}{2}}\bar{z}^{\frac{(1-J)+(1-\Delta)}{2}} \ , \qquad 0\ll z\ll \bar{z} \ll 1\ .
\end{align}
Compared with \eqref{normalization-CPW}, this behavior is the same as the conformal block carrying the unusual quantum number $G_{1-J,1-\Delta}(z,\bar{z})$ up to a normalization coefficient.
Indeed, as shown in \cite{Caron-Huot:2017vep,Isachenkov:2017qgn}, the analytically continued conformal block is a linear combination of two conformal blocks\footnote{The conformal block with the quantum number $(1-J,1-\Delta)$ is also a solution to the Casimir equation \eqref{casimir-eq}. This is due to the D${}_8$ symmetry of the Casimir equation \cite{Caron-Huot:2017vep,Isachenkov:2017qgn,Kravchuk:2018htv}:
\begin{align}
  (\Delta,J)\leftrightarrow(d-\Delta,J)\ ,\qquad (\Delta,J)\leftrightarrow(\Delta,2-d-J)\  , \qquad  (\Delta,J)\leftrightarrow(1-J,1-\Delta)\ . 
\end{align}
} with the quantum number $(\Delta,J)$ and $(1-J,1-\Delta)$,
which is schematically written as (see \eqref{Full-ReggeCPW} for the complete form):
\begin{align}
  G^{\circlearrowleft,\circlearrowright}_{\Delta,J} (z, \bar z)\sim G_{\Delta,J} (z, \bar z)+G_{1-J,1-\Delta} (z, \bar z)  \ .
\end{align}
This observation implies that the Regge conformal block has contributions from the exchange of an operator with the unusual quantum number $(1-J,1-\Delta)$ as well as the one with the ordinary quantum number $(\Delta,J)$ in the $\CO_1 \times \CO_2$ OPE channel (see figure \ref{fig:CPWE-Regge}).
The operator having the unusual quantum number associated with $\CO_{\Delta, J}$ is nothing but the light-ray operator $\bL[\CO_{\Delta, J}]$, which we will review shortly afterwards.

\begin{figure}[htbp]
    \centering
    \begin{tikzpicture}
        \draw (-2,-0.3) node [font=\LARGE]  {$\displaystyle g^\circlearrowleft ( z,\bar{z}) = \sum_{[\Delta, J]}$};
    
        \draw (0.5, 1) node [above, font=\large] {$\CO_{1}$} -- (1.5, 0) -- (0.5, -1) node [below, font=\large] {$\CO_{2}$};
        \draw (1.5, 0) -- node [below, midway, font=\large] {$\CO_{\Delta, J}$} (3.5, 0);
        \draw (4.5, 1) node [above, font=\large] {$\CO_{4}$} -- (3.5, 0) -- (4.5, -1) node [below, font=\large] {$\CO_{3}$};
        
        \draw (6,-0.3) node [font=\LARGE]  {$\displaystyle +\, \sum_{[\Delta, J]}$};
        
        \draw (7.5, 1) node [above, font=\large] {$\CO_{1}$} -- (8.5, 0) -- (7.5, -1) node [below, font=\large] {$\CO_{2}$};
        \draw (8.5, 0) -- node [below, midway, font=\large] {$\bL [\CO_{\Delta, J}]$} (10.5, 0);
        \draw (11.5, 1) node [above, font=\large] {$\CO_{4}$} -- (10.5, 0) -- (11.5, -1) node [below, font=\large] {$\CO_{3}$};
        
    \end{tikzpicture}
    \caption{The conformal block decomposition of the four-point function in the Regge limit.  Light-ray operators $\bL[\CO_{\Delta,J}]$ as well as local primaries $\CO_{\Delta,J}$ appear as intermediate states.}
    \label{fig:CPWE-Regge}
\end{figure}

At first sight, interpreting $G_{1-J, 1-\Delta}$ as a contribution from the OPE exchanging a light-ray operator is counter-intuitive as the two points $1$ and $2$ are far distant from each other in the Regge configuration and the OPE does not appear to converge.
We, however, already know from the discussion in section \ref{ss:Adapted_coord} that both $1$ and $2$ approach to the origin in the Regge limit when described in the adapted coordinates as shown in figure \ref{fig:Regge_adapted1}.
To put it another way, the spacetime configuration of $2$ is equivalent to the light-translated point $2^-$ in a different Minkowski patch, and $1$ and $2^-$ get close to each other in the universal cover $\widetilde \CM_d$ as in figure \ref{fig:Regge_LT}.

The similarity between the ordinary OPE limit and the Regge limit becomes more manifest when viewed in terms of the cross ratios in the conformal block.
Both $z$ and $\bar z$ approach zero in the Regge limit as in \eqref{Regge_cross_ratio} while the ratio $z/\bar z$ kept fixed:
\begin{align}\label{Regge_zzb_ratio}
    \frac{z}{\bar z} = e^{-2t_{\mathrm{R}}}\, \frac{(1 + r\, e^{t_{\mathrm{R}}})^2}{(1+ r\, e^{-t_{\mathrm{R}}})^2} ~ \xrightarrow[t_{\mathrm{R}}\to \infty]{}~ r^2 \ .
\end{align}
Then it follows from \eqref{ReggeCPW_asymptotic} that the dominant contribution arising from the light-ray operator $\bL[\CO_{\Delta, J}]$ can be read off by taking the $r\to 0$ limit in the Regge conformal block.

Light-ray operators have concrete realizations in Lorentzian CFT and are defined through the integral transform named as the light-transform $\bL$ \cite{Kravchuk:2018htv}.
The light transform of a primary operator written in the embedding space is
\begin{align}\label{Light-ray-operator}
    \bL[\CO_{\Delta,J}](P,Z)=\int_{-\infty}^{\infty}\d\alpha\, \CO_{\Delta,J}(Z-\alpha P,-P) \ .
\end{align}
It follows that the transformed operator satisfies the homogeneity condition:
\begin{align}
    \bL[\CO_{\Delta,J}](\lambda P,\alpha Z+\beta P)=\lambda^{-(1-J)}\,\alpha^{(1-\Delta)}\,\bL[\CO_{\Delta,J}](P,Z)\ ,
\end{align}
thus the light transform $\bL$ acting on a primary operator $\CO_{\Delta,J}$ maps its quantum number from $(\Delta,J)$ to $\bL[\Delta, J]\equiv (1-J,1-\Delta)$. According to the representation theory of conformal group, light-ray operators must annihilate the vacuum,
\begin{align}
    \bL[\CO_{\Delta,J}](P,Z)\,|\Omega\rangle =0 \ ,
\end{align}
if $\Delta + J >1$ \cite{Kravchuk:2018htv}.
This property explains why light-ray operators can have continuous spin and are not on the list of the unitary irreducible representations with positive energy \cite{Mack:1975je}.

An illuminating example of light-ray operators is a generalization of the averaged null energy operator (i.e., the light transform of the stress tensor):
\begin{align}\label{P0X0-light-ray}    \bL[\CO_{\Delta,J}](P_0,Z_0)&=\int_{-\infty}^{\infty}\d\alpha\, \CO_{\Delta,v\cdots v}(u=0,v=\alpha,\bm{0}_\perp) \ .
\end{align}
This is a special case of \eqref{Light-ray-operator} with $P$ and $Z$ set to particular values:
\begin{align}\label{P0X0} 
    \begin{alignedat}{2}
        P_0 
            &=(P_0^+, P_0^-,P_0^u,P_0^v,\bm{P}_{0,\perp})
            &&=(0,0,0,-1,\bm{0}_\perp) \ ,\\ 
        Z_0 
            &=(Z_0^+, Z_0^-,Z_0^u,Z_0^v,\bm{Z}_{0,\perp})
            &&=(1,0,0,0,\bm{0}_\perp) \ .
    \end{alignedat}
\end{align}

In section \ref{sec:Regge conformal block via Lorentzian OPE}, we will show the light-ray operators of the form \eqref{P0X0-light-ray} naturally appear in the OPE  $\CO_{1}(x_1)\times \CO_{2}(x_2)$ after taking the Regge limit, and we will reproduce the leading behavior of the Regge conformal block \eqref{ReggeCPW-LT} precisely from merely the OPE consideration, without resorting to the global monodromy analysis of conformal blocks carried out in section \ref{sec:Lorentzian conformal block in Regge regime}.

\section{Regge conformal block via Lorentzian OPE}\label{sec:Regge conformal block via Lorentzian OPE}
The aim of this section is to reproduce the dominant behavior of the time-ordered correlator \eqref{Regge-dom-time-order} from the s-channel OPE directly.
To this end, we adopt the Lorentzian OPE block formalism and use the holographic representation, which we review in section \ref{sec:OPE block formalism}.
We then take the Regge limit of the OPE block and show that it simplifies considerably in section \ref{sec:OPE block in the Regge limit}.
A geometric interpretation of the Regge OPE block and its relation to light-ray operators are given in section \ref{ss:Regge_OPEB_light_ray}.
Finally in section \ref{sec:Plugging into four-point function}, we put the Regge OPE block so-obtained into the four-point function and reproduce the leading behavior of the conformal block \eqref{Regge-dom-time-order} in the Regge limit.

\subsection{OPE block formalism}\label{sec:OPE block formalism}
The OPE of two scalar primaries $\CO_{1}(x_1)\,\CO_{2}(x_2)$ can be decomposed into a summation of the bi-local operators $\CB_{\Delta,J}(x_1,x_2)$ called OPE blocks:\footnote{Our normalization of three-point coefficients is different from those in \cite{Chen:2019fvi}:
\begin{align}
    c_{12,[\Delta, J]}^\text{(CCKN)} = (-1)^J\,c_{12,[\Delta, J]}|_\text{here} \ . 
\end{align}
Hence the normalizations of the OPE blocks are also different:
\begin{align}
    \CB_{\Delta, J}^\text{(CCKN)} = (-1)^J\,\CB_{\Delta, J}|_\text{here} \ . 
\end{align}
We choose this normalization so as to simplify the expression of the OPE block in the coincident limit $x_1 \to x_2$.
}
\begin{align}\label{gen-OPE}
    \begin{aligned}
        \CO_{1}(x_1)\,\CO_{2}(x_2)=\sum_{[\Delta,J]} c_{12, [\Delta,J]}\,\CB_{\Delta,J}(x_1,x_2) \ .
    \end{aligned}
\end{align}
The OPE block $\CB_{\Delta,J}$ in (\ref{gen-OPE}) includes all the contributions from the conformal family labelled by conformal dimension $\Delta$ and spin $J$.
When acted on the CFT vacuum, the OPE block is shown to be convergent and completely fixed by conformal symmetry \cite{Mack:1976pa,Pappadopulo:2012jk}.
The general structures of the OPE block on non-vacuum states, however, remain to be investigated and there might be additional terms contributing to the right hand in \eqref{gen-OPE} as discussed in the Introduction. In what follows, we will be only concerned with the vacuum OPE block, which should be present on any state due to the operator-state correspondence.
We will be mainly focused on the spacelike OPE block here and defer the timelike case to section \ref{sec:timelike-block}.

The Lorentzian OPE block was derived a long time ago by \cite{Ferrara:1971vh,Ferrara:1972uq,Ferrara:1973vz,Dobrev:1977qv}, and has attracted renewed interests in connection with its holographic description on the AdS$_{d+1}$ spacetime in literature \cite{deBoer:2016pqk,Czech:2016xec,daCunha:2016crm} where the scalar block $(J=0)$ has been studied extensively.
A more complete analysis including higher spin cases has been undertaken in a recent paper \cite{Chen:2019fvi}, where the OPE block is shown to take different forms depending on the causal relation of the two scalar primaries for which the OPE is taken.

The (vacuum) OPE block can be fixed by acting both sides on the vacuum and inserting the complete orthonormal basis using the momentum shadow projector\footnote{
We use a special normalization of the Wightman states $|\,\CO(p)\,\rangle$ to simplify the momentum shadow projector.
It is different from the normalization of two-point functions \eqref{2-pt-phys-ast} which is used throughout the rest of this paper.
} \cite{Gillioz:2016jnn,Gillioz:2018mto}:
\begin{align}
    \bm{1} = \sum_{[\Delta, J]}\, \int [\rD ^d p]_\rL\, |\,\tilde\CO_{\bar \Delta}^{\mu_1\cdots\mu_J}(-p)\,\rangle\, \langle\,\CO_{\Delta,\, \mu_1\cdots\mu_J}(p)\,| \ ,
\end{align}
which has a measure defined by \eqref{inv-FT} and is conformally invariant as the shadow operator $\tilde\CO(p)$ has conformal dimension $\bar \Delta = d - \Delta$.
The resulting OPE block takes the form:
\begin{align}
    \CB_{\Delta,J}(x_1,x_2) \propto \int [\rD^d p]_\rL\,\langle\, \CO_{1}(x_1)\,\CO_{2}(x_2)\,\tilde\CO_{d-\Delta}^{\mu_1\cdots\mu_J}(-p)\,\rangle\,\CO_{\Delta,\,\mu_1\cdots\mu_J}(p) \ .
\end{align}
One can proceed with this representation and rewrite the three-point function by introducing a Feynman parametrization $\xi$.
By exchanging the order of integration between $p$ and $\xi$ one ends up with an integral representation of the spacelike OPE block \cite{Chen:2019fvi}:
\begin{align}\label{Spinning_OPEB_1}
    \CB_{\Delta,J}(x_1, x_2) = b_{12,[\Delta,J]} \,\frac{1}{(x_{12}^2)^{\frac{\Delta^+_{12}}{2}}}\,\int_0^1 \d \xi\, \xi^{\frac{\Delta^-_{12}}{2}-1}(1-\xi)^{-\frac{\Delta^-_{12}}{2}-1}\,\Phi_{\Delta,J}(x^\mu(\xi),\eta(\xi))\ .
\end{align}
The coefficient $b_{12, [\Delta, J]}$ is proportional to the light-ray three-point constant in \eqref{sec:Short-handed-notations}:\footnote{Our $b_{12, [\Delta, J]}$ is related to the corresponding coefficient $b_{12, [\Delta, J]}^\text{(CCKN)}$ in \cite{Chen:2019fvi} by 
\begin{align}
    b_{12, [\Delta, J]} 
        = 
        \frac{\pi^\frac{d}{2}\,\Gamma(1-\bar\Delta)}{\Gamma\left(1-\frac{d}{2}+\Delta\right)}\,
        b_{12, [\Delta, J]}^\text{(CCKN)} \ .
\end{align}
This choice will simplify the asymptotic behavior of $\Phi_{\Delta, J}$ in $\eta \to 0$ as in \eqref{Bulk-scalar-bdy-con}.
}
\begin{align}
    b_{12,[\Delta,J]} = \i\,\frac{(\Delta + J -1)}{2^{J+1}\,\pi}\, L_{12,[\Delta,J]}\ .
\end{align}
The newly introduced parameters
\begin{align}\label{geodesic-parameter}
    x^\mu (\xi)=\xi\, x_1^\mu +(1-\xi)\, x_2^\mu \ , \qquad  \eta(\xi)=\sqrt{\xi(1-\xi)\,x^2_{12}} \ ,
\end{align}
have a clear physical interpretation as a point on the geodesic interpolating between $x_1$ and $x_2$ in the Poincar\'e patch of the AdS spacetime:
\begin{align}\label{AdS_Poincare}
    \d s^2 = \frac{\d \eta^2 + g_{\mu\nu}\,\d x^\mu\d x^\nu}{\eta^2} \ .
\end{align}
Moreover $\Phi_{\Delta,J}$ can be regarded as a massive higher-spin field propagating on the AdS spacetime, which can be decomposed into $J+1$ terms,
\begin{align}\label{Lorentzian_higher_spin_HKLL}
    \Phi_{\Delta, J} \left( x^\mu(\xi), \eta(\xi)\right)
        = \sum_{l=0}^{J}\,\Phi^{(l)}_{\mu_1\cdots\mu_l}\left( x^\mu(\xi), \eta(\xi)\right)\,w^{\mu_1}(\xi)\cdots w^{\mu_l}(\xi) \ ,
\end{align}
with different numbers of the vector $w^\mu$ defined by
\begin{align}\label{w-vector}
    w^\mu(\xi)=2\,\xi(1-\xi)\,x_{12}^\mu \ .
\end{align}
We do not bother to write the complete expression of $\Phi_{\Delta,J}$ as it is unnecessary in the following discussion. The interested reader is referred to \cite{Chen:2019fvi} for the detail.
In momentum space the $l^\text{th}$ term takes up to a constant the form:
\begin{align}\label{r-th-Bulk-field}
    \Phi^{(l)}_{\mu_1\cdots\mu_l}\left( x^\mu, \eta\right) 
        \sim \eta^{\Delta+J - 2l}\,\int [\rD ^d p]_{\mathrm{L}}\,e^{\i\, p\cdot x}\,
            \tilde J_{\Delta-d/2+J-l}\left(\sqrt{-p^2}\,\eta\right)\, p^{\mu_{l+1}}\cdots p^{\mu_J}\,\CO_{\Delta, \mu_1\cdots \mu_J}(p) \ , 
\end{align}
where we introduced the renormalized Bessel function of the first kind by
\begin{align}\label{renormalized-Bessel J}
    \tilde J_\nu (x) \equiv \Gamma(\nu+1)\,\left( \frac{x}{2}\right)^{-\nu}\,J_\nu (x)=\sum_{n=0}^{\infty}\frac{\Gamma(\nu+1)}{n!\,\Gamma(n+\nu+1)}\left(-\frac{x^2}{4}\right)^{n} \ .
\end{align}

We will distinguish the $l = J$ term as a ``conserved" field $\Phi_{\text{con},\, \mu_1\cdots\mu_J} \equiv \Phi^{(J)}_{\mu_1\cdots\mu_J}$ 
as it is the only remaining term in \eqref{Lorentzian_higher_spin_HKLL} that survives when $\CO_{\Delta, J}$ is a conserved current, which must have conformal dimension $\Delta = d + J - 2$ (the other terms vanish due to the conservation law, $p^\mu\, \CO_{\Delta, \mu \nu\cdots}(p) = 0$).
In our normalization, the conserved field becomes
\begin{align}\label{Sarkar-Xiao_field}
    \begin{aligned}
       &\Phi_{\text{con},\, \mu_1\cdots\mu_J} \left( x^\mu, \eta\right)  \\
        &\qquad= \frac{1}{\eta^{J-\Delta}}\int [\rD ^d p]_{\mathrm{L}}\,e^{\i\, p\cdot x}\,
        \tilde J_{\Delta-d/2}\left(\sqrt{-p^2}\,\eta\right)\,\CO_{\Delta,\mu_1\cdots\mu_J}(p)\\
        &\qquad= \frac{\Gamma(1-d/2+\Delta)}{\pi^{d/2}\,\Gamma(1-\bar\Delta)}\,\frac{1}{\eta^J}\int_{t'^2 + \bm{y}'^2 \le \eta^2} \d t'\,\d^{d-1} \bm{y}'\, \left( \frac{\eta}{\eta^2 - t'^2 - \bm{y}'^2} \right)^{\bar\Delta}\, \CO_{\Delta, \mu_1\cdots\mu_J} \left(t + t', \bm{x} + \i\,\bm{y}' \right) \ ,
      \end{aligned}
\end{align}
where we write the spacetime representation in the second line, which is seen as a massless higher-spin field in the AdS$_{d+1}$ spacetime that is holographically dual to a conserved current in CFT$_d$ if $\Delta = d+ J -2$ \cite{Sarkar:2014dma}.
For the record we stress that we do not restrict our consideration to the conserved case and the expression \eqref{Sarkar-Xiao_field} is valid for the OPE block with general $\Delta$ and $J$.

\subsection{OPE block in the Regge limit}\label{sec:OPE block in the Regge limit}
Having the time-ordered correlator in the Regge configuration \eqref{Regge_parametrization} in mind, let us consider the Regge configuration of the operator product $\CO_{1}(x_1)\, \CO_{2}(x_2)$ by locating $x_1$ and $x_2$ to the positions,
\begin{align}\label{Regge_OPE_config}
  x_2 =-x_1= ( -\rho , \bar\rho ,\bm{0}_\perp)=(-r\, e^{-t_{\mathrm{R}}},  r\, e^{t_{\mathrm{R}}}  ,\bm{0}_\perp) \ .
\end{align}
In the integral representation of the OPE block \eqref{Spinning_OPEB_1} there are several variables that depend on $x_1$ and $x_2$, whose non-vanishing components in the parametrization \eqref{Regge_OPE_config} become
\begin{align}\label{Regge-geodesic-parameter}
\begin{alignedat}{2}
    x^u (\xi) &= -(1-2\xi)\,\rho \ , \qquad & x^v (\xi) &= (1-2 \xi)\,\bar\rho\ , \\
    \eta(\xi) &= 2\sqrt{\xi(1-\xi)\,\rho\bar\rho}\ , \\
    w^u (\xi) &= 4\,\xi (1-\xi)\,\rho\  , \qquad & w^v (\xi) &= -4\,\xi(1-\xi)\,\bar\rho\ .
    \end{alignedat}
\end{align}

Now we take the same limit \eqref{Regge_limit} as before, then $x^u(\xi)$ and $w^u(\xi)$ go to zero while $x^v(\xi)$ and $w^v(\xi)$ grow exponentially in the limit.
At the same time the radial parameter $\eta(\xi)$ approaches a fixed value:
\begin{align}
    \eta(\xi) \to 2r\sqrt{\xi(1-\xi)} \ .
\end{align}
It follows from the expression \eqref{r-th-Bulk-field} that $\Phi^{(l)}_{\mu_1\cdots\mu_l}$ itself remains finite in the limit \eqref{Regge_limit}.
Thus the leading contribution to the bulk field $\Phi_{\Delta, J}$ in \eqref{Lorentzian_higher_spin_HKLL} arises from terms with $w^v(\xi)$s as many as possible.
This is nothing but the conserved field $\Phi_{\text{con},\, \mu_1\cdots \mu_J}$ with all indices contracted with $w^v(\xi)$:
\begin{align}\label{Regge-dom-Bulk-field}
    \Phi_{\Delta, J} \left( x^\mu(\xi), \eta(\xi)\right)
        = \Phi_{\text{con},\, v\cdots v}\left( x^\mu(\xi), \eta(\xi)\right)\,w^v(\xi)\cdots w^v(\xi)+ (\text{sub-leading terms}) \ .
\end{align}
By plugging \eqref{Regge-geodesic-parameter} and \eqref{Regge-dom-Bulk-field} into \eqref{Spinning_OPEB_1}, we find the asymptotic form of the OPE block:
\begin{align} \label{Spinning-OPE-v component}
    \begin{aligned}
        \CB_{\Delta,J}(x_1, x_2) &= (-1)^J\,b_{12, [\Delta, J]}\,\frac{1}{(x_{12}^2)^\frac{\Delta_{12}^+}{2} } \\
         & \qquad\times \left(4\bar\rho\right)^J\,\int_0^1 \d \xi \, \xi^{\frac{\Delta_{12}^-}{2}+J-1} (1-\xi)^{-\frac{\Delta_{12}^-}{2}+J -1}\, \Phi_{\text{con},\, v \cdots v} \left( x^\mu(\xi), \eta(\xi)\right)\\
         &\qquad \quad \qquad + O\left(\bar\rho^{J-2}\right) \ .
        \end{aligned}
\end{align}

The leading term can be made into a simpler form by changing the integral variable from $\xi$ to a new variable $\alpha = (1-2\xi)\,\bar\rho$.
The second line in \eqref{Spinning-OPE-v component} becomes 
\begin{align} \label{changed integrand}
    2\bar\rho^{J-1}\,\int_{- \bar\rho}^{\bar\rho}  \d \alpha \,\left(1 - \frac{\alpha}{\bar\rho} \right)^{\frac{\Delta_{12}^-}{2}-1 +J} \left( 1+ \frac{\alpha}{\bar\rho}\right)^{\frac{-\Delta_{12}^-}{2}-1 +J} \, \Phi_{\text{con},\, v \cdots v} \left( x^\mu(\xi), \eta(\xi)\right) \ ,
\end{align}
where 
\begin{align}
   x^u(\xi)=-\frac{\rho}{\bar{\rho}}\,\alpha\ , \qquad x^v(\xi)=\alpha  \ ,\qquad \eta(\xi) =  \sqrt{\rho\bar\rho\left( 1 - \frac{\alpha^2}{\bar\rho^2}\right)} \ .
\end{align}
Then the OPE block in the Regge limit, which we denote by $\CB_{\Delta,J}^\text{(Regge)}$, reduces to 
\begin{align} \label{regge OPE block-J}
    \begin{aligned}
        \CB_{\Delta,J}^\text{(Regge)}(x_1, x_2) &= \, \frac{(-1)^J\, 2\, b_{12, [\Delta, J]}}{(2r)^{\Delta_{12}^+} }  \\
        & \quad\times \, \left(r\, e^{t_{\mathrm{R}}}\right)^{J-1}\,\int_{- \infty}^\infty \d \alpha  \, \Phi_{\text{con},\, v \cdots v} \left( u = 0, v = \alpha , {\bm x}_\perp= \bm{0}_\perp , \eta = r\right) \ .
    \end{aligned}
\end{align}

Similar results were obtained in \cite{Hartman:2016lgu,Afkhami-Jeddi:2017rmx} which employed a proposed form of the timelike OPE block for a pair of identical operators in \cite{deBoer:2016pqk,Czech:2016xec}, which we will review in section \ref{ss:Timelike_two_rep}, took the Regge limit and analytically continued it to the spacelike configuration in deriving \eqref{regge OPE block-J} with $\Delta_1 = \Delta_2$.
While their results are consistent with ours there are subtleties in their derivation such that they do not use a standard $\i\,\epsilon$-prescription in the analytic continuation, but use the analyticity of the OPE block with respect to the positions of the operators (see the discussion at the end of section \ref{ss:Regge_timelike_OPEB}).

Our derivation, on the other hand, started with the spacelike OPE block for a general pair of scalar primaries in \cite{Chen:2019fvi} which is also valid for non-identical operators and more suitable for the Regge configuration than the timelike one.
For completeness we will show in section \ref{ss:Regge_timelike_OPEB} that even if we start with a timelike
configuration for $x_1$ and $x_2$ we can use the timelike OPE block in \cite{Chen:2019fvi}, whose derivation is based on a proper $\i\,\epsilon$-prescription, to show that in the Regge limit the OPE block ends up with the same form as \eqref{regge OPE block-J}.
Furthermore we will prove the equivalence of the two types of timelike OPE blocks in section \ref{ss:normalization_timelike_OPEB}.
Hence this line of argument may assure the validity of the results in \cite{Hartman:2016lgu,Afkhami-Jeddi:2017rmx} based on the non-standard analytic continuation for the timelike OPE block.

\subsection{A holographic view of Regge OPE block and light-ray operator}\label{ss:Regge_OPEB_light_ray}
In the previous section we showed in \eqref{regge OPE block-J} that the Regge OPE block has a simple holographic description by a massless higher-spin field $\Phi_{\mathrm{con},\mu_1\cdots \mu_J}(x^\mu,\eta)$ smeared over a null direction.
To gain more insight into the holographic picture, let us define a new field:
\begin{align}\label{bulk-scalar}
    \begin{aligned}
    \tilde{\Phi}(x^\mu,\eta)
        &= \eta^J\,\Phi_{\mathrm{con},v\cdots v}(x^\mu,\eta)\\
        &=\eta^\Delta\,\int [\rD ^d p]_{\mathrm{L}}\,e^{\i\, p\cdot x}\,
        \tilde J_{\Delta-d/2}\left(\sqrt{-p^2}\,\eta\right)\,\CO_{\Delta,v\cdots v}(p) \ .
    \end{aligned}
\end{align}
This reminds us of the so-called HKLL representation of the AdS scalar field of mass $m^2 = \Delta (\Delta -d)$ \cite{Kabat:2012hp}.
Indeed \eqref{bulk-scalar} equals to the HKLL scalar if $\CO_{\Delta,v\cdots v}(p)$ is replaced with a scalar primary $\CO_{\Delta}(p)$.
Hence it satisfies the equation of motion the AdS scalar field of mass $m^2 = \Delta (\Delta -d)$ satisfies.
In the Poincar\'e coordinates \eqref{AdS_Poincare} it reads
\begin{align}
&   \left(\partial_\eta^2+\frac{1-d}{\eta}\,\partial_\eta -\frac{\Delta(\Delta-d)}{\eta^2}-4\partial_{u}\partial_{v}+\partial_{\bm{x}_\perp}^2  \right)\tilde{\Phi}(x^\mu,\eta)=0 \ . \label{bulk-scalar-EoM}
\end{align}
It follows from \eqref{bulk-scalar} and \eqref{renormalized-Bessel J} that $\tilde\Phi$ is subject to the boundary condition in the $\eta \to 0$ limit:
\begin{align}\label{Bulk-scalar-bdy-con}
    \tilde{\Phi}(x^\mu,\eta)
        = \eta^{\Delta}\,\CO_{\Delta,v\cdots v}(x)+O(\eta^{\Delta+1}) \ ,
\end{align}
which reproduces the normalizable boundary condition for the AdS scalar field near the AdS boundary at $\eta = 0$ as expected.

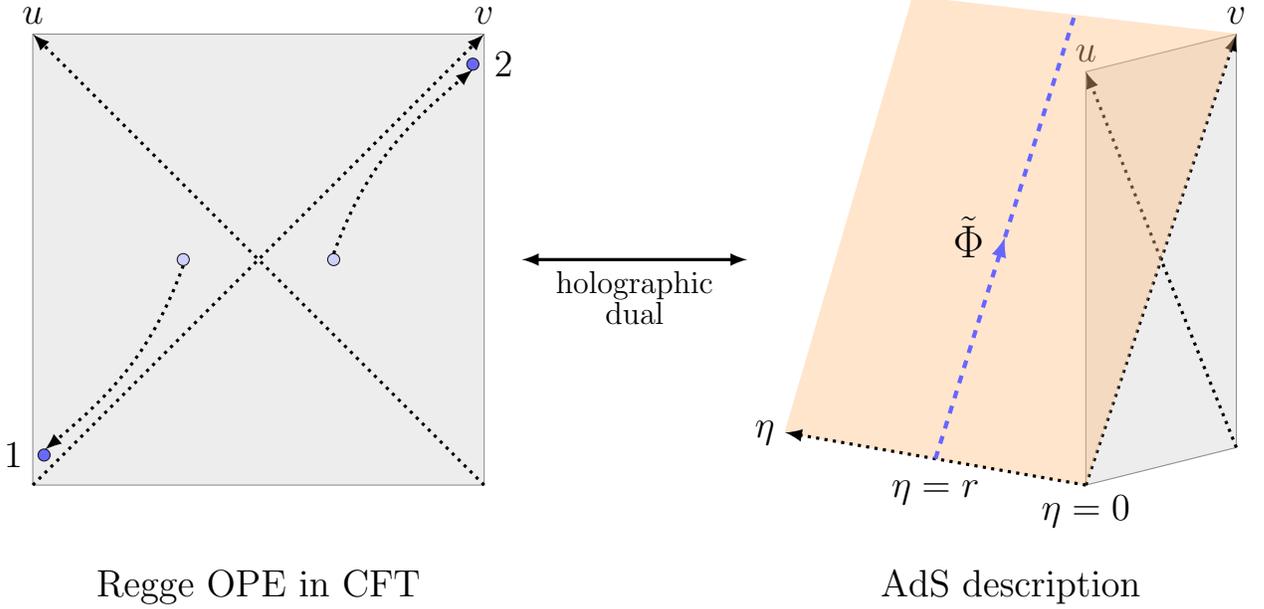
\begin{figure}[ht!]
	\centering
	\begin{tikzpicture}
    \filldraw[gray!20,opacity=0.7] (-1,-3) -- (1,-2.5) -- (1,3) -- (-1,2.5) --(-1,-3);
    \node[above] at (1,3) {\Large $v$};
    \node[above] at (-1,2.5) {\Large $u$};
    \draw[black,opacity=0.4] (-1,-3) -- (1,-2.5) -- (1,3) -- (-1,2.5) --(-1,-3);
    \draw[->, black, very thick, dotted] (-1,-3) -- (1,3); 
    \draw[->, black, very thick, dotted] (1,-2.5) -- (-1,2.5); 
    \filldraw[orange!50,opacity=0.4] (-1,-3) -- (-5,-2.3) -- (-3.3,3.5) -- (1,3);
    \draw[->,black,very thick, dotted] (-1,-3) -- (-5,-2.3);
    \node[left] at (-5,-2.3) {\Large $\eta$};
    \draw[->,dashed,blue!60,ultra thick] (-3,-2.65) -- (-2.075,0.3);
    \draw[dashed,blue!60,ultra thick] (-2.075,0.3) -- (-1.15,3.25);
    \node[below] at (-3,-2.8) {\Large $\eta=r$};
    \node[left,ultra thick] at    (-2.2,0.3) {\LARGE $\tilde{\Phi}$};
    \node[below] at (-2,-4) {$\Large{\text{AdS description}}$};
    \node[below] at (-1,-3) {\Large $\eta=0$};

    \draw[<->,very thick,black] (-8.5,0) -- node[midway, below] {$\LARGE{\substack{\text{holographic}\\\text{dual}}}$} (-5.5,0); 
      
    \filldraw[gray!20, opacity=0.7] (-15,-3) -- (-9,-3) -- (-9,3) -- (-15,3) --(-15,-3);
    \node[above] at (-9,3) {\Large $v$};
    \node[above] at (-15,3) {\Large $u$};
    \draw[black,opacity=0.4] (-15,-3) -- (-9,-3) -- (-9,3) -- (-15,3) --(-15,-3);
    \draw[->, black, very thick, dotted] (-15,-3) -- (-9,3); 
    \draw[->, black, very thick, dotted] (-9,-3) -- (-15,3); 
    \draw[black, fill=blue!20, opacity=0.9] (-13,0) circle (0.08);
    \draw[black, fill=blue!20, opacity=0.9] (-11,-0) circle (0.08);
    \draw[black, fill=blue!65, opacity=0.9] (-9.15,2.6) circle (0.08);
    \draw[black, fill=blue!65, opacity=0.9] (-14.85,-2.6) circle (0.08);
    \node[right] at (-9,2.6) {\Large $2$};
    \node[left] at (-15,-2.6) {\Large $1$};
    \draw[->,black,dotted,very thick] (-11,0.08) to[out=70,in=-140] (-9.15,2.53);
    \draw[->,black,dotted,very thick] (-13,-0.08) 
      to[out=-110,in=40] (-14.84,-2.53);
    \node[below] at (-12,-4) {$\Large{\text{Regge OPE in CFT}}$};
	\end{tikzpicture}
	\caption{[Left] The Regge OPE of two local scalars $\CO_{1}(x_1)\,\CO_{2}(x_2)$ in CFT.
	[Right] The holographic description as a bulk operator smeared in the $v$ direction at fixed radial coordinate $\eta=r$.
	It can be interpreted as a bulk shockwave geometry when $\Delta=d, J=2$ \cite{Afkhami-Jeddi:2017rmx}.
	Our argument is a natural extension of this correspondence to general spinning and non-conserved operators. In the limit $r\to0$, the bulk counterpart approaches a light-ray operator near the AdS boundary.
	In this sense, the bulk smeared operator can be seen as a holographic dual to a light-ray operator.}
	\label{fig:Light-ray and shockwave}
\end{figure}

It will be useful to embed the AdS coordinates $(u,v,\bm{x}_\perp,\eta)$ to the embedding space $\in\BR^{d-1,1}$,
\begin{align}
    X^2=-X^+ X^- -X^u X^v + (\bm{X}^\perp)^2=-1 \ .
\end{align}
The Poincar\'e coordinates correspond to the particular parametrization:
\begin{align}
    X^A=(X^+,X^-,X^u,X^v,\bm{X}^\perp)=\frac{1}{\eta}\,(1,\eta^2+x^\mu x_\mu,u,v,\bm{x}_\perp)\ .
\end{align}
It is clear that the AdS field $\tilde\Phi$ is a scalar function $\tilde\Phi = \tilde\Phi(X)$ in the embedding space, and we find a concise representation of the Regge OPE block \eqref{regge OPE block-J} as a null-averaged scalar field in AdS (see figure\ref{fig:Light-ray and shockwave}):
\begin{align}\label{regge OPE block_embedding}
        \CB^\text{(Regge)}_{\Delta,J}(x_1, x_2) &= \, \frac{(-1)^J\, 2\, b_{12, [\Delta, J]}}{(2r)^{\Delta_{12}^+} } \, e^{(J-1)\,t_{\mathrm{R}}}\,\int_{- \infty}^\infty \d X^v_0  \, \tilde\Phi (X_0)|_{\bm{X}_{0,\perp}=\bm{0}_\perp} \ ,
\end{align}
where we defined 
\begin{align}\label{X0}
    X_0=\frac{1}{r}\,(1,r^2+\bm{x}_\perp^2 ,0,v,\bm{x}_\perp)\ .
\end{align}
This expression will be useful in deriving the Regge behavior of conformal blocks from the OPE block formalism in section \ref{sec:Plugging into four-point function}.

The asymptotic behavior of $\tilde\Phi$ in \eqref{Bulk-scalar-bdy-con} leads to the boundary condition for the Regge OPE block in the $r\to 0$ limit, which also simplifies in the embedding space notation:
\begin{align}\label{OPEB_BC}
    \CB^\text{(Regge)}_{\Delta,J}(x_1, x_2) \xrightarrow[r\to 0]{} \, \frac{(-1)^J\, 2\, b_{12, [\Delta, J]}}{(2r)^{\Delta_{12}^+} }\, e^{(J-1)\,t_{\mathrm{R}}}\,r^{\Delta-1} \, \bL[\CO_{\Delta, J}] \left(P_0, Z_0\right) \ .
\end{align}
Notice that the appearance of the light-ray operator in the $r\to 0$ limit is in accord with the boundary condition for the unusual conformal block $G_{1-J, 1-\Delta}$ (see the discussion around \eqref{Regge_zzb_ratio}), and will be the key to reproduce the Regge conformal block from the OPE block formalism.
In addition, given the identification of $r$ as the holographic coordinate in AdS spacetime the relation \eqref{OPEB_BC} may allow us to view \eqref{regge OPE block_embedding} as a gravity/holographic dual to light-ray operators, which is foreseen by \cite{Hartman:2016lgu} in relation to ANEC.

The Regge OPE block \eqref{regge OPE block_embedding} and its asymptotic behavior \eqref{OPEB_BC} lead us to propose an alternative form of the OPE block and speculate an operator identity relating the ordinary and the alternate blocks in section \ref{sec:Light-ray channel OPE block}.

\subsection{Regge behavior from OPE block }\label{sec:Plugging into four-point function}

Armed with the results obtained so far,
we will show the leading behavior \eqref{Regge-dom-time-order} of the time-ordered correlator in the Regge limit can be reproduced by the OPE block formalism.

We apply the OPE block representation \eqref{gen-OPE} to the operator product $\CO_{1}(x_1)\, \CO_{2}(x_2)$ in the time-ordered correlator.
In the Regge limit \eqref{Regge_limit} the OPE block is dominated by the term \eqref{regge OPE block-J}, so we find
\begin{align}\label{def-of-F}
\begin{aligned}
    \langle\Omega |\, \CO_{4}(x_4)\,\CO_{1}&(x_1)\, \CO_{2}(x_2)\,\CO_{3}(x_3)\,  | \Omega \rangle\\
         \xrightarrow[\text{Regge limit}]{} 
        & \sum_{[\Delta,J]}c_{12,[\Delta,J]}\,\langle\Omega |\,\CO_{4}(x_4)\,\CB^\text{(Regge)}_{\Delta,J}(x_1, x_2)\,\CO_{3}(x_3)\,| \Omega \rangle \\
        =  &\sum_{[\Delta,J]}\,c_{12,[\Delta,J]}\,b_{12,[\Delta,J]}\, \frac{2\,(-1)^J}{\,(2r)^{\Delta_{12}^+} }\,e^{(J-1)\,t_{\mathrm{R}}}\,\bm{F}(\bm{x}_\perp=\bm{0}_\perp, r) \ ,
    \end{aligned}
\end{align}
where $\bm{F}(\bm{x}_\perp, r)$ is a function of $\bm{x}_\perp$ and $r$ defined by
\begin{align}\label{func-F}
     \bm{F}(\bm{x}_\perp, r)
        = \frac{1}{r}\,\int_{- \infty}^\infty \d \alpha  \, \langle\Omega |\,\CO_{4}(x_4)\,\tilde\Phi \left( u = 0, v = \alpha , {\bm x}_\perp , \eta = r\right)\,\CO_{3}(x_3)\,| \Omega \rangle \ .
\end{align}
We will keep the $\bm{x}_\perp$ dependence explicit in $\bm{F}(\bm{x}_\perp,r)$ for the time being so as to make manifest the symmetry of the function.

Integrating of the equation of motion \eqref{bulk-scalar-EoM} for $\tilde\Phi$ with respect to $v$ results in the differential equation $\bm{F}(\bm{x}_\perp,r)$ satisfies:
\begin{align}\label{EoM-for-F}
\left(\partial_r^2+\frac{3-d}{r}\partial_r -\frac{(\Delta-1)(\Delta+1-d)}{r^2}+\partial_{\bm{x}_{\perp}}^2\right)\bm{F}(\bm{x}_\perp,r)=0 \qquad (0<r<1)\ . 
\end{align}
In the derivation we performed a partial integration and used the fact that the integrand dumps as $\sim v^{-\Delta}$ for large $v$ (see the equations (22)-(24) in \cite{Kabat:2018pbj}).
It is worthwhile to emphasize that this is the equation of motion for a massive scalar field with mass $m^2=(\Delta-1)(d-\Delta-1)$ propagating in a $(d-1)$-dimensional hyperbolic space $\BH_{d-1}$, which would be easily seen by comparing \eqref{EoM-for-F} with the equation \eqref{bulk-scalar-EoM} we started with.

The dependence of $\bm{F}(\bm{x}_\perp,r)$ on the coordinates $\bm{x}_\perp$ and $r$ becomes clear when it is written in the embedding space:
\begin{align}\label{Emb-F}
    \bm{F}(\bm{x}_\perp,r)
        =\int_{-\infty}^{\infty}\d X^v_0\,\langle\Omega |\,\CO_{4}(P_4)\, \tilde{\Phi}(X_0)\,\CO_{3}(P_3)\,| \Omega \rangle \ ,
\end{align}
where $P_3$ and $P_4$ are the embedding space coordinates corresponding to the configuration \eqref{Regge_parametrization},
\begin{align}
    P_3 = (1, 1, -1, 1, \b0) \ , \qquad P_4 = (1, 1, 1, -1, \b0) \ .
\end{align}
The integrand is the three-point function of two CFT scalar primaries and one AdS scalar field, so it can only depend on three scalar invariants of the vectors $P_3,P_4$ and $X_0$ given by \eqref{X0}:
\begin{align}
    \begin{aligned}
         -2X_0\cdot P_3 &=-X^v_0+2\times \frac{1+\bm{x}_\perp^2+r^2}{2r}\ ,\\
         -2X_0\cdot P_4 &=X^v_0+2\times \frac{1+\bm{x}_\perp^2+r^2}{2r}\ , \\  -2P_3\cdot P_4 &=4\ .
   \end{aligned}
\end{align}
After integrating over $X^v_0$, the dependence of $\bm{F}(\bm{x}_\perp,r)$ on the variables $\bm{x}_\perp$ and $r$ are only through a particular combination:
\begin{align}\label{dep-of-F}
    \bm{F}(\bm{x}_\perp,r)=\left(\text{some function depending only on}\,\,\frac{1+\bm{x}_\perp^2+r^2}{2r}\right) \ .
\end{align}
The most convenient coordinate system respecting the condition (\ref{dep-of-F}) is the hyperbolic coordinates $(\varrho,\bm{\Omega}_{d-2})$ of $\BH_{d-1}$ which manifests the spherical symmetry:
\begin{align}
    \d s^2_{\BH_{d-1}}
        = \d \varrho^2+\sinh^2\varrho\left( \d\theta^2 + \sin^2\theta\,\d \Omega^2_{d-3}\right) \ .
\end{align}
where $\d \Omega^2_{d-3}$ is a metric on a unit $(d-3)$-sphere.
The new coordinates are related to the original variables by the coordinate transformation:
\begin{align}\label{adp-coord-F}
    \begin{aligned}
        \cosh\varrho 
            & =\frac{1+\bm{x}_\perp^2+r^2}{2r}\ , \\
       \sinh\varrho\,\cos\theta 
            & =\frac{1-\bm{x}_\perp^2-r^2}{2r} \ ,\\
       \sinh\varrho\, \sin\theta\,\Omega^i_{d-3}
            & =\frac{x^i_\perp}{r} \ , \qquad\qquad  (i=1,\cdots ,d-3) \ .
    \end{aligned}
\end{align}
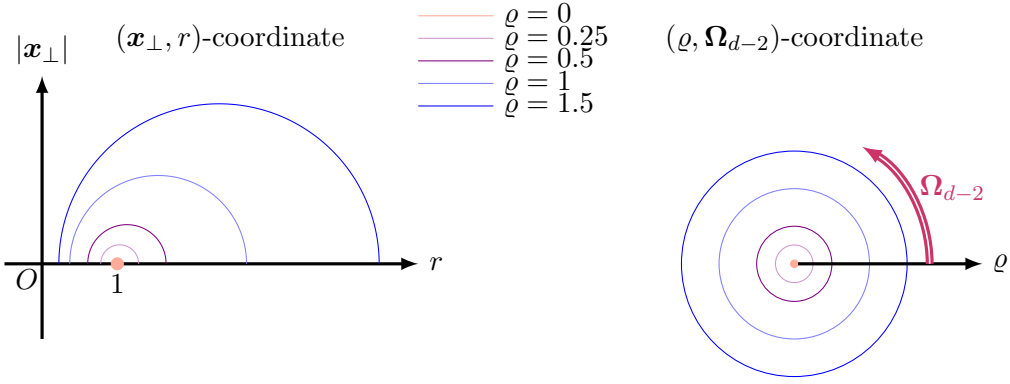
\begin{figure}[ht!]
	\centering
    \begin{tikzpicture}
    \draw[->,very thick] (-0.5,0)--(5,0)node[right]{$r$};
    \draw[->,very thick] (0,-1)--(0,2.5)node[above]{$|\bm{x}_\perp|$}; 
    \node at (-0.2,-0.2) {$O$};
    \draw[blue!50,samples=100,domain=0:pi,variable=\t] plot({cosh(1)+sinh(1)*cos(\t r)},{sinh(1)*sin(\t r)});
    \draw[blue!100,samples=100,domain=0:pi,variable=\t] plot({cosh(1.5)+sinh(1.5)*cos(\t r)},{sinh(1.5)*sin(\t r)});
    \draw[violet!100,samples=100,domain=0:pi,variable=\t] plot({cosh(0.5)+sinh(0.5)*cos(\t r)},{sinh(0.5)*sin(\t r)});
    \draw[violet!40,samples=100,domain=0:pi,variable=\t] plot({cosh(0.25)+sinh(0.25)*cos(\t r)},{sinh(0.25)*sin(\t r)});
    \filldraw[red!60!orange!40] (1,0) node[below, black] {$1$} circle (0.08); 
    \draw[red!60!orange!40] (5,3.3) -- (6,3.3);
    \node[right] at (6,3.3) {$\varrho=0$};
    \draw[violet!40] (5,3) -- (6,3);
    \node[right] at (6,3) {$\varrho=0.25$};
    \draw[violet!100] (5,2.7) -- (6,2.7);
    \node[right] at (6,2.7) {$\varrho=0.5$};
    \draw[blue!50] (5,2.4) -- (6,2.4);
    \node[right] at (6,2.4) {$\varrho=1$};
     \draw[blue!100] (5,2.1) -- (6,2.1);
    \node[right] at (6,2.1) {$\varrho=1.5$};
    \node at (2.5,3) {$(\bm{x}_\perp,r)$-coordinate};
    
    \node at (10,3) {$(\varrho,\bm{\Omega}_{d-2})$-coordinate};
     \filldraw[red!60!orange!40] (10,0) circle (0.05);
     \draw[violet!40] (10,0) circle (0.25);
     \draw[violet!100] (10,0) circle (0.5);
     \draw[blue!50] (10,0) circle (1);
     \draw[blue!100] (10,0) circle (1.5);
     \draw[->][very thick,black] (10.05,0) -- (12.5,0);
     \node[right] at (12.5,0) {$\varrho$};
     \draw[->, very thick,double, purple!80] (11.8,0) arc[radius=1.8, start angle=0, end angle=60];
     \node[purple!80] at (12.1,1) {$\bm{\Omega}_{d-2}$};
\end{tikzpicture}
	\caption{Two coordinates of a hyperbolic space $\BH_{d-1}$ ; $(\bm{x}_\perp,r)$ [Left] and $(\varrho,\bm{\Omega}_{d-2})$ [Right].
	A constant $\frac{1+\bm{x}^2_\perp+r^2}{2r}=\cosh\varrho$ slice in the $(\bm{x}_\perp,r)$-coordinate corresponds to a $(d-2)$-sphere with radius $\varrho$ in the $(\varrho,\bm{\Omega}_{d-2})$-coordinate.}
	\label{fig:hyperbiloc-coordinate}
\end{figure}
It follows from (\ref{dep-of-F}) that $\bm{F}(\bm{x}_\perp,r)$ is spherically symmetric (see figure \ref{fig:hyperbiloc-coordinate}),
\begin{align}
    \bm{F}(\bm{x}_\perp,r)\equiv \bm{F}(\varrho)\ ,
\end{align}
which results in a great simplification of the differential equation (\ref{EoM-for-F}):\footnote{This differential equation is the same as (3.15) in \cite{Kulaxizi:2017ixa} up to the contact term $\delta(\rho)$ we ignored by restricting $r$ to the limited value $0<r<1$. } 
\begin{align}
   \left[\partial^2_{\varrho}+(d-2)\coth\rho\,\partial_{\varrho}-(\Delta-1)(\Delta+1-d)\right]\bm{F}(\varrho)=0\qquad (0<\varrho) \ . \label{F-EoM-rho}
\end{align}

To solve the differential equation \eqref{F-EoM-rho} we need an appropriate boundary condition to be imposed.
Bearing in mind the asymptotic behavior \eqref{OPEB_BC} of the OPE block, we find an equivalent boundary condition for $\bm{F}(\bm{x}_\perp,r)$ that directly follows from \eqref{Emb-F} and \eqref{Bulk-scalar-bdy-con}:
\begin{align}
        \bm{F}(\bm{x}_\perp=\bm{0}_\perp,r)         
            \xrightarrow[r\to 0]{}&~
                r^{\Delta-1}\, \langle\Omega |\,\CO_{4}(P_4)\, \bL[\CO_{\Delta,J}](P_0,Z_0)\,\CO_{3}(P_3)\,| \Omega \rangle \label{small-r-F-ast}\\
            =&~
            r^{\Delta-1}\times c_{43,[\Delta,J]}\,L_{43,[\Delta,J]}\,\frac{(-1)^J}{2^{\Delta^+_{43}+2 J-1}} \label{Bdy-condition-F} \ ,
\end{align}
where we use \eqref{scalar2-light-ray-3pt} for the three-point function involving the light-ray operator \eqref{P0X0-light-ray}.
By solving the differential equation \eqref{F-EoM-rho} with the boundary condition\footnote{When $\bm{x}_{\perp}=\bm{0}_\perp$, the relation \eqref{adp-coord-F} between two coordinate systems for the $(d-1)$-dimensional hyperbolic space $\BH_{d-1}$, $(r,\bm{x}_\perp)$ and $(\varrho,\bm{\Omega}_{d-2})$, reduces to $e^{\varrho}=1/r$.
} \eqref{Bdy-condition-F}, we finally obtain the explicit form:
\begin{align}\label{sol-F-1}
\begin{aligned}
    & \bm{F}(\bm{x}_\perp,r)\,|_{\bm{x}_\perp=\bm{0}_\perp}=   c_{43,[\Delta,J]}\,L_{43,[\Delta,J]}\,\frac{(-1)^J}{2^{\Delta^+_{43}+2 J-1}}\,{}_2 F_1\left[\frac{d}{2}-1,\Delta-1,\Delta+1-\frac{d}{2},r^2\right]\  . 
    \end{aligned}
\end{align}
Finally, putting \eqref{def-of-F} and \eqref{sol-F-1} altogether, we end up with the Regge limit of the OPE block:
\begin{align}\label{ReggeOPEfinal-TO}
    \begin{aligned}
  &c_{12,[\Delta,J]}\,\langle\Omega |\,\CO_{4}(x_4)\,\CB^\text{(Regge)}_{\Delta,J}(x_1, x_2)\,\CO_{3}(x_3)\,| \Omega \rangle\\
  &=-\frac{\i}{\pi}  \frac{ c_{12,[\Delta,J]}\,c_{43,[\Delta,J]}  }{2^{\Delta^+_{12}+\Delta^+_{43}+3J-2}\,\kappa_{\Delta + J}} \, e^{(J-1)t_{\mathrm{R}}}\,r^{\Delta-\Delta^+_{12}-1}  {}_2 F_1\left[\frac{d}{2}-1,\Delta-1,\Delta+1-\frac{d}{2},r^2\right] \ .
    \end{aligned}
\end{align}
This is exactly the same leading behavior as the time-ordered correlator \eqref{Regge-dom-time-order} in the Regge limit.\footnote{The anti-time-ordered correlator can be obtained similarly just by making two replacements below:
\begin{itemize}
    \item $\bar{\rho}^{J-1}=(r\, e^{t_{\mathrm{R}}})^{J-1}\leftrightarrow (-\bar{\rho})^{J-1}  =(-1)^{J-1}(r\, e^{t_{\mathrm{R}}})^{J-1}$ in \eqref{regge OPE block-J}\ ,
    \item $\CO_{4}(P_4)\leftrightarrow\CO_{3}(P_3)$ or equivalently $c_{43,[\Delta,J]}\leftrightarrow c_{34,[\Delta,J]}=(-1)^J c_{43,[\Delta,J]}$ in \eqref{small-r-F-ast}.
\end{itemize}
In total, the leading behavior of the  anti-time-ordered correlator so obtained using the OPE block formalism differs from the time-ordered one only by a sign factor $(-1)$, which is consistent with the result based on the monodromy calculation in the previous section.}

We emphasize that the derivation of the Regge behavior of the conformal block via the OPE block  is simpler than the traditional one given in appendix \ref{sec:Regge-deriv} in the sense that the Regge behavior is captured already in the OPE as the operator relation and no monodromy analysis in the cross ratio space is needed.

\section{Timelike OPE block} \label{sec:timelike-block}

The arguments we have given in the previous section rested on a spacelike configuration of a pair of operators.
Meanwhile there is no difficulty in applying the same reasoning for a timelike configuration and examining the Regge limit of timelike OPE blocks.
The two approaches, spacelike and timelike, are complementary to each other and only differ by an phase factor.

There are two ways to represent timelike OPE blocks: one the analytic continuation of the spacelike OPE blocks \cite{Chen:2019fvi} and the other a different representation proposed by \cite{Czech:2016xec,deBoer:2016pqk} that is intrinsically associated with a timelike configuration. 
While both representations have appropriate OPE limits and satisfy the Casimir equation, the two expressions look quite different and the precise relation between them has never been clarified yet.
In what follows, we prove the equivalence of the two representations by comparing their asymptotic behaviors in the OPE and Regge limits.
We also show that the timelike OPE block can reproduce the Regge behavior of conformal blocks in a similar manner to the spacelike case.

\subsection{Two representations of timelike OPE block}\label{ss:Timelike_two_rep}
Among the two representations of timelike OPE blocks we first study the one obtained by an appropriate analytic continuation of the spacelike OPE block used in section \ref{sec:OPE block formalism}.
There are a variety of timelike OPE blocks depending on the spacetime configuration for a pair of operators as well as their ordering.
We here consider the OPE block of the order $\CO_1 (x_1)\, \CO_2 (x_2)$ for a spacetime configuration $1<2$, namely $x_2$ is in the forward lightcone of $x_1$.
In this specific configuration, the analytic continuation from the Euclidean coordinates to the Lorentzian ones is specified by the following $\i\, \epsilon$-prescription: 
\begin{align}\label{x12-analytic}
    \begin{aligned}
            x_{12}^2 
            &=  ({\bm x}_1 - {\bm x}_2 )^2 - (t_1 -t_2)^2 +  \i\, \epsilon\, (t_1 -t_2)  \\
            &= - \left\{ (t_1 -t_2)^2 - ({\bm x}_1 - {\bm x}_2 )^2 \right\} - \i\, \epsilon\, (t_2-t_1)   \\
            &\rightarrow |x_{12}^2|\, e^{-\i \pi} \ .
    \end{aligned}
\end{align}
This prescribes the timelike OPE block by replacing all $x_{12}^2$s to $|x_{12}^2 |\, e^{- \i \pi}$ in the spacelike OPE block \cite{Chen:2019fvi}:
\begin{align}\label{TL-Spinning_OPEB_1}
        \CB^{\text{T}}_{\Delta,J}(x_1, x_2) = b_{12,[\Delta,J]} \,\frac{e^{\frac{\i \pi}{2}\Delta^+_{12}}}{|x_{12}|^{\Delta^+_{12}}}\,\int_0^1 \d \xi\, \xi^{\frac{\Delta^-_{12}}{2}-1}(1-\xi)^{-\frac{\Delta^-_{12}}{2}-1}\,\Phi^{\mathrm{T}}_{\Delta,J}(x^\mu(\xi),\chi(\xi))\ ,
\end{align}
where $x^\mu(\xi)$ is the same vector in \eqref{geodesic-parameter} and we define
\begin{align}\label{timelike-Regge}
 \chi(\xi)=\sqrt{\xi(1-\xi)\, |x^2_{12}| } \ .
\end{align}
The field $\Phi^{\mathrm{T}}_{\Delta,J}$ has the expansion
\begin{align}\label{TL-Lorentzian_higher_spin_HKLL}
    \Phi^{\mathrm{T}}_{\Delta,J}\left(x^\mu(\xi),\chi(\xi)\right)
        = \sum_{l=0}^{J}\,\Phi^{\mathrm{T},\, (l)}_{\mu_1\cdots\mu_l}\left( x^\mu(\xi), \chi(\xi)\right)\,w^{\mu_1}(\xi)\cdots w^{\mu_l}(\xi) \ ,
\end{align}
with $w^\mu (\xi) = 2\, \xi (1-\xi)\, x_{12}^\mu$, whose $l^\text{th}$ term takes the form,
\begin{align}\label{TL-l-th-term}
\begin{aligned}
    \Phi^{\mathrm{T},\, (l)}_{\mu_1\cdots\mu_l}&\left( x^\mu(\xi) , \chi(\xi)\right)\\
       & \sim e^{-\frac{\i\pi}{2}(\Delta+J-2l)}\,\chi^{\Delta+J - 2l}\,\int [\rD ^d p]_{\mathrm{L}}\,e^{\i\, p\cdot x}\,
            \tilde I_{\Delta+J-d/2 - l}\left(\sqrt{-p^2}\, \chi \right)\, p^{\mu_{l+1}}\cdots p^{\mu_J}\,\CO_{\Delta, \mu_1\cdots \mu_J}(p) \ , 
            \end{aligned}
\end{align}
with the renormalized modified Bessel function defined by 
\begin{align}
    \tilde I_\nu (x) \equiv \Gamma(\nu+1)\,\left( \frac{x}{2}\right)^{-\nu}\,I_\nu (x)=\sum_{n=0}^{\infty}\frac{\Gamma(\nu+1)}{n!\,\Gamma(n+\nu+1)}\left(\frac{x^2}{4}\right)^{n} \ .
\end{align}
Notice that the scalar and conserved current blocks are special in the sense that they have natural holographic descriptions such that $\Phi^{\mathrm{T}}_{\Delta, J}(x^\mu (\xi), \chi(\xi))$ in the integrand obeys the equation of motion of a massless field propagating on a $(d+1)$-dimensional hyperboloid, not on the $\mathrm{AdS}_{d+1}$ spacetime, as pointed out in \cite{Chen:2019fvi}.

Let us confirm whether the timelike OPE block given by \eqref{TL-Spinning_OPEB_1} has an appropriate OPE limit.
From \eqref{TL-Lorentzian_higher_spin_HKLL} and \eqref{TL-l-th-term}, the $l^\text{th}$ term in the timelike OPE block behaves as $|x_{12}|^{\Delta+J-l}$ in the coincident limit, so the leading contribution $\Phi^{\mathrm{T}}_{\text{con}, \mu_1 \cdots \mu_J} \equiv \Phi^{\mathrm{T},(J)}_{\mu_1 \cdots \mu_J}$ comes from $l=J$ term in \eqref{TL-Lorentzian_higher_spin_HKLL}:
\begin{align}\label{TL-simplify-cont}
    \Phi^{\mathrm{T}}_{\text{con},\mu_1\cdots\mu_J}  \left( x^\mu(\xi), \chi(\xi)\right)          ~\xrightarrow[x_1\to x_2]{}~ \chi^{\Delta-J}(\xi)\,\CO_{\Delta,\mu_1\cdots\mu_J}(x_2)\ .
\end{align}
Plugging \eqref{TL-simplify-cont} into \eqref{TL-Spinning_OPEB_1} and performing some integrations, the OPE limit of \eqref{TL-Spinning_OPEB_1} can be read off:
\begin{align}\label{TL-coin-limit-OPE}
    &\CB_{\Delta,J}^{\mathrm{T}}(x_1,x_2)~\xrightarrow[x_1\to x_2]{}~ \frac{e^{\frac{\i\pi}{2}(\Delta^+_{12}-\Delta+J)}}{|x_{12}^2|^{\frac{\Delta^+_{12}-\Delta+J}{2}}}\,x_{12}^{\mu_1}\cdots x_{12}^{\mu_J}\,\CO_{\Delta,\mu_1\cdots \mu_J}(x_2)\ .
\end{align}
This is equivalent to the OPE limit of the spacelike OPE block given in \eqref{coin-limit-OPE} with $x_{12}^2$ replaced by the analytic continuation $|x_{12}^2|\,e^{-\i\pi}$.

Now we turn to the second representation of the timelike OPE block proposed in \cite{deBoer:2016pqk,Czech:2016xec}:
\begin{align}\label{timelikeOPE-SP}
    \begin{aligned}
      \CB^{\diamondsuit}_{\Delta,J}(x_1,x_2)
        &= \tilde{b}_{12,[\Delta,J]} \int_{x_0\in \diamondsuit_{12}}\d^d x_{0}\, \langle \tilde{0} |\, \CO_{1}(x_1)\,\CO_{\bar{\Delta},\mu_1\cdots \mu_J}(x_0)\,\CO_{2}(x_2)\, |\tilde{0}\rangle\, \CO_{\Delta}^{\mu_1\cdots \mu_J}(x_0)\ ,
    \end{aligned}
\end{align}
where the superscript $\diamondsuit$ is meant to distinguish it from the first representation \eqref{TL-Spinning_OPEB_1}.
The integration is restricted to the causal diamond for the pair of timelike-separated points:
\begin{align}
    &x_0\in \diamondsuit_{12}\leftrightarrow \{\, x_0\, |\,  
     1< 0< 2 \, \cup \,  2< 0< 1 \,\}\ ,
\end{align}
and we denote by $\langle\tilde{0}|\cdots |\tilde{0}\rangle$ a normalized three-point structure (see appendix \ref{app:Normalization} for the notation):
\begin{align}\label{normalized-3py-str}
    \begin{aligned}
        &\, \langle \tilde{0} |\, \CO_{1}(x_1)\,\CO_{\Delta,J}(x_3,z)\,\CO_{2}(x_2)\, |\tilde{0}\rangle\,=\frac{(-z\cdot H(x_{13},x_{23}))^J}{|x_{12}^2|^{\frac{\Delta^+_{12}-\Delta+J}{2}} |x_{13}^2|^{\frac{\Delta^-_{12}+\Delta-J}{2}}|x_{23}^2|^{\frac{-\Delta^-_{12}+\Delta-J}{2}}} \ .
    \end{aligned}
\end{align}

We choose the normalization constant $\tilde{b}_{12,[\Delta,J]}$ so that the OPE block has the asymptotic form in the OPE limit:
\begin{align}\label{TL-coin-general}
    \CB^{\diamondsuit}_{\Delta,J}(x_1,x_2)~\xrightarrow[x_1\to x_2]{} ~ \,\frac{1}{|x_{12}^2|^{\frac{\Delta^+_{12}-\Delta+J}{2}}}\,x_{12}^{\mu_1}\cdots x_{12}^{\mu_J}\,\mathcal{O}_{\Delta,\mu_1\cdots\mu_J}(x_2) \ .
\end{align}
Compared with the asymptotic form of the first representation, $\CB^{\diamondsuit}$ has the same OPE limit as \eqref{TL-coin-limit-OPE} up to a phase factor.
The two representations should be equivalent up to a constant, thus one can determine the precise relation between them by comparing their OPE limits:
\begin{align}\label{Relation_bet_TOPEB}
    \CB^\mathrm{T}_{\Delta,J} (x_1,x_2) = e^{\frac{\i \pi}{2} (\Delta_{12}^+ - \Delta + J)}\, \CB^{\diamondsuit}_{\Delta,J} (x_1,x_2) \, .
\end{align}
In section \ref{ss:normalization_timelike_OPEB} we will fix the normalization constant $\tilde{b}_{12,[\Delta,J]}$ by comparing the asymptotic behaviors of the two representations in the Regge limit.

\subsection{Regge limit of timelike OPE block}\label{ss:Regge_timelike_OPEB}

We switch gear and begin to examine the Regge limit of the timelike OPE block.
While there are two types of timelike OPE blocks, $\CB^{\mathrm{T}}$ and $\CB^\diamondsuit$, they are equivalent up to a constant as in \eqref{Relation_bet_TOPEB}, so we will be only concerned with the Regge limit of $\CB^{\mathrm{T}}$.

\begin{figure}[ht!]
	\centering
	\begin{tikzpicture}
    \filldraw[gray!20] (0,-3) -- (3,0) -- (0,3) -- (-3,0);
    \draw[dashed,black] (-0.5,-3.5) -- (3.5,0.5);
    \draw[dashed,black] (0.5,3.5) -- (-3.5,-0.5); 
    \draw[dashed,black] (3.5,-0.5) -- (-0.5,3.5);
    \draw[dashed,black] (-3.5,0.5) -- (0.5,-3.5);
    \draw[->] (-2,-2) to (2,2);
    \node[above right] at (2,2) {\Large $v$};
    \draw[->] (2,-2) to (-2,2);
    \node[above left] at (-2,2) {\Large $u$};
    \draw[black, fill=blue!20] (0,0.8) circle (0.08);
    \draw[black, fill=blue!20] (0,-0.8) circle (0.08);
    \node[below] at (-1.5,-1.6) {$1$};
    \node[above] at (1.5,1.6) {$2$};
    \draw[black, fill=blue!65] (-1.35,-1.5) circle (0.08);
    \draw[black, fill=blue!65] (1.35,1.5) circle (0.08);
    \draw[->,dotted][very thick,black] (0.07,0.8) to[out=0,in=-130] (1.35,1.43);
    \draw[->,dotted][very thick,black] (-0.07,-0.8) to[out=180,in=50] (-1.35,-1.43); 
	\end{tikzpicture}
	\caption{The Regge-like limit of a pair of timelike-separated points.}
	\label{fig:timelike Regge}
\end{figure}
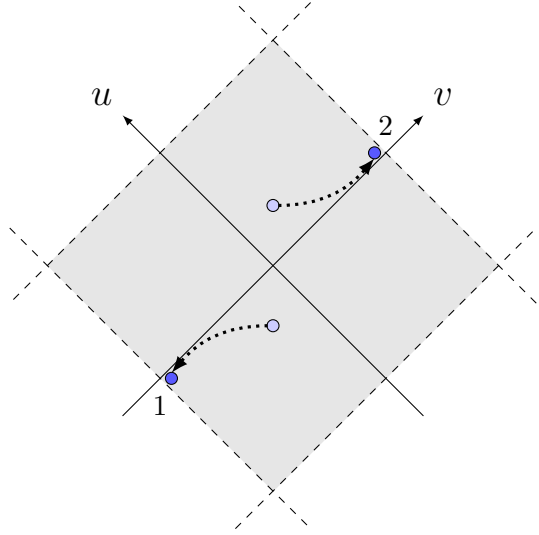

To start with, it is useful to employ the following coordinates for a pair of timelike-separated points,
\begin{align}\label{timelike-Regge-parametrization}
    x_2 = - x_1 = ( \rho, \bar \rho, \b0) = (r\, e^{-t_{\mathrm{R}}}, r\, e^{t_{\mathrm{R}}}, \b0 )  \ ,
\end{align}
and let them move to the same points as in the Regge limit as shown in figure \ref{fig:timelike Regge}:
\begin{align}\label{regge limit in timelike}
    \rho \rightarrow 0 \, , \quad \bar \rho \rightarrow \infty \, , \quad \rho\, \bar \rho: \text{fixed} \ .
\end{align}
Proceeding with the same argument as in section \ref{sec:OPE block in the Regge limit} the leading contribution to the timelike OPE block \eqref{TL-Spinning_OPEB_1} in the Regge limit is seen to arise from $\Phi^{\mathrm{T}}_{\text{con},\, v \cdots v}$ given by 
\begin{align}\label{timelike-conserved-field}
    \begin{aligned}
        \Phi^{\mathrm{T}}_{\text{con},\, v \cdots v} \left( x^\mu, \chi \right)  
        &= \frac{e^{-\frac{\i\pi}{2}(\Delta-J)} }{\chi^{J-\Delta}}\int [\rD ^d p]_{\mathrm{L}}\,e^{\i\, p\cdot x}\,
        \tilde I_{\Delta-d/2}\left(\sqrt{-p^2}\,\chi\right)\,\CO_{\Delta,v \cdots v}(p) \ .
      \end{aligned}
\end{align}
By making a change of variable from $\xi$ to  $\alpha = (1-2\xi) \bar \rho$ and taking the limit \eqref{regge limit in timelike}, we find
\begin{align} \label{regge OPE block-J timelike}
    \begin{aligned}
        \CB_{\Delta,J}^\text{\text{T},(Regge)}(x_1, x_2) &= \, \frac{(-1)^J\, 2\, b_{12, [\Delta, J]}}{(2r)^{\Delta_{12}^+} }\, e^{\frac{\i\pi}{2}(J - \Delta + \Delta^+_{12})} \\
        & \quad\times \, \left(r\, e^{t_{\mathrm{R}}}\right)^{J-1}\,\int_{- \infty}^\infty \d \alpha  \,  \Phi^{\mathrm{T}}_{\text{con}, v\cdots v} \left(u = 0,\, v = \alpha , \, \bm{x}_\perp = \bm{0}_\perp, \,  \chi = r \right)  \ .
    \end{aligned}
\end{align}

Let us pause here to notice that 
\begin{align}
    \begin{aligned}
    \tilde{\Phi}^\mathrm{T}  (x^\mu, \chi )        &\equiv  \chi^J\,  \Phi^{\mathrm{T}}_{\text{con}, v\cdots v} \left( x^\mu, \chi \right)  \\
        &= e^{ \frac{\i \pi}{2}J}\, (e^{- \frac{\i \pi }{2}} \chi)^\Delta \int [\rD ^d p]_{\mathrm{L}}\,e^{\i\, p\cdot x}\,
        \tilde J_{\Delta-d/2}\left(\sqrt{-p^2}\, e^{-\frac{\i\pi}{2}} \chi\right)\,\CO_{\Delta,v \cdots v}(p)
    \end{aligned}
\end{align}
is the same field as \eqref{bulk-scalar} up to the phase factor with the replacement $\eta \rightarrow e^{-\frac{\i\pi}{2}} \chi = - \i\, \chi$, which implies that formally $\tilde\Phi^{\text{T}}$ satisfies the equation of motion \eqref{bulk-scalar-EoM} with the identification $\eta \rightarrow  - \i\, \chi$. It means that $\tilde{\Phi}^\mathrm{T}$ can be obtained by analytically continuing the spacelike counterpart $\tilde\Phi$ with an additional phase factor:
\begin{align}\label{relation btw timelike and spacelike}
    \tilde{\Phi}^\mathrm{T}(x^\mu, \chi ) = e^{\frac{\i \pi}{2} J }\, \tilde \Phi ( u,  v , \bm{x}_\perp , \eta = -\i\,  \chi) \ .
\end{align}

To examine the leading behavior of the four-point function in the Regge-like limit \eqref{timelike-Regge-parametrization},let us introduce the following function as in section \ref{sec:Plugging into four-point function},
\begin{align}
     \bm{F}^\mathrm{T}(\bm{x}_\perp, r)
        \equiv \frac{1}{r}\,\int_{- \infty}^\infty \d \alpha  \, \langle\Omega |\,\CO_{4}(x_4)\,\tilde \Phi^{\mathrm{T}} \left( u = 0, v = \alpha , {\bm x}_\perp , \chi = r\right)\,\CO_{3}(x_3)\,| \Omega \rangle \ .
\end{align}
Given the relation \eqref{relation btw timelike and spacelike}, we can verify that $\bm{F}^\mathrm{T}$ satisfies the equation of motion \eqref{EoM-for-F} with the replacement $r \rightarrow  - \i\, r$ and can identify it with the spacelike counterpart \eqref{Emb-F} as 
\begin{align}\label{F in timelike}
    \bm{F}^\mathrm{T}(\bm{x}_\perp , r) = e^{\frac{\i \pi}{2}J }\, \bm{F} ( \bm{x}_\perp, \, - \i\,  r) \ .
\end{align}
As in section \ref{sec:Plugging into four-point function}, we set $\bm{x}_\perp = \bm 0$ to reproduce the Regge behavior of conformal blocks. In taking the $r \to 0$ limit in \eqref{F in timelike},  $F^\mathrm{T} (\bm{x}_\perp ,r) $ correctly reproduces \eqref{Regge-dom-time-order} in the timelike configuration with the additional phase factor $\exp\left[\frac{\i\pi}{2}(\Delta^+_{12}-\Delta+J)\right]$, which can be fixed by comparing the small $r$ limits of the spacelike OPE block \eqref{OPEB_BC} and the timelike one:
\begin{align}\label{OPEB-TL-Regge-gen}
    \begin{aligned}
    &\CB^{\text{T},(\text{Regge})}_{\Delta,J}(x_1,x_2)\\
    &\qquad \xrightarrow[r\to0]{} \, e^{\frac{\i\pi}{2}(\Delta^+_{12}-\Delta+J)} \,\frac{(-1)^J\, 2\, b_{12,[\Delta,J]}}{(2r)^{\Delta^+_{12}}}\,e^{(J-1)t_{\mathrm{R}}}\,r^{\Delta-1}\,\bL[\CO_{\Delta, J}](P_0, Z_0) \ .
    \end{aligned}
\end{align}

While our derivation is based on the proper $\i\,\epsilon$-prescription, the expression \eqref{OPEB-TL-Regge-gen} follows more directly from \eqref{OPEB_BC} by analytically continuing the coordinates from the spacelike configuration to the timelike one:
\begin{align}
    x_2 = - x_1 = (- r\, e^{ - t_{\mathrm{R}}} , r\, e^{t_{\mathrm{R}}}, \bm{0}_\perp) \rightarrow  (r\, e^{ - t_{\mathrm{R}}}, r\, e^{t_{\mathrm{R}}},\bm{0}_\perp ) \, ,
\end{align}
which is equivalent to 
\begin{align}
    r\to e^{-\frac{\i\pi}{2}}\, r\ , \qquad t_{\mathrm{R}}\to t_{\mathrm{R}} + \frac{\i\,\pi}{2} \, .
\end{align}
This is (the inverse of) the analytic continuation employed by \cite{Afkhami-Jeddi:2017rmx,Hartman:2016lgu} in deriving the Regge behavior of conformal blocks by stating from the timelike OPE block.

\subsection{More on the equivalence between two timelike OPE blocks}\label{ss:normalization_timelike_OPEB}

The relation \eqref{Relation_bet_TOPEB} between $\CB^\mathrm{T}$ and $\CB^\diamondsuit$ was derived by comparing their OPE limits with an appropriate choice of the normalization constant $\tilde b_{12, [\Delta, J]}$ for $\CB^\diamondsuit$, which has been left undetermined so far, but will be fixed shortly below.

To this end, let us consider the coincident limit  $x_1\to x_2$ while keeping their causal order $x_1<x_2$.
The integration region $\diamondsuit_{12}$ shrinks to zero in the limit, so we can approximate $\mathcal{O}_{\Delta,\mu_1\cdots\mu_J}(x_0)\simeq  \mathcal{O}_{\Delta,\mu_1\cdots\mu_J}(x_2)$ in the integrand of \eqref{timelikeOPE-SP}, which means that we ignore all the contributions from its descendants.
Then from \eqref{TL-coin-general} we expect the integral to reduce to the primary contribution in the OPE:
\begin{align}\label{coin-tensor-str}
    \begin{aligned}
       & (-1)^J\, \tilde{b}_{12,[\Delta,J]}\,\CO_{\Delta,\mu_1\cdots \mu_J}(x_2)\int_{x_0\in \diamondsuit_{12}}\d^d x_{0}\, \frac{H^{\mu_1} (x_{10},x_{20}) \cdots H^{\mu_J} (x_{10},x_{20})}{|x_{12}^2|^{\frac{\Delta^+_{12}-\bar{\Delta}+J}{2}}\,|x_{10}^2|^{\frac{\Delta^-_{12}+\bar{\Delta}-J}{2}}\,|x_{20}^2|^{\frac{-\Delta^-_{12}+\bar{\Delta}-J}{2}}}\\
       &\qquad\qquad =\frac{1}{|x_{12}^2|^{\frac{\Delta^+_{12}-\Delta+J}{2}}}\,x_{12}^{\mu_1}\cdots x_{12}^{\mu_J}\,\mathcal{O}_{\Delta,\mu_1\cdots\mu_J}(x_2) \ .
    \end{aligned}
\end{align}
Since $\CO_{\Delta, \mu_1\cdots\mu_J}$ is a symmetric traceless tensor, to determine $\tilde{b}_{12,[\Delta,J]}$ it is enough to evaluate \eqref{coin-tensor-str} by substituting $\CO_{\Delta, \mu_1\cdots\mu_J} \to z_{\mu_1}\cdots z_{\mu_J}$ with a null polarization vector $z$:
\begin{align}\label{contract-z-coin}
    \frac{(z\cdot x_{12})^J}{|x_{12}^2|^{\frac{\Delta^+_{12}-\Delta+J}{2}}} = (-1)^J\,\tilde{b}_{12,[\Delta,J]} \int_{x_0\in \diamondsuit_{12}}\d^d x_{0}\, \frac{(z\cdot H (x_{10},x_{20}))^J }{|x_{12}^2|^{\frac{\Delta^+_{12}-\bar{\Delta}+J}{2}}\,|x_{10}^2|^{\frac{\Delta^-_{12}+\bar{\Delta}-J}{2}}\,|x_{20}^2|^{\frac{-\Delta^-_{12}+\bar{\Delta}-J}{2}}}\ .
\end{align}

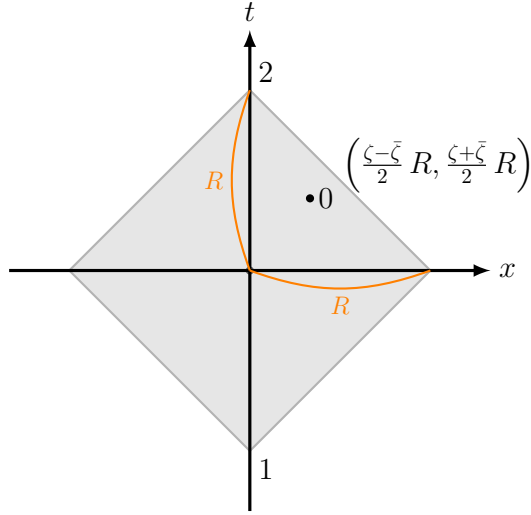
\begin{figure}[t!]
	\centering
	\begin{tikzpicture}[thick,scale=0.8, transform shape]
       \filldraw[gray!20] (0,-3) -- (3,0) -- (0,3) -- (-3,0);
        \draw[thick,black!30] (0,-3) -- (3,0);
        \draw[thick,black!30] (0,3) -- (-3,0); 
         \draw[thick,black!30] (3,0) -- (0,3);
            \draw[thick,black!30] (-3,0) -- (0,-3);
            \draw[very thick,->] (0,-4) to (0,4);
          \node[above] at (0,4) {\Large $t$};
           \draw[very thick,->] (-4,0) to (4,0);
            \node[right] at (4,0) {\Large $x$};
         \filldraw[black,opacity=0.6] (0,0) circle (0.05);
         \node[above right] at (0,3) {\Large $2$};
         \node[below right] at (0,-3) {\Large $1$};
         \draw[fill=black] (1,1.2) node[right]  {\Large $0$} circle (0.05);
         \node[right] at (1.4,1.75) {\Large $\left(\frac{\zeta-\bar{\zeta}}{2}\,R,\frac{\zeta+\bar{\zeta}}{2}\,R\right)$};
          \draw[-,orange] (0,0) to[out=110,in=-110] node [left, midway, font=\large] {$R$} (0,3);
           \draw[-,orange] (0,0) to[out=-20,in=-160] node [below, midway, font=\large] {$R$} (3,0);
	\end{tikzpicture}
	\caption{The causal diamond $\diamondsuit_{12}$ corresponding to the configuration \eqref{0simpleconfig} and \eqref{1simpleconfig} when $d=2$.}
	\label{fig:TL-coin-OPE}
\end{figure}

Without loss of generality, we can choose such a null vector and set the positions of the external operators as
\begin{align}\label{0simpleconfig}
    x_2=-x_1=(R,\vec{0})\ , \qquad z=(z^0,z^1,\bm{z}_\perp)=(-1,-1,\bm{0}_\perp)\ .
\end{align}
We then parametrize the coordinate $x_0$ in the diamond $\diamondsuit_{12}$ by 
\begin{align}\label{1simpleconfig}
     x_0=\left(\frac{\zeta-\bar{\zeta}}{2}\,R,\frac{\zeta+\bar{\zeta}}{2}\,R\,\vec{\Omega}_{d-2}\right)\ , \qquad  \vec{\Omega}_{d-2}\in\mathbb{S}^{d-2}\ , \qquad \zeta,\,\bar{\zeta}\in [-1,1]\ .
\end{align}
See figure \ref{fig:TL-coin-OPE} for $d=2$ case.
In this parametrization, the volume element becomes 
\begin{align}
        \sqrt{-g} \, \d\zeta\,\d\bar{\zeta}\,\d\Omega_{d-2}  = \frac{R^d}{2^d}\,|\zeta+\bar{\zeta}|^{d-2}\,\d\zeta\,\d\bar{\zeta}\,\d\Omega_{d-2} \ .
\end{align}
Note that we divide the measure by two to take into account the fact that this parametrization covers the causal diamond twice for $d\ge 3$ as $x_0$ is invariant under the replacement:
\begin{align}
    \zeta \to -\bar \zeta \ , \qquad \bar\zeta \to - \zeta \ , \qquad \vec\Omega_{d-2} \to - \vec\Omega_{d-2} \ .
\end{align}

By further decomposing the spherical coordinates $\vec{\Omega}_{d-2}$ into 
\begin{align}
      \vec{\Omega}_{d-2}=(\Omega^1_{d-2},\bm{\Omega}^\perp_{d-2})=(\cos\theta,\sin\theta\,\bm{\Omega}^\perp_{d-3})\ ,\qquad 0\le \theta \le \pi \ ,
\end{align}
\eqref{contract-z-coin} is reduced to 
\begin{align}\label{CF from coin}
        \tilde{b}_{12,[\Delta,J]}^{-1}=  I(d, J, \Delta, \Delta_{12}^-) \ ,
\end{align}
where $I(d, J, \Delta, \Delta_{12}^-)$ is the triple integral defined by\footnote{This is valid for $d\ge 3$, but it can be analytically continued to $d=2$.}
\begin{align}\label{triple integral}
\begin{aligned}
   I(d, J, \Delta, \Delta_{12}^-) &\equiv
   \frac{\pi^{\frac{d}{2}-1}}{2^{2\Delta+J-1}\,\Gamma\left(\frac{d}{2}-1\right)}\,\int_{-1}^{1}\d\zeta\int_{-1}^{1}\d\bar{\zeta}\, |\zeta+\bar{\zeta}|^{d-2}\,\int_{0}^\pi\d\theta\, \sin^{d-3}\theta\,
     \\
    &\times
    [(1+\zeta)(1-\bar{\zeta})]^{\frac{\Delta-J-d+\Delta^-_{12}}{2}}\,
    [(1-\zeta)(1+\bar{\zeta})]^{\frac{\Delta-J-d-\Delta^-_{12}}{2}}\,[2-(\zeta^2+\bar{\zeta}^2)-(\zeta^2-\bar{\zeta}^2)\cos\theta]^J
     \ .
    \end{aligned}
\end{align}

Now it remains to perform the integration in \eqref{triple integral}.
It is however not known to us if it can be given a closed form for general $d, J,\Delta$, and $\Delta^-_{12}$.
Nevertheless, there are two cases where we can carry out the integration explicitly:
\begin{itemize}
    \item[1.] $J=0\, , \,\Delta^-_{12}=0$:\\
    When $J=0$ and the external operators are identical, we find
    \begin{align}\label{remaining J=0 and identical}
        I(d, 0, \Delta, 0)
           = \frac{\pi^{\frac{d}{2}-1}}{2}\,\frac{\Gamma\left(\frac{\Delta}{2}\right)^2\,\Gamma\left(\frac{\Delta+2-d}{2}\right)^2}{\Gamma(\Delta)\,\Gamma\left(\Delta+1-\frac{d}{2}\right)}\ ,
    \end{align}
    by expanding $|\zeta+\bar{\zeta}|^{d-2}$ and performing term-by-term integrations.
    \item[2.] $J=0\, , \,d=2$:\\
     When $d=2,J=0$ the integral factorizes and using the formula
     \begin{align}\label{Beta-func-formula}
         \int_\alpha^\beta\d t\,(t-\alpha)^{\gamma-1}\,(\beta-t)^{\delta-1}
        = (\beta-\alpha)^{\gamma+\delta-1}\,\frac{\Gamma(\gamma)\,\Gamma(\delta)}{\Gamma(\gamma+\delta)}\ ,
     \end{align}
     we find\footnote{Here, we use the shorthanded notation $\Gamma(x\pm y)\equiv \Gamma(x+y)\,\Gamma(x-y)$.} 
     \begin{align}\label{remaining J=0 and d=2}
        I(2, 0, \Delta, \Delta_{12}^-)      
        =
         \frac{1}{2}\,
         \frac{\Gamma\left(\frac{\Delta\pm \Delta^-_{12}}{2}\right)^2}{\Gamma(\Delta)^2}\ .
     \end{align}
\end{itemize}

While we are left with the analytically intractable integral $I(d,J,\Delta,\Delta_{12}^-)$ in general, there still be an alternative way to fix $\tilde{b}_{12,[\Delta,J]}$ which is worth to the examination.
This is archived by the Regge limit of $\CB^\diamondsuit$ followed by taking $r \rightarrow 0$.
The behavior of $\CB^\text{T}$ is already fixed by \eqref{OPEB-TL-Regge-gen}, thus
we can compare it with the small $r$ behavior of $\CB^\diamondsuit$ to determine $\tilde{b}_{12,[\Delta,J]}$.
In what follows we will conduct this calculation explicitly and find the analytic formula for $I(d,J,\Delta,\Delta_{12}^-)$.

In the Regge parametrization \eqref{timelike-Regge-parametrization} the causal diamond $\diamondsuit_{12}$ becomes
\begin{align}
    \begin{aligned}
        &|v|\leq r\, e^{t_{\mathrm{R}}}\ ,\qquad \bm{x}^2_\perp\leq r_+r_-\ , \qquad u_-\leq\, u\leq \, u_+\ ,
    \end{aligned}
\end{align}
where we introduced short-handed symbols:
\begin{align}
    \begin{aligned}
         &r_\pm= r\pm v\, e^{-t_{\mathrm{R}}}\ , \qquad u_\pm=\pm r\, e^{t_{\mathrm{R}}}\mp \frac{\bm{x}^2_\perp}{r\, e^{t_{\mathrm{R}}}\mp v}\ .
    \end{aligned}
\end{align}
In the timelike Regge limit \eqref{regge limit in timelike},
the dominant contribution of 
\begin{align*}
H^{\mu_1}(x_{10},x_{20})\cdots
H^{\mu_J}(x_{10},x_{20})\,\CO_{\Delta,\mu_1\cdots \mu_J}(x_0)    
\end{align*}
comes from 
\begin{align}
\begin{aligned}
  (H^{v}(x_{10},x_{20}))^J\,\CO_{\Delta,v\cdots v}(u=0,v,\bm{x}_{\perp})\ ,  \qquad  x_0=(u,v,\bm{x}_\perp) \ ,
   \end{aligned}
\end{align} 
all the others being subleading.
Then, the integral in \eqref{timelikeOPE-SP} becomes 
\begin{align}
    \begin{aligned}
        &(-1)^J\int_{x_0\in \diamondsuit_{12}}\d^d x_{0}\, \frac{H^{\mu_1} (x_{10},x_{20}) \cdots H^{\mu_J} (x_{10},x_{20}) \, \CO_{\Delta,\mu_1\cdots \mu_J}(x_0)}{|x_{12}^2|^{\frac{\Delta^+_{12}-\bar{\Delta}+J}{2}}\,|x_{10}^2|^{\frac{\Delta^-_{12}+\bar{\Delta}-J}{2}}\,|x_{20}^2|^{\frac{-\Delta^-_{12}+\bar{\Delta}-J}{2}}}\\
        &\simeq \frac{1}{2}\int_{-r\, e^{t_{\mathrm{R}}}}^{r\, e^{t_{\mathrm{R}}}}\d v  \int_{\bm{x}_\perp^2\leq r_+r_- } \d^{d-2}\bm{x}_\perp \frac{(-1)^J\,[2r\, e^{t_{\mathrm{R}}}(r_+r_- -\bm{x}^2_\perp)]^J \,\CO_{\Delta , v\cdots v}(u=0,v,\bm{x}_\perp) }{(2r)^{\Delta^+_{12}-\bar{\Delta}+J}\,e^{(\bar{\Delta}-J)t_{\mathrm{R}}}\,r_-^{\frac{-\Delta^-_{12}+\bar{\Delta}+J}{2}}\,r_+^{\frac{\Delta^-_{12}+\bar{\Delta}+J}{2}}} \\
        &\qquad\qquad\times   \int_{u_-}^{u_+}\d u\,
       (u-u_-)^{\frac{-\Delta^-_{12}-\bar{\Delta}-J}{2}} (u_+-u)^{\frac{+\Delta^-_{12}-\bar{\Delta}-J}{2}} \\
         &= \frac{(-1)^J\, \Gamma\left(\frac{-\bar{\Delta}-J+2\pm \Delta^-_{12}}{2}\right)}{2^{J}\, \Gamma(\Delta-d-J+2)}\times \frac{r^{1-J}\,e^{(J-1)t_{\mathrm{R}}}}{(2r)^{\Delta^+_{12}}}\\
         &\qquad\qquad\times \int_{-r\, e^{t_{\mathrm{R}}}}^{r\, e^{t_{\mathrm{R}}}}\d v  \int_{\bm{x}_\perp^2\leq r_+r_- } \d^{d-2}\bm{x}_\perp \frac{(r_+r_- -\bm{x}^2_\perp)^{-\bar{\Delta}+1}}{r_-^{\frac{-\Delta^-_{12}-\bar{\Delta}-J+2}{2}}r_+^{\frac{\Delta^-_{12}-\bar{\Delta}-J+2}{2}}}\,  \CO_{\Delta , v\cdots v}(u=0,v,\bm{x}_\perp)\ ,
    \end{aligned}
\end{align}
where we used the integration formula \eqref{Beta-func-formula} for $u$.
In the Regge limit, $t_{\mathrm{R}}\to \infty$, it simplifies to
\begin{align}
    \begin{aligned}
         &\frac{(-1)^J\,\Gamma\left(\frac{-\bar{\Delta}-J+2\pm\Delta^-_{12}}{2}\right)}{2^J\,\Gamma(-\bar{\Delta}-J+2)}\times \frac{r^{-\bar{\Delta}-1}\,e^{(J-1)t_{\mathrm{R}}}}{(2r)^{\Delta^+_{12}}}\int_{-\infty}^\infty \d v\int_{\bm{x}_\perp^2\leq r^2 } \d^{d-2}\bm{x}_\perp\, (r^2-\bm{x}^2_\perp)^{-\bar{\Delta}+1}\,\CO_{\Delta , v\cdots v}(u=0,v,\bm{x}_\perp) \ .
    \end{aligned}
\end{align}
Subsequently taking $r \rightarrow 0$, we finally obtain
\begin{align}
    \begin{aligned}
         & \frac{(-1)^J\,\pi^{\frac{d}{2}-1}\,\Gamma\left(\frac{-\bar{\Delta}-J+2\pm\Delta^-_{12}}{2}\right)\,\Gamma(2-\bar{\Delta})}{2^{J}\,\Gamma(-\bar{\Delta}-J+2)\,\Gamma\left(\Delta+1-\frac{d}{2}\right)}\times \frac{r^{\Delta-1}\, e^{(J-1)t_{\mathrm{R}}}}{(2r)^{\Delta^+_{12}}}\,\bL[\CO_{\Delta, J}](P_0, Z_0) \ .
    \end{aligned}
\end{align}
Comparing with \eqref{OPEB-TL-Regge-gen}, we can read off
\begin{align}\label{CF from regge}
\begin{aligned}
    &\tilde{b}_{12,[\Delta,J]}=\frac{2\,\Gamma(\Delta+J)\,\Gamma(\Delta+\bar{J})\,\Gamma\left(\Delta+1-\frac{d}{2}\right)}{\pi^{\frac{d}{2}-1}\,\Gamma\left(\frac{\Delta+J\pm \Delta^-_{12}}{2}\right)\,\Gamma\left(\frac{\Delta+\bar{J}\pm\Delta^-_{12}}{2}\right)\,\Gamma(\Delta+2-d)} \ .
\end{aligned}    
\end{align}
This is manifestly symmetric under the exchange $J\leftrightarrow\bar{J}=2-d-J$, which can be related to the spin shadow symmetry of the conformal block $G_{\Delta,J}(z,\bar{z})=G_{\Delta,\bar{J}}(z,\bar{z})$.

It follows from the equations \eqref{CF from coin} and \eqref{CF from regge} that the triple integral \eqref{triple integral} should have the following analytic form:
\begin{align}
    \begin{aligned}
    &I(d, J, \Delta, \Delta_{12}^-) = \frac{\pi^{\frac{d}{2}-1}}{2}\,\frac{\Gamma(\Delta+2-d)\,\Gamma\left(\frac{\Delta+J\pm\Delta^-_{12}}{2}\right)\,\Gamma\left(\frac{\Delta+\bar{J}\pm\Delta^-_{12}}{2}\right)}{\Gamma\left(\Delta+1-\frac{d}{2}\right)\,\Gamma(\Delta+J)\,\Gamma(\Delta+\bar{J})}\ .
    \end{aligned}
\end{align}
This identity is consistent with the special cases \eqref{remaining J=0 and identical} and \eqref{remaining J=0 and d=2}.
Moreover one can easily check the validity for general cases numerically.

For later use, we generalize the timelike OPE block \eqref{timelikeOPE-SP} to continuous spins.
The tensor construction is taken over by the conformal integral for the polarization vector $z$.
We defer the detail to appendix \ref{sec:continuous spin OPE block}, showing only the final result:
\begin{align}
    \CB^\diamondsuit_{\Delta, J} (x_1, x_2) = b^\diamondsuit_{12,[\Delta,J]}\,\int_{x_0\in \diamondsuit_{12}} \d^d x_0\, D^{d-2}z\, \langle \tilde 0 |\, \CO_1(x_1)\,\CO_2( x_2)\,\CO_{\bar\Delta, \bar J}(x_0, z)\,|\tilde 0\rangle\, \CO_{\Delta, J}(x_0, z) \, ,
\end{align}
where 
\begin{align}
        b^\diamondsuit_{12,[\Delta,J]} 
            = \frac{(-2)^{\bar{J}}\,\Gamma\left (-\bar{J}\right)}{\pi^{\frac{d}{2}-1}\Gamma\left(J+\frac{d}{2}-1\right)}\, \tilde{b}_{12,[\Delta,J]} \ .
\end{align}

\section{Light-ray channel OPE block}\label{sec:Light-ray channel OPE block}

Let us recapitulate our results obtained so far before moving onto a further discussion on their implications.
In section \ref{sec:Regge conformal block via Lorentzian OPE} we considered the Regge limit of the spacelike OPE block and ended up with the simple holographic description \eqref{regge OPE block-J} of the Regge OPE block as a higher-spin AdS field smeared over a null geodesic, which is valid for general $\Delta$ and an integer $J$.
We also showed that the Regge OPE block is subject to the ``holographic" boundary condition \eqref{OPEB_BC} characterized by the light-ray operator, which may be written as the asymptotic form of the spacelike OPE block in the Regge configuration:
\begin{align}
    \CB_{\Delta, J}(x_1, x_2)
        \quad \underset{t_{\mathrm{R}}\to\infty,~ r\to 0}{\sim}\quad
        e^{(J-1)\,t_{\mathrm{R}}}\,r^{\Delta - \Delta_{12}^- -1} \, \bL[\CO_{\Delta, J}] \left(P_0, Z_0\right) \ .
\end{align}
This relation was the key to reproduce the Regge behavior of the conformal block from the vacuum OPE block promoted to the operator:
\begin{align}\label{OPEB_Regge_limit}
    \langle\,\CO_3(x_3)\,\CB_{\Delta, J}(x_1, x_2)\,\CO_4(x_4)\,\rangle
        \quad\underset{\text{Regge limit}}{\sim}\quad G_{1-J, 1-\Delta}(z, \bar z) \ ,
\end{align}
where $G_{1-J, 1-\Delta}(z, \bar z)$ is the conformal block given by \eqref{ReggeCPW-LT} satisfying the asymptotic boundary condition \eqref{ReggeCPW_asymptotic} corresponding to the light-ray operator of quantum dimensions $(1-J, 1-\Delta)$.

If we are only concerned with a conformal block as a function of the cross ratios, we can no longer distinguish the Regge limit from the ordinary OPE limit as the cross ratios approach zeros in both cases (see \eqref{Regge_cross_ratio} for the Regge limit).
In spacetime picture, this is due to the invariance of the cross ratios under a null translation $\CT$ of a point from one Minkowski patch to another (see figure \ref{fig:Regge_LT}) as explained in section \ref{ss:Adapted_coord}.
Bearing in mind the indistinguishability between the two limits through conformal blocks, 
\eqref{OPEB_Regge_limit} leads us to a speculation, put forward by \cite{Caron-Huot:2017vep,Kravchuk:2018htv}, that the Regge limit of the OPE exchanging an operator $\CO_{\Delta, J}$ can be interpreted as the ordinary OPE limit of a pair of operators at the positions $1$ and $2^-$ exchanging the light-ray operator $\bL[\CO_{\Delta, J}]$.

While the above perspective is plausible and sheds light on the new role of the spacetime structure in Lorentzian CFT, it remains open how to realize this line of thought in a concrete setting.
To tackle this issue in a feasible way we wish to make this statement more precise by promoting the relation \eqref{OPEB_Regge_limit} to an operator identity:
\begin{align}\label{OPE_to_LOPE_temp}
    \CB_{\Delta, J}(x_1, x_2)\quad \xrightarrow[\text{Regge limit}]{} \quad \BB_{\bL [\Delta,J]} (x_1,x_2) \ ,
\end{align}
where $\BB_{\bL [\Delta,J]} (x_1,x_2)$ is an OPE block exchanging the light-ray operator $\bL[\CO_{\Delta, J}]$ in the operator product $\CO_1(x_1)\, \CO_2(x_2)$, which we will call the \emph{light-ray channel OPE block}.
In order for the vev $\langle\,\CO_3(x_3)\,\BB_{\bL [\Delta,J]} (x_1,x_2)\,\CO_4(x_4)\,\rangle$ to reproduce the conformal block $G_{1-J, 1-\Delta}(z, \bar z)$ the light-ray channel OPE block must have a dominate contribution from $\bL[\CO_{\Delta, J}]$ in the Regge limit, which is seen as the OPE limit $1\to 2^-$ in the universal cover of Minkowski patch:
\begin{align}
    \BB_{\bL [\Delta,J]} (x_1,x_2) \quad \underset{x_1 \to x_{2^-}}{\sim}\quad |x_{12^-}|^{-\Delta_{12}^+ + \Delta -J}\,\bL [\CO_{\Delta, J}](x_2, z = x_{12^-}) \ .
\end{align}

The new OPE block must transform in the same way as the original OPE block under the conformal group.
To keep manifest the conformal covariance it would be straightforward to use an analogue of the projector formalism \cite{SimmonsDuffin:2012uy,Ferrara:1972ay,Ferrara:1972uq,Ferrara:1972xe,Ferrara:1973vz} incorporating light-transformed operators.
Such a projector, however, ceases to exist as light-ray operators annihilate the vacuum. 
Without delving into this issue in detail, but inspired by the projector formalism, we propose an integral representation of the light-ray channel OPE block which meets the aforementioned requirements:
\begin{align}\label{Light_OPE_block}
    \BB_{\bL [\Delta,J]}(x_1,x_2) = \Bb_{12, \bL[\Delta,J]}\,\int_{2^- < 0 < 1} \d^d x_0\,D^{d-2} z\, \langle \tilde{0} |\,\CO_1(x_1)\, \CO_{\bL[\bar\Delta, \bar J]}(x_0, z)\,\CO_2(x_2)\,|\tilde{0}\rangle\,\bL [\CO_{\Delta, J}](x_0, z) \ ,
\end{align}
where $\langle \tilde{0} |\,\CO_1(x_1)\, \CO_{\bL[\bar\Delta, \bar J]}(x_0, z)\,\CO_2(x_2)\,|\tilde{0}\rangle$ is the three-point structure normalized as in \eqref{normalized-3py-str} including an operator with the same quantum number $\bar\Delta = d - \Delta$ and $\bar J = 2-d - J$ as the light-ray operator $\bL[\CO_{\bar\Delta, \bar J}]$.

When a pair of points $1$ and $2$ are spacelike, the points $1$ and $2^-$ become timelike as in figure \ref{fig:Regge_LT}.
We then recognize that \eqref{Light_OPE_block} takes the same form as the representation $\CB^\diamondsuit$ of a timelike OPE block proposed by \cite{deBoer:2016pqk,Czech:2016xec} (see also section 5.3 in \cite{Chen:2019fvi} for the related discussion).
The correspondence becomes clearer in moving to the adapted coordinates \eqref{Adapted_coord} where the two points $\check 1$ and $\check 2$ associated to the pair of operators are timelike-separated as in figure \ref{fig:Regge_adapted1}.
In the adapted coordinates, \eqref{Light_OPE_block} becomes
\begin{align}\label{Light_OPE_block_adapted}
    \begin{aligned}
    &\BB_{\bL [\Delta,J]}(x_1,x_2) \\
        &\quad = \Bb_{12, \bL[\Delta,J]}\,|\check{v}_1|^{\Delta_1}\,|\check{v}_2|^{\Delta_2}\,\int_{\check 2 < \check 0 < \check 1} \d^d \check{x}_0\,D^{d-2} z\, \langle \tilde 0 |\,\widecheck{\CO}_1(\check{x}_1)\,\widecheck{\CO}_{\bL [\bar \Delta, \bar J]}(\check{x}_0, z)\,\widecheck{\CO}_2(\check{x}_2)\,|\tilde 0\rangle\,\widecheck{\bL [\CO_{\Delta, J}]}(\check{x}_0, z) \ ,
    \end{aligned}
\end{align}
where the factor $|\check{v}_1|^{\Delta_1}\,|\check{v}_2|^{\Delta_2}$ arises due to the Weyl transformation properties of the external scalar primaries $\CO_{1}(x_1), \, \CO_{2}(x_2)$ as in \eqref{Weyl-adapted}.
For the internal operators $\widecheck{\CO}_{\bL [\bar \Delta, \bar J]}(\check{x}_0, z),\widecheck{\bL [\CO_{\Delta, J}]}(x_0,z)$, no additional factors appear as the integral with respect to $x_0$ is conformally invariant. 
Using the OPE limit of the timelike OPE block \eqref{TL-coin-general}, we find 
\begin{align}
  \BB_{\bL [\Delta,J]}&(x_1,x_2)
    ~\xrightarrow[\check{x}_1 \to \check{x}_2]{}~
    \frac{\Bb_{12, \bL[\Delta,J]}}{b^\diamondsuit_{12,\bL[\Delta,J]}}\,
   |\check{v}_1|^{\Delta_1}\,|\check{v}_2|^{\Delta_2}\,
    |\check{x}_{12}|^{-\Delta_{12}^+ + \Delta -J }\,\widecheck{\bL [\CO_{\Delta, J}]}(\check{x}_2, z = \check{x}_{12}) \ .
\end{align}

Now we put \eqref{Light_OPE_block_adapted} into the Regge configuration \eqref{Regge_parametrization} in the polar coordinates \eqref{Regge_polar}.
In the adapted coordinates, $v_1 = -v_2 = r\, e^{t_{\mathrm{R}}}$ and 
\begin{align}
    \check{x}_{12}^2 = -4\,e^{-2t_{\mathrm{R}}} \ , \qquad \check{x}_{12} = \left( 2\,r\,e^{-t_{\mathrm{R}}}, 2\,e^{-t_{\mathrm{R}}}/r, \b0\right) \ ,
\end{align}
so in the Regge limit, $\check{x}_1, \check{x}_2 \to 0$ $(t_{\mathrm{R}}\to \infty)$, followed by the $r\to 0$ limit,
we find
\begin{align}
    \begin{aligned}
        &\BB_{\bL [\Delta,J]} (x_1,x_2)  \xrightarrow[t_{\mathrm{R}} \to \infty,\, r\to 0]{}
        \frac{\Bb_{12, \bL[\Delta,J]}}{b^\diamondsuit_{12,\bL[\Delta,J]}}\,(2r)^{-\Delta_{12}^+}\,
        2^{ 1 - J}\,
        e^{(J-1)t_{\mathrm{R}}}\,r^{\Delta - 1}\,
        \bL [\CO_{\Delta, J}](P_0, Z_0) \ .
    \end{aligned}
\end{align}
In deriving this, we approximate the position of $\bL[\CO_{\Delta, J}]$ by the origin of the adapted coordinates and rewrite the operator using \eqref{general-gaugeP^+} in terms of the embedding coordinate $\check{P}^A_0,\,\check{Z}^A_0$:
\begin{align}
    \check{P}^A_0=(1,0,0,0,\bm{0}_\perp)\ , \qquad \check{Z}^A_0=(0,0,0,1,\bm{0}_\perp)\ ,
\end{align}
then pull back the operator $\widecheck{\bL[\CO_{\Delta, J}]}(\check{P}_0, \check{Z}_0)$ in the adapted section to $\bL [\CO_{\Delta, J}](P_0, Z_0)$ in the Poincar\'e section according to \eqref{Adapted-embedding-transf} and \eqref{P0X0}.
Compared with \eqref{OPEB_BC} the light-ray channel OPE block coincides with the Regge limit of the OPE block as in \eqref{OPE_to_LOPE_temp} if the coefficient is chosen such that
\begin{align}
        \Bb_{12, \bL[\Delta,J]} = \frac{(-1)^J\,\Gamma(\Delta+J)}{\Gamma\left(\frac{\Delta+J\pm\Delta^-_{12}}{2}\right)}\,b^\diamondsuit_{12,\bL[\Delta,J]} \ .
\end{align}

To encapsulate the result of this section, we proposed the light-ray channel OPE block $\BB_{\bL[\Delta, J]}(x_1, x_2)$ by \eqref{Light_OPE_block} that exchanges a light-ray operator $\bL[\CO_{\Delta, J}]$ in the OPE between a pair of points $1$ and $2$.
The construction of $\BB_{\bL[\Delta, J]}(x_1, x_2)$ is based on the equivalence between the two spacelike-separated points $1$ and $2$ and a pair of timelike-separated points $1$ and $2^-$ in the universal cover $\widetilde \CM_d$ of the Minkowski patch, and $\BB_{\bL[\Delta, J]}(x_1, x_2)$ takes the same form as the timelike OPE block $\CB^\diamondsuit_{\bL[\Delta, J]}(x_1, x_{2^-})$ (or equivalently $\CB^\diamondsuit_{\bL[\Delta, J]}(\check x_1, \check x_2)$ in the adapted coordinates).
We have also shown \eqref{OPE_to_LOPE_temp} relating the spacelike OPE block $\CB_{\Delta, J}$ in one Minkowski patch to the light-ray channel OPE block $\BB_{\bL[\Delta, J]}$ that is a timelike OPE block in a different Minkowski patch, confirming the assertion foreseen by \cite{Caron-Huot:2013fea,Kravchuk:2018htv} about the equivalence between the Regge limit in one Minkowski patch and the OPE limit in another.

\section{Discussion}\label{sec:Discussions}
The relation \eqref{OPE_to_LOPE_temp} between the spacelike OPE block $\CB_{\Delta,J}(x_1,x_2)$ and the light-ray channel OPE block $\BB_{\bL [\Delta,J]}(x_1,x_2)$ we established is more concrete but weaker than the original speculation advocated in \cite{Caron-Huot:2017vep,Kravchuk:2018htv} that the Regge limit of the OPE exchanging an operator $\CO_{\Delta, J}$ can be interpreted as the OPE limit of a pair of operators at the positions $1$ and $2^-$ exchanging the light-ray operator $\bL[\CO_{\Delta, J}]$.
Having the speculation in mind we want to promote the relation \eqref{OPE_to_LOPE_temp} to an operator identity relating the two blocks $\CB_{\Delta,J}(x_1,x_2)$ and $\BB_{\bL [\Delta,J]}(x_1,x_2)$ that can hold even away from the Regge limit.

To this end let us examine the conformal block $\langle\,\CO_1(x_1)\,\CO_2(x_2)\,\CO_3(x_3)\,\CO_4(x_4)\,\rangle$ with different operator ordering than before.
Using the OPE block the Regge limit becomes
\begin{align}\label{OPE_no_monodromy}
    \langle\,\CB_{\Delta, J}(x_1, x_2)\,\CO_3(x_3)\,\CO_4(x_4)\,\rangle
        \quad\underset{\text{Regge limit}}{\sim}\quad G_{\Delta, J}(z, \bar z) \ ,
\end{align}
as the contour in the space of the cross ratios does not cross the branch cut in this case.
The right-hand side is the same conformal block as in the OPE channel $1\to 2$, but in the present case it is physically more natural to null-translate the point $2$ to $2^-$ in a different patch and interpret $G_{\Delta, J}(z, \bar z)$ as a block in the timelike OPE channel $1 \to 2^-$ exchanging the operator $\CO_{\Delta, J}$.

Depending on the operator ordering we find the two relations in the Regge limit; one \eqref{OPEB_Regge_limit} for the OPE block in the middle of the correlator, the other \eqref{OPE_no_monodromy} for the OPE block in the left in the correlator.
The OPE block as an operator, on the other hand, should be independent of the position inserted inside correlators.
We presume as a simplest possibility the operator identity for the OPE blocks which is compatible with the two different behaviors:
\begin{align}\label{OPE_to_LOPE_regge}
    \CB_{\Delta, J}(x_1, x_2)\quad \underset{\text{Regge limit}}{\sim}\quad \CB^\diamondsuit_{ \Delta,J}(x_1,x_{2^-}) + \BB_{\bL [\Delta,J]}(x_1,x_2) \ .
\end{align}
The presence of the light-ray channel OPE block in the right hand side is consistent with the relation \eqref{OPE_to_LOPE_temp} as it dominates over the first term in the Regge limit.
In addition the light-transformed operator annihilates the vacuum $\bL [\CO] |\Omega\rangle = 0$ \cite{Kravchuk:2018htv}, so \eqref{OPE_to_LOPE_regge} is also in accordance with \eqref{OPE_no_monodromy}.

Provided the relation \eqref{OPE_to_LOPE_regge} holds in the Regge limit, one may well argue that it should be a general operator identity:
\begin{align}\label{OPE_to_LOPE}
    \CB_{\Delta, J}(x_1, x_2) ~\sim ~ \CB^\diamondsuit_{\Delta,J}(x_1,x_{2^-})+  \BB_{\bL [\Delta,J]}(x_1,x_2) \ .
\end{align}
Indeed one may be able to derive the operator identity from the presumed relation \eqref{OPE_to_LOPE_regge} by resorting to the fact that both sides satisfy the same conformal Casimir equation with \eqref{OPE_to_LOPE_regge} as the boundary condition in the Regge limit.
There are a few ways to confirm an operator identity of this type.
For instance one can check if both sides have the same asymptotic behavior in the ordinary OPE limit $x_1 \to x_2$.
Another nontrivial check would be to see if \eqref{OPE_to_LOPE} could reproduce the transformation law of the conformal block $G_{\Delta,J}$ given in \eqref{Full-ReggeCPW}.
We leave further investigations of these issues for a future work.

The light-ray channel OPE block \eqref{Light_OPE_block} we propose has satisfactory features that it transforms properly under the conformal group and projects the OPE to a light-ray operator channel, reproducing the Regge behavior of the conformal block.
It is, however, not given in a covariant form as we restricted our consideration to the Minkowski patch including the two points $1$ and $2^-$ in section \ref{sec:Light-ray channel OPE block}.
To treat the two points on an equal footing, it would be more natural to use the time-ordered correlator $\langle \,\CO_1(x_1)\,\CO_2(x_2)\,\CO_{\bL[\bar\Delta, \bar J]}(x_0, z)\,\rangle$ as the integration kernel:
\begin{align}\label{Light_OPE_block_time-ordered}
    \BB_{\bL [\Delta,J]}^\text{(cov)} (x_1,x_2) \propto\,\int \d^d x_0\,D^{d-2} z\, \langle \,\CO_1(x_1)\,\CO_2(x_2)\,\CO_{\bL[\bar\Delta, \bar J]}(x_0, z)\,\rangle\,\bL [\CO_{\Delta, J}](x_0, z) \ .
\end{align}
The time-ordered correlator is related to the Wightman structures as\footnote{Following \cite{Kravchuk:2018htv} we distinguish between correlators $\langle \Omega | \cdots |\Omega\rangle$ and structures $\langle 0| \cdots | 0 \rangle$, the latter representing the tensor structures without the OPE coefficients.
}
\begin{align}
    \begin{aligned}
        \langle \,\CO_1(x_1)\,\CO_2(x_2)\,\CO_{\bL[\bar\Delta, \bar J]}(x_0, z)\,\rangle &=  \langle 0 |\,\CO_1(x_1)\,\CO_{\bL[\bar\Delta, \bar J]}(x_0, z)\,\CO_2(x_2)\,|0\rangle\,\Theta(2^- < 0 < 1) \\
            & \quad +  \langle 0 |\,\CO_2(x_2)\,\CO_{\bL[\bar\Delta, \bar J]}(x_0, z)\,\CO_1(x_1)\,|0\rangle\,\Theta(1^- < 0 < 2) \ ,
    \end{aligned}
\end{align}
thus $\BB_{\bL [\Delta,J]}^\text{(cov)} (x_1,x_2)$ consists of two timelike OPE blocks, one the same as $\BB_{\bL [\Delta,J]} (x_1,x_2)$ and the other with the role of $1$ and $2$ exchanged.
We anticipate the additional block is subdominant in the Regge limit and $\BB_{\bL [\Delta,J]}^\text{(cov)} (x_1,x_2)$ also satisfies the same relation as  \eqref{OPE_to_LOPE_temp}.

Related to the covariance issue of the block is whether there exists an analogue of the projector formalism incorporating light-ray operators.
A naive application of the shadow projector \cite{SimmonsDuffin:2012uy,Ferrara:1972ay,Ferrara:1972uq,Ferrara:1972xe,Ferrara:1973vz} fails to work as light-ray operators annihilate the vacuum \cite{Kravchuk:2018htv}.
Nonetheless one can formally write a fully Weyl invariant shadow projector using the principal series representations with complex conformal dimension $\Delta = d/2 + \i\, \mu$ and complex spin $J = (2-d)/2 + \i\,\nu$ where $\mu, \nu \in \BR$ (see e.g.\,\cite{Kravchuk:2018htv,Chen:2019gka}).
It would be worthwhile to explore the relation between the principal series representation and light-ray operators in such a formal shadow projector and see if the resulting OPE block agrees with the light-ray channel OPE block proposed in this paper.

\acknowledgments
We would like to thank H.\,Y.\,Chen and J.\,Sakamoto for useful discussions and especially H.\,Y.\,Chen for careful reading of the manuscript and giving us invaluable comments.
The work of N.\,K. was supported in part by the Program for Leading Graduate Schools, MEXT, Japan and by JSPS Research Fellowship for Young Scientists, and also supported by World Premier International Research Center Initiative (WPI Initiative), MEXT, Japan.
The work of T.\,N. was supported in part by the JSPS Grant-in-Aid for Scientific Research (C) No.19K03863 and the JSPS Grant-in-Aid for Scientific Research (A) No.16H02182.
The work of Y.\,O. was supported by FoPM, WINGS Program, the University of Tokyo.


\appendix

\section{Notations and normalization}\label{sec:Notations and Normalization}
In this appendix, we explain and summarize our notations and normalization of correlation functions used in the main text.

\subsection{Notations}
\begin{itemize}
    \item Coordinate system:
\begin{align}
    \begin{aligned}
        \d  s^2 =-\d  t^2+(\d x^1)^2+\d \bm{x}_{\perp}^2=-\d  u\, \d  v   + \d   \bm{x}_{\perp}^2
    \end{aligned}
\end{align}
where 
\begin{align}
    u=t-x^1\ ,\qquad v=t+x^1\ , \qquad \bm{x}_{\perp}\in\BR^{d-2}\ .
\end{align}
\item Causal relation:
\begin{align}\label{causal-rel}
    \begin{aligned}
        1\approx 2&: x_1 \  \mathrm{and \ }x_2\ \mathrm{are \ spacelike \ separated}, & & x_{12}^2>0  \\
        1>2&:  x_1 \  \mathrm{is \ in\ the\ forward \ lightcone\ of \ }x_2, & & x_{12}^2<0\ ,\quad  t_1 > t_2 
    \end{aligned}
\end{align}
\item Short-hand notations:
\begin{align}\label{sec:Short-handed-notations}
\begin{aligned}
    &x_{ij}=x_i-x_j \ ,\qquad \Delta_{ij}^\pm =\Delta_i\pm \Delta_j\ , \qquad \bar{\Delta}= d-\Delta\ , \qquad \bar{J} =2-d-J\ , \qquad \tau =\Delta-J\ , \\
    &(\alpha)_\beta=\frac{\Gamma(\alpha+\beta)}{\Gamma(\alpha)} \ , \qquad  \Gamma(x\pm y)=\Gamma(x+y)\,\Gamma(x-y)\ ,\qquad  L_{12,[\Delta,J]}  =-2\pi\i\,\frac{\Gamma(\Delta+J-1)}{\Gamma\left(\frac{\Delta+J\pm\Delta^-_{12}}{2}\right)}\ ,\\
    &\kappa_{\Delta + J}= \frac{\Gamma\left( \frac{\Delta+J\pm\Delta^-_{12}}{2}\right)\,\Gamma\left(\frac{\Delta+J\pm\Delta^-_{43}}{2}\right)}{2\pi^2\,\Gamma(\Delta+J)\, \Gamma(\Delta+J-1)}\ ,\qquad b_{12,[\Delta,J]} 
      = \i\,\frac{(\Delta + J -1)}{2^{J+1}\,\pi}\, L_{12,[\Delta,J]}\ .
    \end{aligned}
\end{align}
\item Inverse Fourier transformation of a primary operator
\begin{align}\label{inv-FT}
    &\CO_{\Delta}^{\mu_1\cdots \mu_J}(x)=\int [\rD^d p]_{\mathrm{L}}\, e^{-\i\, p\cdot x}\,\CO_{\Delta}^{\mu_1\cdots \mu_J}(p)\ , \qquad [\rD^d p]_{\mathrm{L}}\equiv \frac{\d^d p}{(2\pi)^d}\,\Theta(p^0)\,\Theta(-p^2)\ .
    \end{align}
\end{itemize}

\subsection{Normalization}\label{app:Normalization}
We summarize our normalizations of correlation functions in Lorentzian signature with all operators mutually spacelike or in Euclidean signature.

\subsubsection*{Two-point function}
We choose the following normalization for two-point functions:
\begin{align}\label{2-pt-Enb}
&    \langle\,\CO_{\Delta,J}(P_1,Z_1)\,\CO_{\Delta,J}(P_2,Z_2)\,\rangle=\frac{\left[2(P_1\cdot Z_2)(P_2\cdot Z_1)-2(P_1\cdot P_2)(Z_1\cdot Z_2)  \right]^J}{(-2P_1\cdot P_2)^{\Delta+J}}\ .
\end{align}
In physical space, it reduces to 
\begin{align}\label{2-pt-phys-ast}
    &  \langle\,\CO_{\Delta,J}(x_1,z_1)\,\CO_{\Delta,J}(x_2,z_2)\,\rangle=\frac{\left[(z_1\cdot z_2)x^2_{12}-2(z_1\cdot x_{12})(z_2\cdot x_{12})\right]^J}{(x^2_{12})^{\Delta+J}}\ ,
\end{align}
or equivalently
\begin{align}\label{2-pt-phys}
    &  \langle\,\CO_{\Delta,\mu_1\cdots \mu_J}(x_1)\,\CO_{\Delta,\nu_1\cdots \nu_J}(x_2)\,\rangle=\frac{\Pi^{\rho_1\cdots\rho_J}_{\nu_1\cdots\nu_J}\, I_{\mu_1\rho_1}(x_{12})\cdots I_{\mu_J\rho_J}(x_{12}) }{(x^2_{12})^{\Delta}}\ , \qquad I_{\mu}^{\nu}(x)=\delta_{\mu}^{\nu}-2\frac{x^\nu x_\mu}{x^2}\ ,
\end{align}
where $\Pi^{\rho_1\cdots\rho_J}_{\nu_1\cdots\nu_J}$ is the projector onto the rank-$J$ symmetric and traceless subspace \cite{Costa:2011dw}.
This projector has the following property:
\begin{align}\label{Gegenbaur-pol-ST}
    x^{\nu_1}\cdots x^{\nu_J}\, \Pi^{\rho_1\cdots\rho_J}_{\nu_1\cdots\nu_J}\,y_{\rho_1}\cdots y_{\rho_J}
        =c_{d,J}\,(x^2 y^2)^{J/2}\,C_J^{d/2-1}\left(\frac{x\cdot y}{\sqrt{x^2y^2}}\right)\ , \qquad c_{d,J}=\frac{\Gamma(J+1)}{2^J (d/2-1)_J} \ ,
\end{align}
where $C_J^{d/2-1}(x)$ is the Gegenbauer polynomial:
\begin{align}\label{Gegenbaur-pol-asy}
   C_J^{d/2-1}(x)=\frac{\Gamma(J+d-2)}{\Gamma(J+1)\,\Gamma(d-2)}\ 
   {}_2F_1\left[-J,J+d-2,\frac{d-1}{2},\frac{1-x}{2}\right] =\frac{1}{c_{d,J}}x^J + O(x^{J-1})\ .
\end{align}

\subsubsection*{Three-point function}
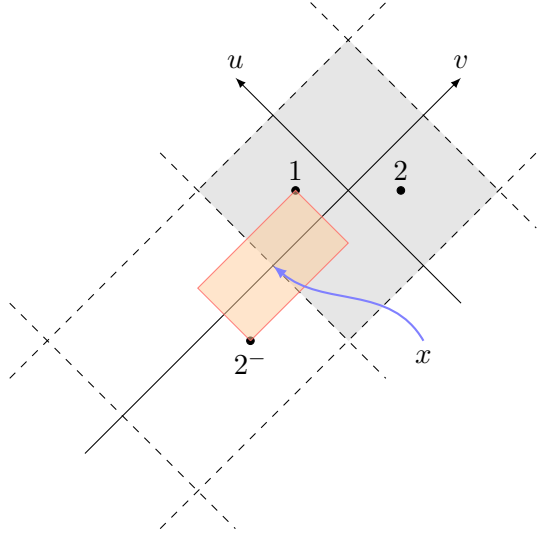
\begin{figure}[ht!]
	\centering
	\begin{tikzpicture}
     \filldraw[gray!20] (0,-2) -- (2,0) -- (0,2) -- (-2,0);
     \draw[dashed,black] (-4.5,-2.5) -- (0.5,2.5);
       \draw[dashed,black] (-2.5,-4.5) -- (2.5,0.5);
    \draw[dashed,black] (-0.5,2.5) -- (2.5,-0.5);
    \draw[dashed,black] (-2.5,0.5) -- (0.5,-2.5);
    \draw[dashed,black] (-4.5,-1.5) -- (-1.5,-4.5);
     \filldraw[black] (0.7,0) circle (0.05);
        \filldraw[black] (-0.7,0) circle (0.05);
          \node[above] at (-0.7,0) {$1$};
          \node[above] at (0.7,0) {$2$};
          \draw[->] (-3.5,-3.5) to (1.5,1.5);
          \node[above] at (1.5,1.5) {$v$};
           \draw[->] (1.5,-1.5) to (-1.5,1.5);
            \node[above] at (-1.5,1.5) {$u$};
            \node[below] at (-1.3,-2) {$2^-$};
            \filldraw[black] (-1.3,-2) circle (0.05);
            \filldraw[orange!40,opacity=0.5]  (-1.3,-2) -- (0,-0.7) -- (-0.7,0) -- (-2,-1.3);
             \draw[red!60,opacity=0.8]  (-1.3,-2) -- (0,-0.7) -- (-0.7,0) -- (-2,-1.3) -- (-1.3,-2);
            \draw[->][thick,blue!50] (1,-2) to[out=120,in=-40] (-1,-1); 
             \node[below] at (1,-2) {$x$};
	\end{tikzpicture}
	\caption{The causal diamond $2^-<x<1$ is shown in orange color.}
	\label{fig:config-of-2^-<x<1}
\end{figure}

We normalize a scalar-scalar-spin-$J$ three-point function to be\footnote{The apparent asymmetry in the interchange of the operators $1$ and $2$ in the right hand side of (\ref{3-pt-Emb}) is compensated by the property of the three point coupling constant: $c_{12,[\Delta,J]}=(-1)^J c_{21,[\Delta,J]}$.}
\begin{align}\label{3-pt-Emb}
    \langle\,\CO_{1}(P_1)\,\CO_{2}(P_2)\,\CO_{\Delta,J}(P_3,Z_3)\,\rangle
        = c_{12,[\Delta,J]}\,\frac{(2P_1\cdot C_3\cdot P_2)^J}{P_{12}^{ \frac{\Delta^+_{12}-\Delta+J}{2} }  P_{13}^{ \frac{\Delta^-_{12}+\Delta+J}{2} }  P_{23}^{ \frac{-\Delta^-_{12}+\Delta+J}{2} }} \ ,
\end{align}
where we introduced the following short-hand notations:
\begin{align}
     P_{i j}=-2P_i\cdot P_j\ , \qquad  C_i^{AB}=Z_i^A P_i^B-P_i^A Z_i^B \ .
\end{align}
In physical space, the equation (\ref{3-pt-Emb}) reduces to
\begin{align}\label{3-pt-phys}
   &  \langle\,\CO_{1}(x_1)\,\CO_{2}(x_2)\,\CO_{\Delta,J}(x_3,z_3) \, \rangle=c_{12,[\Delta,J]}\,\frac{(z_3\cdot x_{13}x_{23}^2 -z_3\cdot x_{23}x_{13}^2 )^J}{(x_{12}^2)^{\frac{\Delta^+_{12}-\Delta+J}{2}}\,(x_{13}^2)^{\frac{\Delta^-_{12}+\Delta+J}{2}} \,  (x_{23}^2)^{\frac{-\Delta^-_{12}+\Delta+J}{2}}}\ , 
   \end{align}
or equivalently
\begin{align}\label{3-pt-phys-ast}
     \langle\,\CO_{1}(x_1)\,\CO_{2}(x_2)\,\CO_{\Delta,\mu_1\cdots \mu_J}(x_3) \, \rangle
        &= c_{12,[\Delta,J]}\,\frac{\Pi^{\nu_1\cdots\nu_J}_{\mu_1\cdots\mu_J}\,H_{\nu_1}(x_{13},x_{12})\cdots H_{\nu_J}(x_{13},x_{12}) }{(x_{12}^2)^{\frac{\Delta^+_{12}-\Delta+J}{2}}\,(x_{13}^2)^{\frac{\Delta^-_{12}+\Delta+J}{2}} \,  (x_{23}^2)^{\frac{-\Delta^-_{12}+\Delta+J}{2}}}\ ,
\end{align}
where
\begin{align}
    H^\mu(x,y)=\frac{x^\mu}{x^2}-\frac{y^\mu}{y^2} \ .
\end{align}
The scalar-scalar-light-ray three-point function in the configuration $2^-<x<1$ (see figure \ref{fig:config-of-2^-<x<1}) behaves in the same way as $\langle\,\CO_{1}\,\CO_{2}\,\CO_{1-J,1-\Delta}\rangle$  up to a normalization factor \cite{Kravchuk:2018htv}:\footnote{
This additional factor $(-2)^{1-J-\Delta}$ arises from the difference of the normalizations of three-point functions:
\begin{align}
     \langle\,\CO_{1}\,\CO_{2}\,\CO_{\Delta,J}\,\rangle\,|_{\mathrm{ours}}=(-2)^{-J}\, \langle\,\CO_{1}\,\CO_{2}\,\CO_{\Delta,J}\rangle\,|_{\mathrm{KS}}\ .
\end{align}
}
\begin{align}\label{scalar2-light-ray-3pt}
    \begin{aligned}
       & \langle\Omega|\,\CO_{1}(P_1)\,\bL[\CO_{\Delta,J}](P,Z)\,\CO_{2}(P_2)\,|\Omega\rangle \\
         &=(-2)^{1-J-\Delta}\,c_{12,[\Delta,J]}\,L_{12,[\Delta,J]}\,\frac{[(-2P_2\cdot P)(Z\cdot P_1)-(-2P_1\cdot P)(Z\cdot P_2)]^{1-\Delta}}{(-2P_1\cdot P_2)^{\frac{\Delta^+_{12}-\Delta+ J}{2}}(2P_1\cdot P)^{\frac{\Delta^-_{12}+2-\Delta- J}{2}}(-2P_2\cdot P)^{\frac{-\Delta^-_{12}+2-\Delta- J}{2}}}\ .
    \end{aligned}
\end{align}
where $L_{12,[\Delta,J]}$ is given in \eqref{sec:Short-handed-notations}.

\subsubsection*{Relation to OPE block}
The normalization of two- and three-point functions (\ref{2-pt-phys-ast}), (\ref{3-pt-phys}) are related to each other via OPE because we can also compute three-point functions by taking the OPE of two out of the three operators:
\begin{align}\label{3p2pOPE}
    \langle\,\CO_{1}(x_1)\,\CO_{2}(x_2)\,\CO_{\Delta,J}(x_3,z_3) \, \rangle
        = c_{12,[\Delta,J]}\,\langle\,\CB_{\Delta,J}(x_1,x_2)\,\CO_{\Delta,J}(x_3,z_3)\,\rangle \ .
\end{align}
Just to check the overall normalization, it is enough to see the leading behavior in the limit $x_1\to x_2$.
Here, to confirm the consistency of our normalization, we assume all components of $x_{12}^\mu$ are small and evaluate both sides of the equation (\ref{3p2pOPE}) up to the leading order in $x_{12}^\mu$.

First, the left hand side (LHS) of (\ref{3p2pOPE}) or (\ref{3-pt-phys}) reduces to
\begin{align}\label{3pt_OPE_limit}
    &(\mathrm{LHS\ of\ (\ref{3p2pOPE})})\xrightarrow[x_{1}\to x_2]{} c_{12,[\Delta,J]}\,\frac{\left[(z_3\cdot x_{12})x_{23}^2 -2(x_{12}\cdot x_{23})(z_3\cdot x_{23}) \right]^J}{(x_{12}^2)^{\frac{\Delta^+_{12}-\Delta+J}{2}}\,x_{23}^{2(\Delta+J)}}\ .
\end{align}
Then, to see the leading contribution of the right hand side (RHS) in (\ref{3p2pOPE}), we need the limiting behavior of $\CB_{\Delta,J}(x_1,x_2)$ in the limit $x_{12}^\mu\to0^\mu$.
The dominant contribution comes from the $l=J$ term in (\ref{Lorentzian_higher_spin_HKLL}), since the $l^\text{th}$ term behaves as $\sim  |x_{12}|^{\Delta+J-l}(1+(\text{sub-leading}))$ in this limit.
In addition to this, $x^\mu(\xi) \to x_2^\mu$ and the leading term (\ref{Sarkar-Xiao_field}) simplifies:
\begin{align}\label{simplify-cont}
    \Phi_{\text{con},\mu_1\cdots\mu_J}\,  \left( x^\mu(\xi), \eta(\xi)\right)          ~\xrightarrow[x_1\to x_2]{}~ \eta^{\Delta-J}(\xi)\,\CO_{\Delta,\mu_1\cdots\mu_J}(x_2)\ .
\end{align}
In deriving the limit, we started with the momentum representation of $\Phi_{\text{con},\, \mu_1\cdots\mu_J}$, evaluated to the leading order in $\eta(\xi)=\sqrt{\xi(1-\xi)x_{12}^2}$ and used the inverse Fourier transformation (\ref{inv-FT}). 
Plugging (\ref{simplify-cont}) into (\ref{Lorentzian_higher_spin_HKLL}) and conducting some calculations, the coincident limit $x_1\to x_2$ of the OPE block reads:
\begin{align}\label{coin-limit-OPE}
    \CB_{\Delta,J}(x_1,x_2)~\xrightarrow[x_1\to x_2]{}~ \frac{1}{(x_{12}^2)^{\frac{\Delta^+_{12}-\Delta+J}{2}}}\,x_{12}^{\mu_1}\cdots x_{12}^{\mu_J}\,\CO_{\Delta,\mu_1\cdots \mu_J}(x_2)\ .
\end{align}
Combining (\ref{coin-limit-OPE}) with (\ref{2-pt-phys-ast}), we find
\begin{align}
\begin{aligned}
    (\mathrm{RHS\ of\ (\ref{3p2pOPE})})
    &\simeq c_{12,[\Delta,J]}\, \frac{1}{(x_{12}^2)^{\frac{\Delta^+_{12}-\Delta+J}{2}}}\,\langle\,\CO_{\Delta,J}(x_2,x_{12})\,\CO_{\Delta,J}(x_3,z)\,\rangle\\
    &   
    = c_{12,[\Delta,J]}\,\frac{\left[(z_3\cdot x_{12})x_{23}^2 -2(x_{12}\cdot x_{23})(z_3\cdot x_{23}) \right]^J}{(x_{12}^2)^{\frac{\Delta^+_{12}-\Delta+J}{2}}x_{23}^{2(\Delta+J)}}\ ,
    \end{aligned}
\end{align}
which precisely matches with the OPE limit of the three-point function \eqref{3pt_OPE_limit}, verifying the consistency of our normalization.

\subsection*{Four-point function and conformal block expansion}
Consider a Lorentzian four-point function with all operators are spacelike separated:
\begin{align}
    & \langle\, \CO_{1}(x_1)\,\CO_{2}(x_2)\,\CO_{3}(x_3)\,\CO_{4}(x_4)\, \rangle \ .
\end{align}
By using the OPE twice: $x_1\to x_2, x_4\to x_3$ and the orthogonality of two-point functions, we obtain the conformal block expansion of the four point function:
\begin{align}\label{cpwe-L-spacelike}
\begin{aligned}
    g(z,\bar{z})&=\sum_{[\Delta,J]}c_{12,[\Delta,J]}\, c_{43,[\Delta,J]}\,G_{\Delta,J}(z,\bar{z})  \ , \\
    G_{\Delta,J}(z,\bar{z})&=\frac{1}{T_{\{\Delta_i\}}(x_i)}\,\langle\,\CB_{\Delta,J}(x_1,x_2)\,\CB_{\Delta,J}(x_4,x_3)\,\rangle\ .
    \end{aligned}
\end{align}
In Euclidean signature $z$ and $\bar{z}$ are complex conjugate with each other. On the other hand, in Lorentzian signature, $z$ and $\bar{z}$ are real and independent of each other.
We denote Lorentzian four-point function and conformal block in the same way as in Euclidean when all the four operators are spacelike-separated.

We can fix the normalization of the conformal block by taking the double coincident limit $x_1\to x_2,x_4\to x_3$:
\begin{align}\label{cpwe-asym-1}
    \begin{aligned}
    G_{\Delta,J}&(z,\bar{z})\\
        &\simeq \frac{x_{12}^{\Delta_{12}^+}\,x_{34}^{\Delta_{43}^+}}{\left( x_{24}/x_{14}\right)^{\Delta_{12}^-}\left( x_{13}/x_{14}\right)^{\Delta_{43}^-} }
        \frac{1}{x_{12}^{\Delta^+_{12}-\tau}x_{43}^{\Delta^+_{43}-\tau}}\,
        x_{12}^{\mu_1}\cdots x_{12}^{\mu_J}\,
        x_{43}^{\nu_1}\cdots x_{43}^{\nu_J}\,
        \langle\,\CO_{\Delta,\mu_1\cdots \mu_J}(x_2)\,\CO_{\Delta,\nu_1\cdots \nu_J}(x_3)\,\rangle \\
        &\simeq x_{12}^{\tau}\,x_{43}^{\tau}\times x_{12}^{\mu_1}\cdots x_{12}^{\mu_J}\,
        x_{43}^{\nu_1}\cdots x_{43}^{\nu_J}\,
        \frac{\Pi^{\rho_1\cdots\rho_J}_{\nu_1\cdots\nu_J}\, I_{\mu_1\rho_1}(x_{23})\cdots I_{\mu_J\rho_J}(x_{23}) }{(x^2_{23})^{\Delta}}\\
        &=c_{d,J}\, \frac{x_{12}^{\Delta}\,x_{43}^{\Delta}}{x_{23}^{2\Delta}}\, C^{d/2-1}_J\left(\frac{x_{12}\cdot I(x_{23})\cdot x_{43}}{\sqrt{x_{12}^2\, x_{43}^2}}\right)\ ,
    \end{aligned}
\end{align}
where we used (\ref{coin-limit-OPE}), (\ref{2-pt-phys}) and (\ref{Gegenbaur-pol-ST}) successively.

In the limit, the cross ratios (\ref{cross_ratio}) behaves as:
\begin{align}
    \mathfrak{u}=z\,\bar{z}\simeq \frac{x_{12}^2x_{43}^2}{x_{23}^4}\to 0\ , \qquad 1-\mathfrak{v}\simeq z+\bar{z}\simeq 2\,\frac{x_{12}\cdot I(x_{23})\cdot x_{43}}{x_{23}^2}\ .
\end{align}
Thus, (\ref{cpwe-asym-1}) can be written in terms of the cross ratios:
\begin{align}
    G_{\Delta,J}(z,\bar{z})=(z\bar{z})^{\frac{\Delta}{2}}\left[ c_{d,J}\, C^{d/2-1}_J\left(\frac{z+\bar{z}}{2\sqrt{z\bar{z}}}\right)+O(z\bar{z})\right]\ , \qquad c_{d,J}=\frac{\Gamma(J+1)}{2^J (d/2-1)_J} \ .
\end{align}
In particular, when $0\ll z\ll\bar{z}\ll 1$ the asymptotic form of the conformal block can be read from (\ref{Gegenbaur-pol-asy}):
\begin{align}\label{bdy-Casimir}
   G_{\Delta,J}(z,\bar{z})\to 2^{-J}z^{\frac{\Delta-J}{2}}\bar{z}^{\frac{\Delta+J}{2}}\ ,\qquad 0\ll z\ll\bar{z}\ll 1 \ .
\end{align}

\subsubsection*{Conformal Casimir equation}
The conformal block $G_{\Delta,J}(z,\bar{z})$ satisfies the conformal Casimir equation of the form:
\begin{align}\label{casimir-eq}
    &\mathcal{D}_2\, G_{\Delta,J}(z,\bar{z})=c_2(\Delta,J)\,G_{\Delta,J}(z,\bar{z})\ , \qquad c_2(\Delta,J)=\frac{1}{2}\left[\Delta(\Delta-d)+J(J+d-2) \right]\ ,
\end{align}
where $\CD_2$ is the second order differential \cite{Dolan:2003hv} defined by
\begin{align}
    \begin{aligned}
    \mathcal{D}_2
        &=z^2(1-z)\partial_z^2-\left(1-\frac{\Delta^-_{12}+\Delta^-_{43}}{2}\right)z^2\partial_z -\frac{\Delta^-_{12}\,\Delta^-_{43}}{4}z +(z\leftrightarrow\bar{z})\\
        &\qquad +(d-2)\frac{z\bar{z}}{z-\bar{z}}\left[(1-z)\partial_{z}-(1-\bar{z})\partial_{\bar{z}}\right]  \ .
    \end{aligned}
\end{align}
This holds true regardless of the spacetime signature.

We can use (\ref{bdy-Casimir}) as a boundary condition (or normalization) for this differential equation in Lorentzian signature with all operators spacelike-separated.

\subsubsection*{Lightcone limit}
Combining (\ref{bdy-Casimir}) and (\ref{casimir-eq}), we can solve the lightcone limit ($z\to0$) of the conformal block to all order in $\bar{z}$:
\begin{align}
    &\lim_{z\to0}G_{\Delta,J} (z, \bar z)=2^{-J}z^{\frac{\Delta-J}{2}}\bar{z}^{\frac{\Delta+J}{2}}\, {}_2F_1\left[\frac{\Delta+J-\Delta^-_{12}}{2},\frac{\Delta+J-\Delta^-_{43}}{2},\Delta+J,\bar{z}\right]\ .\label{lightcone}
\end{align}

\section{Wightman functions with different operator orderings
}\label{sec:Analytic continuation of the second and third ordering}
In the second (third) ordering of \eqref{Wightman_orderings}, the naive $\i\,\epsilon$ prescription is
\begin{align}
     t_{41} \to t_{41}\mp \i\, \epsilon \ , \qquad t_{23} \to t_{23} \pm \i\,\epsilon \,     \qquad\qquad (\epsilon >0) \ , \label{wrong-shifts-1}
 \end{align}
which yields shifts in $\rho,\bar{\rho}$,
\begin{align}\label{wrong-shifts-2}
     \rho\to\rho\pm \i\, \epsilon \ , \qquad \bar{\rho}\to\bar{\rho}\pm \i\, \epsilon \qquad\qquad (\epsilon >0) \ .
\end{align}
The resulting correlator, however, is no longer of Euclidean type as the cross ratios $z$ and $\bar{z}$ given by \eqref{cross_ratio} are not complex conjugate to each other.
To remedy the situation, we use the two-to-one correspondence between $\bar{\rho}$ and $\bar{z}$ (\ref{cross_ratio}).

\begin{figure}[ht!]
	\centering
	\begin{tikzpicture}
        \draw[] (2.05,1.4) -- (2.05,1.1) -- (2.73,1.1);
        \draw[fill=black] (0,0) circle (0.07);
        \draw[fill=black] (3,0) circle (0.07);
        \filldraw[gray] (1.8,-0.5) circle (0.05);
        \filldraw[gray] (1.8,0.5) circle (0.05);
        \draw[->][orange!70] (1.75,0.5) to[out=180,in=60] (0.3,0.03);
        \draw[->][blue!70] (1.75,-0.5) to[out=180,in=-70] (0.7,-0.03);
        \draw[->] (-1,0) to (4.5,0);
        \draw[->] (0,-1.5) to (0,1.5);
        \draw[double distance=0.05pt,green!50!black!80] (3.1,0) to (4.33,0);
	    \node[above] at (2.4,1) {$z,\bar{z}$}; 
	    \node at (-0.2,-0.2) {$0$};
	    \node[below] at (2.85,0) {$1$};
	    \node[below] at (1.8,-0.5) {$\bar{z}$};
	    \node[above] at (1.8,0.5) {$z$};
	
	    \draw[fill=black] (7,0) circle (0.07);
	    \node at (6.8,-0.2) {$0$};
	    \draw[dashed,green!50!black!80, thick] (7,0) circle (1.3);
	    \draw[->] (5,0) to (11,0);
	    \draw[->] (7,-1.6) to (7,1.6);
	    \node[above] at (8.4,0) {$1$};
	    \draw[] (10.05,1.4) -- (10.05,1.1) -- (10.73,1.1);
	    \node[above] at (10.4,1) {$\rho,\bar{\rho}$};
        \node[below] at (7.8,-0.5) {$\bar{\rho}$};
	    \node[above] at (7.8,0.5) {$\rho$};
	    \draw[->][orange!70] (7.75,0.5) to[out=190,in=60] (7.3,0.03);
	    \draw[->][blue!70] (7.85,-0.5) to[out=20,in=-90] (8,-0.03);
	    \filldraw[gray] (7.8,-0.5) circle (0.05);
        \filldraw[gray] (7.8,0.5) circle (0.05);
	\end{tikzpicture}
	\caption{These two figures illustrate the paths of $(z,\bar{z})$ and $(\rho,\bar{\rho})$ in the second ordering of \eqref{Wightman_orderings} during the analytic continuation to $\bar{\rho}>1$ regime. In this procedure, both $\rho$ and $\bar{\rho}$ remain in the unit sphere. Thus
	neither $z$ nor $\bar{z}$ crosses the branch cut on positive real half line $[1,\infty)$ anymore.
	The final value of $\bar{\rho}$ is the inverse of that in figure \ref{fig:z-1-crossratios}, whereas the final configurations of $z$, $\bar{z}$ and $\rho$ are the same as those in figure \ref{fig:z-1-crossratios}. That is because of the two-to-one correspondence between $\bar{\rho}$ and $\bar{z}$: $\bar{z}(1/\bar{\rho})=\bar{z}(\bar{\rho})$.}
	\label{fig:z-2-crossratios}
\end{figure}
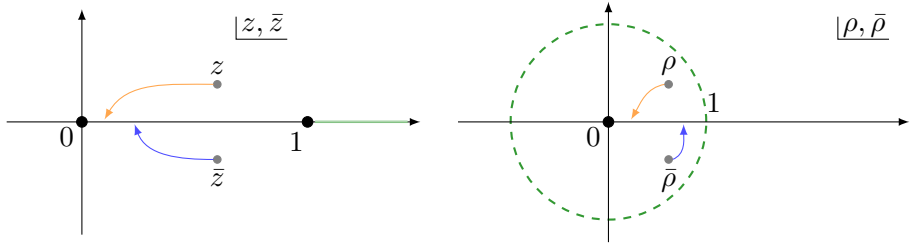

First, we consider the following analytic continuation for $\rho$ and $\bar{\rho}$\,: 
\begin{align}\label{correct-indirect}
     \rho \to r\, e^{-t_{\mathrm{R}}}  \pm \i\,\epsilon\ , \qquad \bar{\rho} \to  \frac{1}{r}\,e^{-t_{\mathrm{R}}} \mp \i\,\epsilon\ .
\end{align}
The cross ratios $z,\bar{z}$ transform as
\begin{align}
     &z(\rho)=\frac{4\,\rho}{(1+\rho)^2}=\frac{4\,r\, e^{t_{\mathrm{R}}}\pm\i\,\epsilon}{(1+r\, e^{t_{\mathrm{R}}}\pm\i\,\epsilon)^2}\ , \qquad \bar{z}(\bar{\rho})=\frac{4\,\bar{\rho}}{(1+\bar{\rho})^2}=\frac{4\, e^{-t_{\mathrm{R}}}/r\mp\i\,\epsilon}{(1+ e^{-t_{\mathrm{R}}}/r\mp\i\,\epsilon)^2} \ .
\end{align}
In the configuration of our interest ($0<r\, e^{-t_{\mathrm{R}}}< e^{-t_{\mathrm{R}}}/r <1$), neither $z$ nor $\bar{z}$ crosses the branch cut on $[1,\infty)$, thus the correlators so obtained are identical to the Euclidean correlator as a function of $z$ and $\bar{z}$ (see figure \ref{fig:z-2-crossratios}).

Next, by using the identity $\bar{z}(\bar{\rho})=\bar{z}(1/\bar{\rho})$,
\begin{align}
     \bar{z}(1/\bar{\rho})=\bar{z}(\bar{\rho})=\frac{4/\bar{\rho}}{(1+1/\bar{\rho})^2}=\frac{4\,r\, e^{t_{\mathrm{R}}}\mp \i\,\epsilon}{(1+r\, e^{t_{\mathrm{R}}}\mp\i\,\epsilon)^2} \ ,
\end{align}
we can see that $\bar{z}$ transforms exactly in the same way as the naive $\i \, \epsilon$-prescription we conducted before in (\ref{wrong-shifts-2}).
Therefore we conclude that, from the viewpoint of the cross ratio space, the conformal blocks analytically continued in the procedures (\ref{correct-indirect}) are equivalent to those we want, and the correlator is given by the Euclidean correlator itself.

\section{Derivations of Regge conformal block}
\label{sec:Regge-deriv}
In this appendix, we derive the Regge behavior of conformal blocks $G_{\Delta, J}^{\circlearrowleft,(\circlearrowright)}(z,\bar{z})$ by evaluating the asymptotic behavior following \cite{Caron-Huot:2017vep}. (This method is ensured by the integrability of conformal blocks \cite{Isachenkov:2017qgn}.)
Note that $G^{\circlearrowleft}_{\Delta,J}$ is complex conjugate of $G^{\circlearrowright}_{\Delta,J}$ so we only consider the former.

In this appendix, we use some short-handed notations:
\begin{align}\label{short-handed}
    h=\frac{\tau}{2}=\frac{\Delta-J}{2}\ ,\qquad  {\bar{h}}=\frac{\tau}{2}+J=\frac{\Delta+J}{2}\ , \qquad \blacktriangle^{\pm}_{i j}=\frac{\Delta^{\pm}_{i j}}{2}=\frac{\Delta_i \pm \Delta_j}{2}\ .
\end{align}
For example, with this notation the lightcone conformal block (\ref{lightcone}) reads
\begin{align}
    2^J\,\lim_{z\to 0}G_{\Delta,J}(z,\bar{z})=z^{h}\,\bar{z}^{{\bar{h}}}\,{}_2 F_1({\bar{h}}-\blacktriangle^{-}_{12},{\bar{h}}-\blacktriangle^{-}_{43},2{\bar{h}},\bar{z})\ .
\end{align}
The useful formula is in (15.3.6) of \cite{Abramowitz:1964}:\footnote{This formula (\ref{form-1}) is valid for limited case $a+b-c\neq 1,2,3,\cdots$. However, if we take the positive integer limit of $(a+b-c)$ we can healthfully obtain the correct results.}
\begin{align}\label{form-1} 
\begin{aligned}
   {}_2 F_1(a,b,c,z) &= \frac{\Gamma(c)\, \Gamma( c-a-b)}{\Gamma( c-a)\,\Gamma(c- b)}\, {}_2 F_1(a,b,a+b-c+1,1-z)\\
   &+(1-z)^{c-a-b}\,\frac{\Gamma( c)\, \Gamma(a+b-c)}{\Gamma( a)\,\Gamma(b)}\, {}_2 F_1(c-a,c-b,c-a-b+1,1-z)  \ .
   \end{aligned}
\end{align}
First, we start with the lightcone conformal block (\ref{lightcone}). The analytically continued lightcone block takes the form: 
\begin{align*}
&   2^J \lim_{z\to 0}G^{\circlearrowleft}_{\Delta, J} (z, \bar z)= z^{h}\,\bar{z}^{{\bar{h}}}\,{}_2 F_1({\bar{h}}-\blacktriangle^{-}_{12},{\bar{h}}-\blacktriangle^{-}_{43},2{\bar{h}},\bar{z})\,|_{(1-\bar
{z})\to e^{2\i\pi }(1-\bar{z})}\\
&=2^J\lim_{z\to 0}G_{\Delta, J} (z, \bar z)+2\,\i\, \sin\left[\pi\left(\blacktriangle^-_{12}+\blacktriangle^-_{43}\right)\right]\,   e^{\i\pi \left(\blacktriangle^-_{12}+\blacktriangle^-_{43}\right)}\,(1-\bar{z})^{\blacktriangle^-_{12}+\blacktriangle^-_{43}}\,\frac{\Gamma( 2{\bar{h}})\, \Gamma( -\blacktriangle^-_{12}-\blacktriangle^-_{43})}{\Gamma( {\bar{h}}-\blacktriangle^-_{12})\,\Gamma({\bar{h}}-\blacktriangle^-_{43})}\notag\\
&\qquad\qquad\qquad\qquad\qquad\times z^{h}\,\bar{z}^{{\bar{h}}}\,{}_2 F_1\left({\bar{h}}+\blacktriangle^-_{12},{\bar{h}}+\blacktriangle^-_{43},\blacktriangle^-_{12}+\blacktriangle^-_{43}+1,1-\bar{z}\right)\ ,
\end{align*}
where we used (\ref{form-1}) by setting $(a,b,c)\to\left({\bar{h}}-\blacktriangle^-_{12},{\bar{h}}-\blacktriangle^-_{43},2{\bar{h}}\right)$.

The leading behavior for small $\bar{z}$ is:
\begin{align*}
    & 2^J \lim_{\bar{z}\to 0}\lim_{z\to 0}G^{\circlearrowleft}_{\Delta, J} (z, \bar z)=2\, \i\, z^{h}\,\bar{z}^{{\bar{h}}}\, \bar{z}^{1-2{\bar{h}}} \notag\\
    &\times \sin\left[\pi\left(\blacktriangle^-_{12}+\blacktriangle^-_{43}\right)\right] \,  e^{\i\pi \left(\blacktriangle^-_{12}+\blacktriangle^-_{43}\right)}\,\frac{\Gamma( 2{\bar{h}})\, \Gamma(2{\bar{h}}-1)\,\Gamma( -\blacktriangle^-_{12}-\blacktriangle^-_{43})\,\Gamma(\blacktriangle^-_{12}+\blacktriangle^-_{43}+1)}{\Gamma( {\bar{h}}\pm\blacktriangle^-_{12})\,\Gamma({\bar{h}}\pm\blacktriangle^-_{43})}\ ,
\end{align*}
where, in taking small $\bar{z}$ limit, we used (\ref{form-1}) with $z$ and $ (1-z)$ interchanged and the replacement $(a,b,c)\to\left({\bar{h}}+\blacktriangle^-_{12},{\bar{h}}+\blacktriangle^-_{43},\blacktriangle^-_{12}+\blacktriangle^-_{43}+1\right)$.
Hence, by the use of the Euler reflection formula $\Gamma(z)\,\Gamma(1-z)=\pi/\sin(\pi z)$, the Regge conformal block in the limit $z,\bar{z}\to0,\bar{z}>z$ reads:
\begin{align} \label{sub-Regge-conf-block}
\begin{aligned}
    &2^J  \lim_{\bar{z}\to 0}\lim_{z\to 0}G^{\circlearrowleft}_{\Delta, J} (z, \bar z)= -\frac{\i}{\pi}\frac{e^{ \i\pi(\blacktriangle^-_{12}+\blacktriangle^-_{43})}}{\kappa_{\Delta + J}}\, z^{h}\,\bar{z}^{1-{\bar{h}}} \ ,  \qquad \kappa_{\Delta + J}= \frac{\Gamma( {\bar{h}}\pm\blacktriangle^-_{12})\,\Gamma({\bar{h}}\pm\blacktriangle^-_{43})}{2\pi^2\,\Gamma( 2{\bar{h}})\, \Gamma(2{\bar{h}}-1)}\ .
\end{aligned}
\end{align}

\subsection{Sub-leading terms in $z/\bar{z}$}
Here we want to seek sub-leading contributions in $z/\bar{z}$, which are very important because the Regge limit is the limit where $z,\bar{z}\to0$ keeping $z/\bar{z} <1$ finite.

It is evident that the Regge conformal block must satisfy the Casimir equation as well. 
The Casimir differential operator (\ref{casimir-eq}) simplifies in the Regge limit:
\begin{align}\label{cas-Regge}
    &\mathcal{D}_2\to z^2\partial_z^2+\bar{z}^2\partial_{\bar{z}}^2 +(d-2)\,\frac{z\bar{z}}{z-\bar{z}}\,(\partial_z-\partial_{\bar{z}})\ \ \ \ \ \ \mathrm{as}\ \  \ \ z,\bar{z}\to 0\ \ \ \  \ \mathrm{keeping} \ \ \frac{z}{\bar{z}}<1\ \ \  \mathrm{fixed} \ . 
\end{align}
Let us assume that the Regge conformal block takes the form:
\begin{align}\label{anz-re-bl}
    &G^{\circlearrowleft}_{\Delta,J} (z, \bar z)=-\frac{\i}{\pi}\frac{e^{ \i\pi(\blacktriangle^-_{12}+\blacktriangle^-_{43})}}{\kappa_{\Delta + J}}\, 2^{-J}\,z^{h}\,\bar{z}^{1-{\bar{h}}}\,\mathbf{g}\left(\frac{z}{\bar{z}}\right)\ ,
\end{align}
where $\mathbf{g}(x)$ is some polynomial in $x$ normalized as $\mathbf{g}(x=0)=1$.

By plugging (\ref{cas-Regge}) and (\ref{anz-re-bl}) into (\ref{casimir-eq}) and solving the differential equation in $\mathbf{g}\left(z/\bar{z}\right)$ so obtained with the boundary condition; $\mathbf{g}(x=0)=1$,
we finally get the sub-leading behavior of the Regge conformal block:
\begin{align}\label{Regge-confpw}
    G^{\circlearrowleft}_{\Delta,J} (z, \bar z)
        =-\frac{\i}{\pi}\frac{e^{ \frac{\i\pi}{2}(\Delta^-_{12}+\Delta^-_{43})}}{\kappa_{\Delta + J}}\, 2^{-J}\,z^{\frac{\Delta-J}{2}}\,\bar{z}^{1-\frac{\Delta+J}{2}}\,{}_2 F_1\left[\frac{d}{2}-1,\Delta-1,\Delta+1-\frac{d}{2},\frac{z}{\bar{z}}\right]\ .
\end{align}
Rewriting \eqref{Regge-confpw} in terms of $r$ and $t_{\mathrm{R}}$ in the Regge limit
given by \eqref{Regge_polar} and \eqref{Regge_limit},\footnote{In the Regge limit, the cross ratios behave as $z\simeq 4\,r\, e^{t_{\mathrm{R}}},\bar{z}\simeq 4\, e^{-t_{\mathrm{R}}}/r$.
Thus, we have:
\begin{align}\label{Regge-crossR-2}
    z\,\bar{z}\simeq 16\, e^{-2t_{\mathrm{R}}}\ , \qquad \frac{z}{\bar{z}}\simeq  r^2\ .
\end{align}
} then we have
\begin{align}\label{Regge-confpw-ast}
    G^{\circlearrowleft}_{\Delta,J} (z, \bar z)
        =-\frac{\i}{\pi}\frac{e^{ \frac{\i\pi}{2}(\Delta^-_{12}+\Delta^-_{43})}}{\kappa_{\Delta + J}}\, 2^{2-3J}\, e^{(J-1)t_{\mathrm{R}}}\,r^{\Delta-1}\, {}_2 F_1\left[\frac{d}{2}-1,\Delta-1,\Delta+1-\frac{d}{2},r^2\right]\ .
\end{align}

\subsection{Regge conformal block}
We can obtain the full asymptotic behavior by doing similar calculations as below \eqref{form-1}. We find that,
\begin{align*}
\begin{aligned}
    &2^J \lim_{z\to 0}G^{\circlearrowleft}_{\Delta, J} (z, \bar z)\\
    &=z^{h}\,\bar{z}^{{\bar{h}}}\, {}_2F_1({\bar{h}}-\blacktriangle^-_{12},{\bar{h}}-\blacktriangle^-_{43},2{\bar{h}},\bar{z})\,\left[ 1-2\, \i\,  e^{ \i\pi(\blacktriangle^-_{12}+\blacktriangle^-_{43})}\, \frac{\sin\left[\pi\left({\bar{h}}-\blacktriangle^-_{12}\right)\right]\sin\left[\pi\left({\bar{h}}-\blacktriangle^-_{43}\right)\right]}{\sin(2\pi{\bar{h}})} \right]\\
   &\qquad-2\,\i\,\pi\, e^{ \i\pi(\blacktriangle^-_{12}+\blacktriangle^-_{43})}\,z^{h}\,\bar{z}^{1-{\bar{h}}} \, \frac{\Gamma( 2{\bar{h}})\, \Gamma(2{\bar{h}}-1)}{\Gamma( {\bar{h}}\pm\blacktriangle^-_{12})\,\Gamma({\bar{h}}\pm\blacktriangle^-_{43})}
   \times {}_2 F_1\left(1+\blacktriangle^-_{43}-{\bar{h}},1+\blacktriangle^-_{12}-{\bar{h}},2-2{\bar{h}},\bar{z}\right)\ .
\end{aligned}
\end{align*}
Comparing this with (\ref{lightcone}) and (\ref{short-handed}), we finally obtain
\begin{align}
    2^J \lim_{z\to 0}G^{\circlearrowleft}_{\Delta, J} (z, \bar z)=&2^J\lim_{z\to 0}G_{\Delta, J} (z, \bar z)\left[ 1-2\, \i\,  e^{ \i\pi(\blacktriangle^-_{12}+\blacktriangle^-_{43})}  \frac{\sin\left[\pi\left({\bar{h}}-\blacktriangle^-_{12}\right)\right]\sin\left[\pi\left({\bar{h}}-\blacktriangle^-_{43}\right)\right]}{\sin(2\pi{\bar{h}})} \right]\notag\\
    &- \frac{\i}{\pi}\frac{ e^{ \i\pi(\blacktriangle^-_{12}+\blacktriangle^-_{43})}}{\kappa_{\Delta + J}}  \, 2^{1-\Delta}\,\lim_{z\to 0}G_{1-\Delta,1- J} (z, \bar z) \ . \qquad \label{monodromy-1}
\end{align}
This formula is identical\footnote{Note that the conformal block in \cite{Caron-Huot:2017vep} is normalized in a different way from ours (\ref{bdy-Casimir}):
\begin{align}
    G_{\Delta, J} (z, \bar z)\,|_{\mathrm{CH}} =2^J G_{\Delta, J} (z, \bar z)\,|_{\mathrm{ours}}\ .
\end{align}} to the equation (A.22) 
in \cite{Caron-Huot:2017vep} and holds even if we drop $\lim_{z\to 0}$ in (\ref{monodromy-1}):
\begin{align}\label{Full-ReggeCPW}
\begin{aligned}
    G^{\circlearrowleft}_{\Delta, J} (z, \bar z)&= G_{\Delta, J} (z, \bar z)\left[ 1-2\, \i\,  e^{ \frac{\i\pi}{2}(\Delta^-_{12}+\Delta^-_{43})}\,  \frac{\sin\left[\pi\left(\frac{\Delta+J-\Delta^-_{12}}{2}\right)\right]\sin\left[\pi\left(\frac{\Delta+J-\Delta^-_{43}}{2}\right)\right]}{\sin\left[\pi\left(\Delta+J\right)\right]} \right]\\
    &- \frac{\i}{\pi}\frac{ e^{\frac{\i\pi}{2}(\Delta^-_{12}+\Delta^-_{43})}}{\kappa_{\Delta + J}}\, 2^{1-\Delta-J} \, G_{1-\Delta,1- J} (z, \bar z)\ .
    \end{aligned}
\end{align}

\section{Timelike OPE blocks with continuous spin}
\label{sec:continuous spin OPE block}
In this appendix we generalize \eqref{timelikeOPE-SP} for continuous spin.
Our starting point is the useful formula:
\begin{align}
    f_{\mu_1\cdots \mu_J}(x)\, \Pi^{\mu_1\cdots \mu_J,\nu_1\cdots \nu_J}\, g_{\nu_1\cdots \nu_J}(y)=c_{J}\, \int D^{d-2}z_1\int D^{d-2}z_2\, f(x,z_1)\,g(y,z_2)\,(z_1\cdot z_2)^{2-d-J} \ ,
\end{align}
where the coefficient $c_J$ is given by
\begin{align}
    c_J =(-2)^{2-d}\,\pi^{1-d}\, \left(J+\frac{d}{2}-1\right)\,\sin\left[\pi\left(J+\frac{d}{2}\right)\right]\,\Gamma(-J)\,\Gamma(J+d-2) \ .
\end{align}

By the use of this formula, the integrand of the last line of \eqref{timelikeOPE-SP} can be written as
\begin{align}
    \begin{aligned}
     &(-1)^J \,c_J\,(-2)^{d-2}\int D^{d-2}z_1\int D^{d-2}z_2\, \frac{ (-2z_1\cdot z_2)^{\bar J}\,  (-2H (x_{10},x_{20})\cdot z_1)^J \, \CO_{\Delta,J}(x_0,z_2)}{|x_{12}^2|^{\frac{\Delta^+_{12}-\bar\Delta+J}{2}}\,|x_{10}^2|^{\frac{\Delta^-_{12}+\bar\Delta-J}{2}}\,|x_{20}^2|^{\frac{-\Delta^-_{12}+\bar\Delta-J}{2}}}\\
     &=(-1)^J\, c_J\,(-2)^{d-2}\,\frac{\pi^{\frac{d-2}{2}}\,\Gamma\left(-J-\frac{d}{2}+1\right)}{\Gamma(-J)}\\
     &\qquad\qquad\qquad \times\int D^{d-2}z_2\, \frac{ (-H^2(x_{10},x_{20}))^{\frac{J-\bar{J}}{2}}\,  (-2H (x_{10},x_{20})\cdot z_2)^{\bar J}\, \CO_{\Delta,J}(x_0,z_2)}{|x_{12}^2|^{\frac{\Delta^+_{12}-\bar\Delta+J}{2}}\,|x_{10}^2|^{\frac{\Delta^-_{12}+\bar\Delta-J}{2}}\,|x_{20}^2|^{\frac{-\Delta^-_{12}+\bar\Delta-J}{2}}}\\
     &= \frac{(-2)^{\bar{J}}\,\Gamma(J+d-2)}{\pi^{\frac{d}{2}-1}\,\Gamma\left(J+\frac{d}{2}-1\right)}\int D^{d-2} z_2\,\frac{ (-H (x_{10},x_{20})\cdot z_2)^{\bar{J}} \, \CO_{\Delta,J}(x_0,z_2)}{|x_{12}^2|^{\frac{\Delta^+_{12}-\bar{\Delta}+\bar{J}}{2}}\,|x_{10}^2|^{\frac{\Delta^-_{12}+\bar{\Delta}-\bar{J}}{2}}\,|x_{20}^2|^{\frac{-\Delta^-_{12}+\bar{\Delta}-\bar{J}}{2}}} \ ,
    \end{aligned}
\end{align}
where we used the conformal integral (2.39) in \cite{SimmonsDuffin:2012uy} with $(\Delta,d)$ replaced with $(-J,d-2)$ and 
\begin{align}
    \begin{aligned}
        &-H^2(x_{10},x_{20})=-\left(\frac{x_{10}^\mu}{x^2_{10}}-\frac{x_{20}^\mu}{x^2_{20}}\right)^2=\frac{|x_{12}^2|}{|x_{10}^2||x_{20}^2|}\ .
    \end{aligned}
\end{align}
Notably the integrand in the last line is proportional to the normalized scalar-scalar-full shadow $(\bar{\Delta},\bar{J})$ three-point structure. In the end, we obtain
\begin{align} \label{rewriting the tensor cont. to conformal int.}
    \begin{aligned}
         &\int_{x_0\in \diamondsuit_{12}}\d^d x_{0}\,  \langle \tilde{0} |\, \CO_{1}(x_1)\,\CO_{\bar{\Delta},\mu_1\cdots \mu_J}(x_0)\,\CO_{2}(x_2)\, |\tilde{0}\rangle\, \CO_{\Delta}^{\mu_1\cdots \mu_J}(x_0)\\
         &=\frac{(-2)^{\bar{J}}\,\Gamma(-\bar{J})}{\pi^{\frac{d}{2}-1}\,\Gamma\left(J+\frac{d}{2}-1\right)}\, \int_{x_0\in \diamondsuit_{12}}\d^d x_{0}\,\int D^{d-2} z\,  \langle \tilde{0} |\, \CO_{1}(x_1)\,\CO_{\bar{\Delta},\bar{J}}(x_0,z)\,\CO_{2}(x_2)\, |\tilde{0}\rangle\, \CO_{\Delta,J}(x_0,z)\ .
    \end{aligned}
\end{align}
This resulting expression is analytic in $J$, hence valid for continuous spin $J$.

\bibliographystyle{JHEP}
\bibliography{Regge_OPEB}

\end{document}